\documentclass[letter,fleqn,usenatbib]{mnras}

\usepackage[T1]{fontenc}
\usepackage{ae,aecompl}
\usepackage{natbib}

\usepackage{graphicx}	
\usepackage{amsmath}	
\usepackage{amssymb}	

\newcommand{\msun}{~M$_{\odot}$}

\title[Clumpy galaxy in the FIRE simulations]{Giant clumps in the FIRE simulations: a case study of a massive high-redshift galaxy}

\author[A. Oklop\v ci\'c et al.]{
Antonija Oklop\v ci\'c,$^{1}$\thanks{E-mail: oklopcic@astro.caltech.edu} 
Philip F. Hopkins,$^{1}$ 
Robert Feldmann$^{2}$,
Du\v san Kere\v s$^{3}$,
\newauthor Claude-Andr\'e Faucher-Gigu{\`e}re$^{4}$ and Norman Murray$^{5}$
\\
$^{1}$California Institute of Technology, MC 350-17, 1200 East California Boulevard, Pasadena, California 91125, USA\\
$^{2}$Department of Astronomy and Theoretical Astrophysics Center, University of California Berkeley, Berkeley, CA 94720, USA\\
$^{3}$Department of Physics, Center for Astrophysics and Space Science, University of California at San Diego, 9500 Gilman Drive,\\ La Jolla, CA 92093, USA\\
$^{4}$Department of Physics and Astronomy and CIERA, Northwestern University, 2145 Sheridan Road, Evanston, IL 60208, USA\\
$^{5}$ Canadian Institute for Theoretical Astrophysics, 60 St George Street, University of Toronto, ON M5S 3H8, Canada
}

\date{Accepted XXX. Received YYY; in original form ZZZ}

\pubyear{2016}

\begin{document}
\label{firstpage}
\pagerange{\pageref{firstpage}--\pageref{lastpage}}
\maketitle

\begin{abstract}
The morphology of massive star-forming galaxies at high redshift is often dominated by giant clumps of mass $\sim 10^8-10^9$\msun\ and size $\sim 100-1000$~pc. Previous studies have proposed that giant clumps might have an important role in the evolution of their host galaxy, particularly in building the central bulge. However, this depends on whether clumps live long enough to migrate from their original location in the disc or whether they get disrupted by their own stellar feedback before reaching the centre of the galaxy. We use cosmological hydrodynamical simulations from the FIRE (Feedback in Realistic Environments) project that implement explicit treatments of stellar feedback and ISM physics to study the properties of these clumps. We follow the evolution of giant clumps in a massive ($M_*\sim 10^{10.8}$\msun\ at $z=1$), discy, gas-rich galaxy from redshift $z \gtrsim 2$ to $z=1$. Even though the clumpy phase of this galaxy lasts over a gigayear, individual gas clumps are short-lived, with mean lifetime of massive clumps of $\sim 20$~Myr. During that time, they turn between $0.1\%$ and $20\%$ of their gas into stars before being disrupted, similar to local GMCs. Clumps with $M\gtrsim 10^7$\msun\ account for $\sim$ 20 \% of the total star formation in the galaxy during the clumpy phase, producing $\sim 10^{10}$\msun\ of stars. We do not find evidence for net inward migration of clumps within the galaxy. The number of giant clumps and their mass decrease at lower redshifts, following the decrease in the overall gas fraction and star-formation rate.

\end{abstract}




\section{Introduction}

Most massive ($M_* \sim 10^{10-11}$~\msun) star-forming galaxies at high redshift ($z\sim 1-3$) have much more irregular morphology than star-forming galaxies of similar mass in the local Universe. Although these galaxies often show signs of having extended rotating discs, their structure is dominated by a few giant star-forming clumps \citep[e.g.][]{Cowie95, vdBergh96, Genzel11, Guo12}, especially in the UV light that traces young stars. The size of these clumps is usually between a few hundred parsecs and a kiloparsec, with mass in the range $\sim 10^8 - 10^9$~M$_{\odot}$. Clumps typically comprise a few percent of a galaxy's mass, but they can account for$\sim 10-20$\% of its total star formation \citep{Elmegreen09,Wuyts12,Guo15}.

Clumpy morphology of galaxies at high redshift has been established in numerous observational campaigns in terms of different tracers of star formation activity. Giant clumps have been observed in the rest-frame UV and optical images \citep{Elmegreen07,  Elmegreen09, FSchreiber11, Guo12}, in the rest-frame optical line emission spectra \citep{Genzel08, Genzel11}, resolved maps of molecular gas \citep{Tacconi13} and in the line emission of lensed galaxies \citep{Jones10, Swinbank10, Livermore2012, Livermore2015}. On the other hand, maps of stellar mass distribution of those galaxies do not show very prominent clumps \citep{Wuyts12}.

It is possible that the irregular morphology of some high-redshift galaxies is due to an ongoing merger \citep{Somerville01}. However, the overall abundance of clumpy galaxies is too high compared to the expected merger rate to explain all of them as merging systems \citep{Dekel09b, Stewart09, HopkinsHernquist10, Hopkins10}. Alternatively, clumpy morphology can be the result of disk fragmentation due to gravitational instabilities in gas-rich discs \citep[e.g.][]{Noguchi99, Dekel09a}. The observed structure and kinematics of high-redshift clumpy galaxies suggest that many of them do in fact have underlying rotating discs \citep[e.g.][]{Elmegreen07, Genzel08, Genzel11} and that they are very gas-rich, with gas fraction $\sim 50\%$ \citep{Tacconi10}. Local fragmentation occurs in regions of the disc where self-gravity of gas and stars overcomes the pressure support in the form of velocity dispersion and the shearing effect of differential rotation. Disc stability can be parametrized in terms of the Toomre $Q$ parameter \citep{Toomre64} given by
\begin{equation}
Q = \frac{\sigma \kappa}{\pi G \Sigma} \ \mbox{,}
\label{eq:toomre}
\end{equation}
where $\sigma$ is the 1D velocity dispersion, $\kappa$ is the epicyclic frequency (which is related to the circular frequency $\kappa \propto \Omega$), and $\Sigma$ is the mass surface density.
If Q is above the critical value of order unity, the disc is stable against fragmentation. On the other hand, if $Q<1$, the disc is locally unstable and will undergo gravitational collapse. This scenario is used to explain the formation of giant molecular clouds (GMCs) in the local Universe. Assuming that giant clumps at high redshift form via the same mechanism as GMCs, but in more gas-rich environments, can explain many of their properties, as we discuss below \citep[see also][]{Murray10}.

Discs with large surface densities that drive Q below the critical value can experience gravitational instabilities that lead to disc fragmentation. Fragmentation can then stir up the disk, causing velocity dispersions to increase, and hence increasing $Q$. If $Q$ grows beyond the critical value, the disc becomes stable, further fragmentation is suppressed and the velocity dispersion gradually decreases, consequently bringing down the value of $Q$ along with it. In this way discs can self-regulate and maintain a marginally stable state with $Q\approx 1$. Analysis of gravitational instability in a disc with two components (gas and stars) is slightly more complicated \citep[see e.g.][]{Jog1996,Elmegreen11}, but qualitatively similar to this single-component analysis. 

The characteristic size of fragments is given by $\lambda_{c} \sim G\Sigma/\Omega^2$ \citep{BinneyTremaine08}. For a disc with $Q\approx 1$ this becomes $\lambda_c \sim \sigma/\Omega$. The characteristic radius and mass of fragments (clumps) in that case are\footnote{If the gas component is characterized by $Q=1$, then $\sigma \propto G\Sigma/\Omega$, where $\sigma$ and $\Sigma$ are the gas velocity dispersion and the gas surface density, respectively. We use the expression for the angular frequency $\Omega = v_c/R=\sqrt{GM/R^3}$, where $v_c$ is the circular velocity at radius $R$ and $M$ is the total mass enclosed within that radius. The mass of gas in the galaxy is proportional to $\Sigma R^2$, hence the fraction of the total mass of the galaxy that is in the gas component is $f^\prime_g \propto \Sigma R^2/M$.}
\begin{equation}
R_{cl}\propto \frac{\sigma}{\Omega} \propto \frac{G\Sigma}{\Omega^2} \propto \frac{\Sigma R^2}{M}R \propto f^\prime_g R
\label{eq:toomre_scale1}
\end{equation}
\begin{equation}
M_{cl}\propto \Sigma R_{cl}^2 \propto \Sigma f_g^{\prime 2} R^2 \propto f_g^{\prime 3} M \mbox{.} 
\label{eq:toomre_scale2}
\end{equation}
Here $f^\prime_g$ denotes the ratio of the gas mass to the total (baryonic and dark matter) mass of the galaxy. It is proportional to what is in the rest of the paper referred to as the gas fraction of the galaxy -- the ratio of the gas mass to the baryonic mass of the galaxy -- and which we denote by $f_g$. $R$ and $M$ are the radius and the mass of the galaxy, respectively. Observations indicate that star-forming galaxies at high redshift have much higher gas fractions ($f_g\sim 0.3-0.7$) compared to galaxies in the local universe ($f_g\sim 0.1$). Hence, these relations can explain the difference in size and mass of GMCs (with $R\sim 100$~pc, $M\sim 10^{5-6}$~M$_{\odot}$) in the Milky Way, and giant clumps in high-redshift galaxies (with $R\sim 1$~kpc, $M\sim 10^{8-9}$~M$_{\odot}$), assuming that they form via the same type of gravitational instability. In other words, the giant clumps observed in high-redshift galaxies seem to be a consequence of their gas-rich nature, caused by fresh supplies of gas that are continuously being provided through accretion of gas from the intergalactic medium \citep{Keres05, Keres09, Dekel09a, Brooks09,FaucherGKeresMa11}.

The importance of giant clumps for the evolution of high-redshift galaxies is a subject of ongoing research. In recent years it has been proposed that giant clumps may be responsible for the formation of bulges in their host galaxies, as an alternative to the classical bulge formation scenario involving mergers. In these models, clumps form throughout the disk, lose angular momentum due to dynamical friction and gravitational torques within the disc and gradually sink toward the centre of the galaxy \citep{Noguchi99, Immeli04, Dekel09a}. Typical time-scales over which clumps can migrate to the centre are a few times $10^8$~yrs \citep{Ceverino10, Bournaud11}, which is comparable to a few disc orbital times. 

However, it is uncertain whether individual clumps can survive that long. Because they are sites of intense star formation, clumps are exposed to strong stellar feedback which could destroy them on much shorter time-scales ($10^6 - 10^8$~yr). Analytical studies of this problem give different results depending on the assumptions made \citep{Murray10, KrumholzDekel10,DekelKrumholz2013}. Over the past several years, a number of groups have used numerical simulations, both cosmological and  idealized simulations of isolated galaxies, to study the phenomenon of giant clumps \citep[e.g.][]{Ceverino10, Hopkins12,Mandelker14,  Bournaud14, Tamburello15}. In most cases, gas-rich discs with properties similar to those of observed high-redshift galaxies do break up into large clumps of mass $\sim 10^7-10^9$\msun. However, what happens to the clumps afterwards -- how long they live and whether they migrate to the centre of the galaxy before being disrupted -- heavily depends on the details of the ISM physics and stellar feedback implemented in the simulation. For example, most simulations that model stellar feedback as heating from supernovae only, while ignoring some other modes of feedback such as radiation pressure, produce massive, gravitationally bound clumps that manage to survive for hundreds of millions of years, long enough to sink to the centre of the galaxy. 

On the other hand, the cosmological simulations of \cite{Genel12} that include a phenomenological model of feedback in the form of strong momentum-driven winds have giant clumps that get disrupted on time-scales $\sim 50$~Myr. Improved physical models of feedback have been implemented by \cite{Hopkins12} in simulations of isolated galaxies with properties that match those of high-redshift star-forming disc galaxies. They found that giant clumps get disrupted by stellar feedback in a few times $10^7$~yr, after converting only a few percent of their gas into stars, which suppresses the bulge formation rate.

Stellar feedback seems to be crucial in determining whether or not clumps live long enough to have an important effect on their host galaxy. However, properly accounting for environmental effects such as mergers and gas accretion from large-scale filaments can be just as important, because these processes  affect the overall state of the host galaxy. Isolated, gravitationally unstable discs are inherently artificial and those galaxies quickly exhaust their gas. Since there is no fresh supply of gas, there can be no new clumps formed after the initial clumpy phase. Hence, using simulations with feedback models as realistic as possible and in a full cosmological context is the most promising way to improve our understanding of the nature of giant star-forming clumps and their significance for galaxy evolution.

In this paper, we use the results of the FIRE project, a suite of cosmological zoom-in simulations with explicit treatment of stellar feedback \citep{Hopkins14}. We analyse the formation and evolution of giant clumps in a simulated massive gas-rich galaxy at $2.2 \geq z \geq 1.0$. Section \ref{sec:simulations} describes the simulation used in this work. In Section \ref{sec:clump_identification} we outline the procedure employed to identify clumps in snapshots. In Section \ref{sec:results} we present distributions of various physical properties of giant clumps, such as their mass, size, elongation, gas fraction, and stellar age. We investigate how the occurrence of giant clumps changes with redshift and overall properties of the host galaxy. We re-run our simulation with high time resolution (taking snapshots every $\sim 1$~Myr) over six 50-Myr periods in order to follow the evolution of individual clumps. This allows us to directly measure clump lifetimes and compare them to stellar ages of clumps. We discuss the implications of our results and compare them to previous studies of clumpy galaxies in Section \ref{sec:discussion}. Finally, we summarize the conclusions in Section \ref{sec:conclusions}. In Appendix \ref{sec:massivefire} we compare the properties of clumps found in this simulation with clumps from a few additional cosmological simulations of galaxies at high redshift, run at higher resolution, but over a more limited redshift range.

\section{The Simulation}
\label{sec:simulations}

\subsection{The \textit{FIRE} project}

The simulation analysed in this study is part of the FIRE (Feedback in Realistic Environments) project\footnote{\url{http://fire.northwestern.edu/}}, a suite of high-resolution cosmological zoom-in simulations of galaxy evolution. A detailed description of the numerical methods and physics implemented in the FIRE simulations can be found in \cite{Hopkins14} and references therein. Here we briefly summarize the most important features.

The FIRE simulations were performed with the newly-developed GIZMO code\footnote{A public version of GIZMO is available at \url{http://www.tapir.caltech.edu/~phopkins/Site/GIZMO.html}.}, using a pressure-entropy formulation of the smoothed particle hydrodynamics equations (P-SPH). P-SPH implementation gives good agreement to analytic solutions and grid codes on a broad range of test problems \citep{Hopkins13}, thereby resolving some well-known issues of density-based SPH codes and their differences with grid-based codes \citep{Agertz2007}. The gravity solver used is a modified version of GADGET-3 \citep{Springel05}. The initial conditions for our simulations were generated with the MUSIC code \citep{HahnAbel11} and are part of the AGORA comparison project \citep{Kim14}.

The FIRE simulations include explicit treatment of the multi-phase interstellar medium (ISM) consisting of molecular, atomic, ionized and hot diffuse components. This was made possible by the high resolution and the improved modeling of gas cooling and heating. The resolution of the FIRE simulations is sufficiently high to resolve the turbulent Jeans mass/length of interstellar gas. The initial particle mass in the high resolution region of the simulation we analyse in detail is $3.7\times 10^5$\msun\ for gas and star particles, and $2.3\times 10^6$\msun\ for dark matter. Force softening lengths are adaptive, with the minimum fixed past $z\sim 10$ at 20~pc (physical units) for baryons and 210~pc for dark matter.

Star formation proceeds in regions that are: i) dense (above density threshold $n_{th}=10$~cm$^{-3}$), ii) molecular and iii) self-gravitating. Gas particles that meet all of these criteria form stars at 100\% efficiency per free-fall time. This is motivated by high-resolution simulations of turbulent media (e.g. Padoan \& Nordlund 2011; Ballesteros-Paredes et al. 2011; Padoan, Haugbolle \& Nordlund 2012) that suggest that small-scale star-formation efficiency should be high, however our results are not very sensitive to the exact value we chose. The galaxy-averaged efficiency of $\sim 1\%$ per dynamical time is regulated by stellar feedback and basically independent of the small-scale star-formation law \citep{Hopkins11, Ostriker11, FaucherG13}. 

The FIRE simulations implement multiple modes of stellar feedback: stellar radiation pressure, photoionization and photoelectric heating, energy and momentum feedback from supernovae of type I \& II, stellar winds from AGB and O-type stars. The input parameters for modeling feedback are \textit{not} adjusted `by hand' to produce results that match observations. They are taken directly from stellar population models in STARBURST99 \citep{Leitherer99}, assuming a \cite{Kroupa02} initial-mass function for masses in the range $0.1 -100$~M$_{\odot}$. We do not consider feedback from active galactic nuclei (AGN).

Explicit treatment of the ISM and feedback physics allow the FIRE simulations to successfully reproduce many observed galaxy properties, such as the stellar mass-halo mass relation, Kennicutt-Schmidt law, and the star-forming main sequence \citep{Hopkins14}. They show a good agreement with the observed mass-metallicity relations \citep{Ma15}. Furthermore, our simulations predict neutral hydrogen covering fractions around high-redshift galaxies consistent with observations \citep{FaucherG15,FaucherG16}. The FIRE simulations also produce galactic-scale winds whose velocities and mass loading factors satisfy observational requirements \citep{Muratov15}. These results give us confidence that our simulations can reliably reproduce star formation and ISM properties on galactic scales, which are crucial for investigating the nature of giant star-forming clumps and their impact on the host galaxy.  

\subsection{The \textit{m13} simulation}

In this work, we study in detail the most massive galaxy in the suite of FIRE simulations that is described in \cite{Hopkins14}, named \textit{m13}. The name of the galaxy comes from the fact that it resides in a dark matter halo of mass $\sim 10^{13}$~M$_{\odot}$ at $z=0$, corresponding to a halo hosting a small galaxy group. We choose to focus on this galaxy because it is the only galaxy in the original FIRE sample that is as massive at $z\sim 2$ as most observed clumpy galaxies at this redshift -- its stellar mass\footnote{All global properties of the galaxy are computed in a 10-kpc sphere around the galaxy's center of mass.} at $z= 2$ is $\sim 4\times 10^{10}$~M$_{\odot}$, matching the stellar mass of clumpy galaxies in \cite{Genzel11}. The mass of \textit{m13}'s dark matter halo at $z\sim 2$ is $8.7 \times 10^{11}$\msun. 

We analyse the evolution of the \textit{m13} galaxy from $z=2.2$ to $z=1.0$. During this period, the galaxy changes from being a very clumpy, gas-rich system that forms stars vigorously, to a more regular and moderately star-forming galaxy, similar to those at lower redshift. By studying the transition of one galaxy from its clumpy to non-clumpy phase, we wish to determine what causes this transformation and what long-term effects clumps might have on the evolution of their host galaxy.

In order to simulate a galaxy as massive as those observed at high redshift to $z=0$, we adopted a resolution lower than in the rest of the original FIRE sample. In Appendix \ref{sec:massivefire} we compare the main properties of clumps in \textit{m13} with clumps found in a subset of simulations from the MassiveFIRE sample \citep{Feldmann16}. This suite of simulations includes more massive galaxies than the original FIRE sample, run at 16 times higher resolution. However, they have only been evolved down to $z\sim 1.7$ due to their high computational cost. When possible, we also compare our results to the results of \cite{Hopkins12} who analysed the properties of star-forming clumps in high-resolution simulations of isolated galaxies implementing similar models of stellar feedback. We do caution that the stellar feedback implementation in that study was based on an older code, with a different hydro solver, and less accurate treatment of feedback processes compared to the updated simulations here (as discussed in \citet{Hopkins14}). However this does not appear to change the qualitative conclusions therein.

During the analysed period the galaxy experiences interactions with a few smaller systems, but no major mergers. The most notable merger is that with a galaxy whose baryonic mass is $\sim 20$ times smaller than the baryonic mass of \textit{m13}. The merger begins at $z\sim 2.3$ (that is $\sim 150$ Myr before the first snapshot analysed in our series) and the secondary loses almost all of its gas during the first passage near the centre of \textit{m13}. What is left is a stellar component that is discernible in maps of stellar surface density all the way down to $z=1$, as is orbits around \textit{m13}. 

We record snapshots of \textit{m13} and its surroundings every $\sim 15-40$~Myr, depending on the redshift. Thus we obtain a suite of 111 snapshots, which we refer to as the original series. We analyse these snapshots to determine the properties of clumps and how they change with redshift. In addition to that, we re-simulate six $50$-Myr periods, starting at redshifts of 2.0, 1.9, 1.8, 1.7, 1.6 and 1.5, over which we follow the evolution of the \textit{m13} galaxy with fine time resolution, taking snapshots every $\sim 0.5-5$~Myr. This results in an additional 170 snapshots. The short time-scale between these snapshots allows us to track individual clumps and put constraints on their lifetime.

\section{Clump identification}
\label{sec:clump_identification}

\begin{figure*}
\centering
\includegraphics[width=0.32\textwidth]{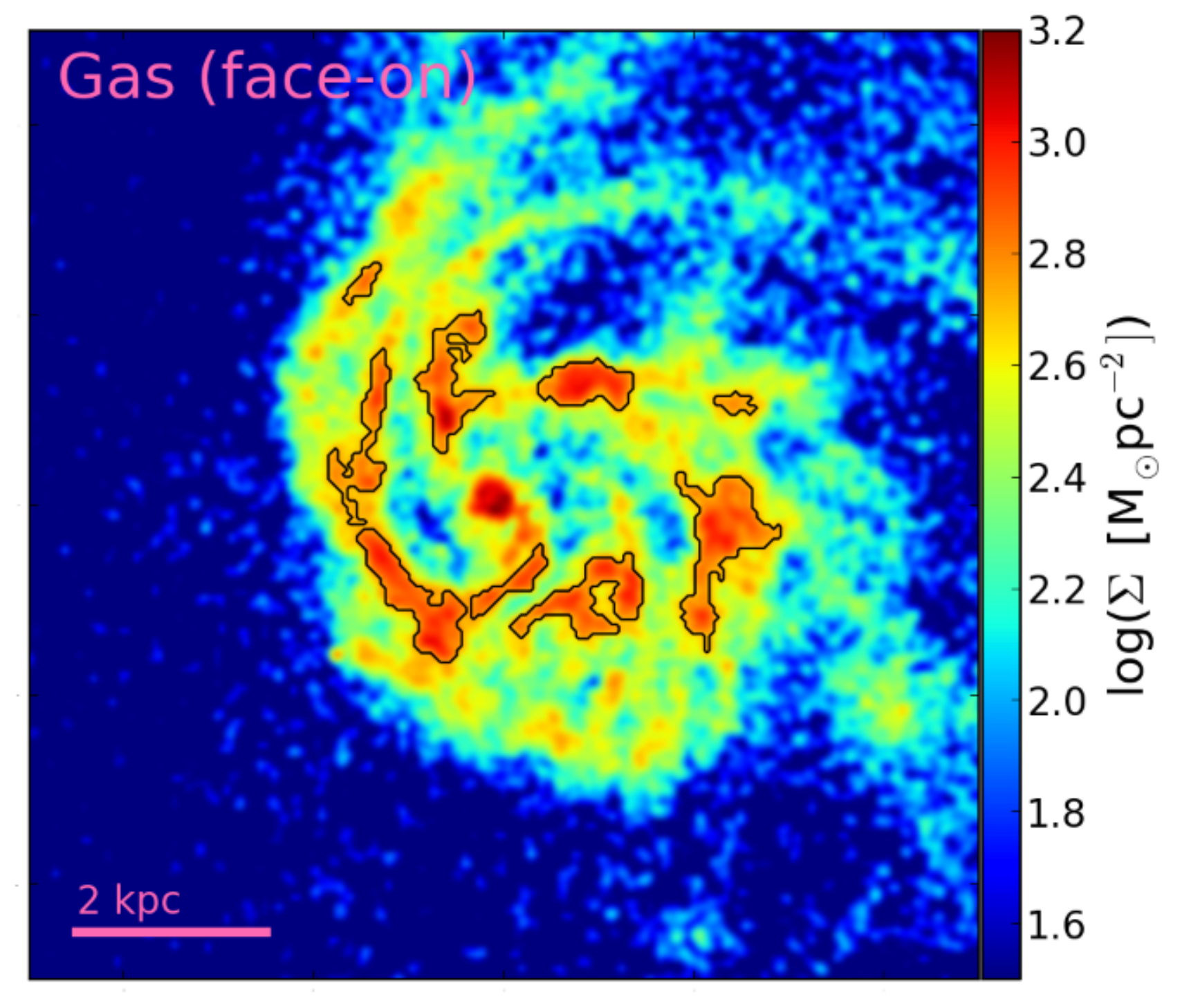}
\includegraphics[width=0.32\textwidth]{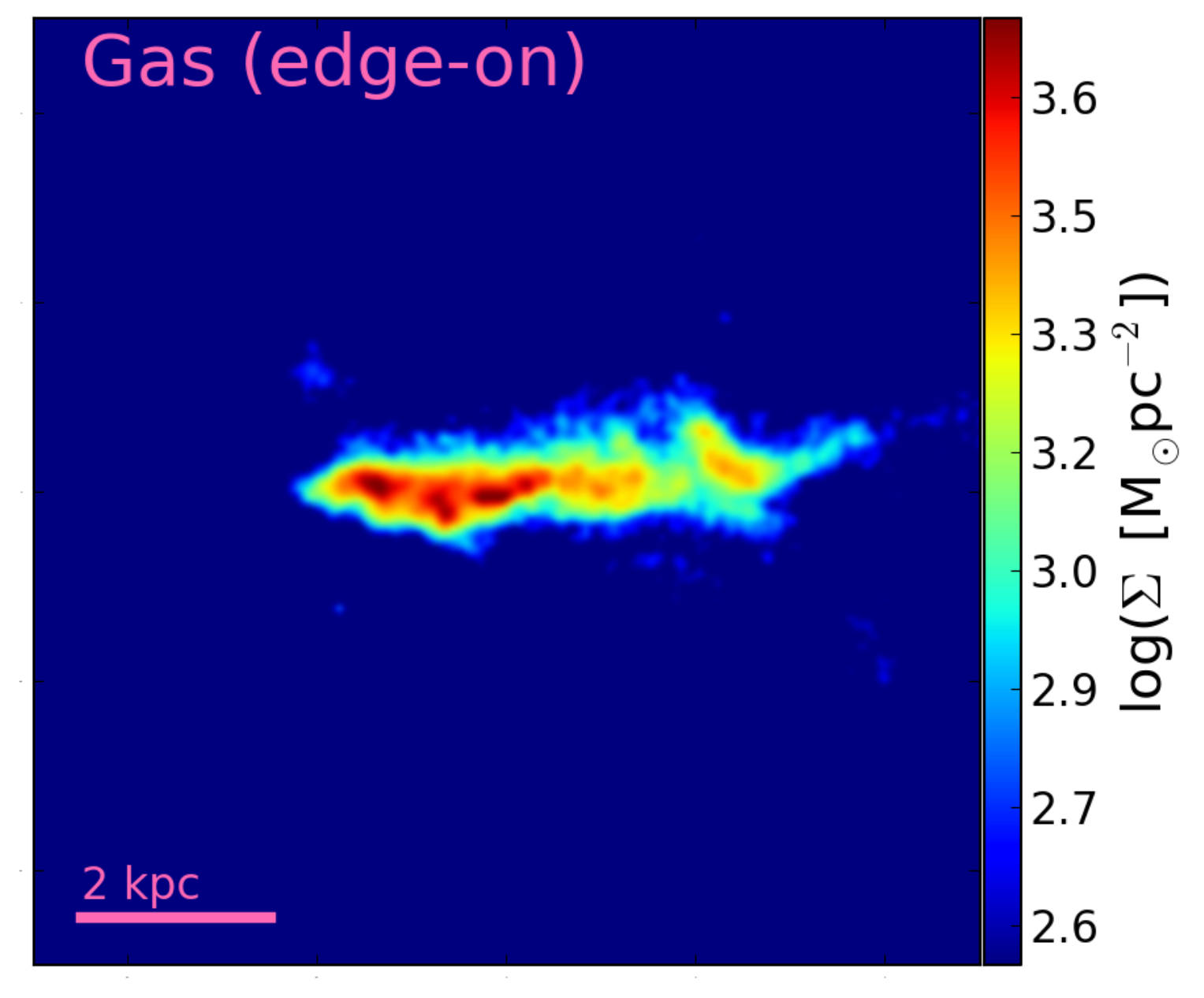}
\includegraphics[width=0.32\textwidth]{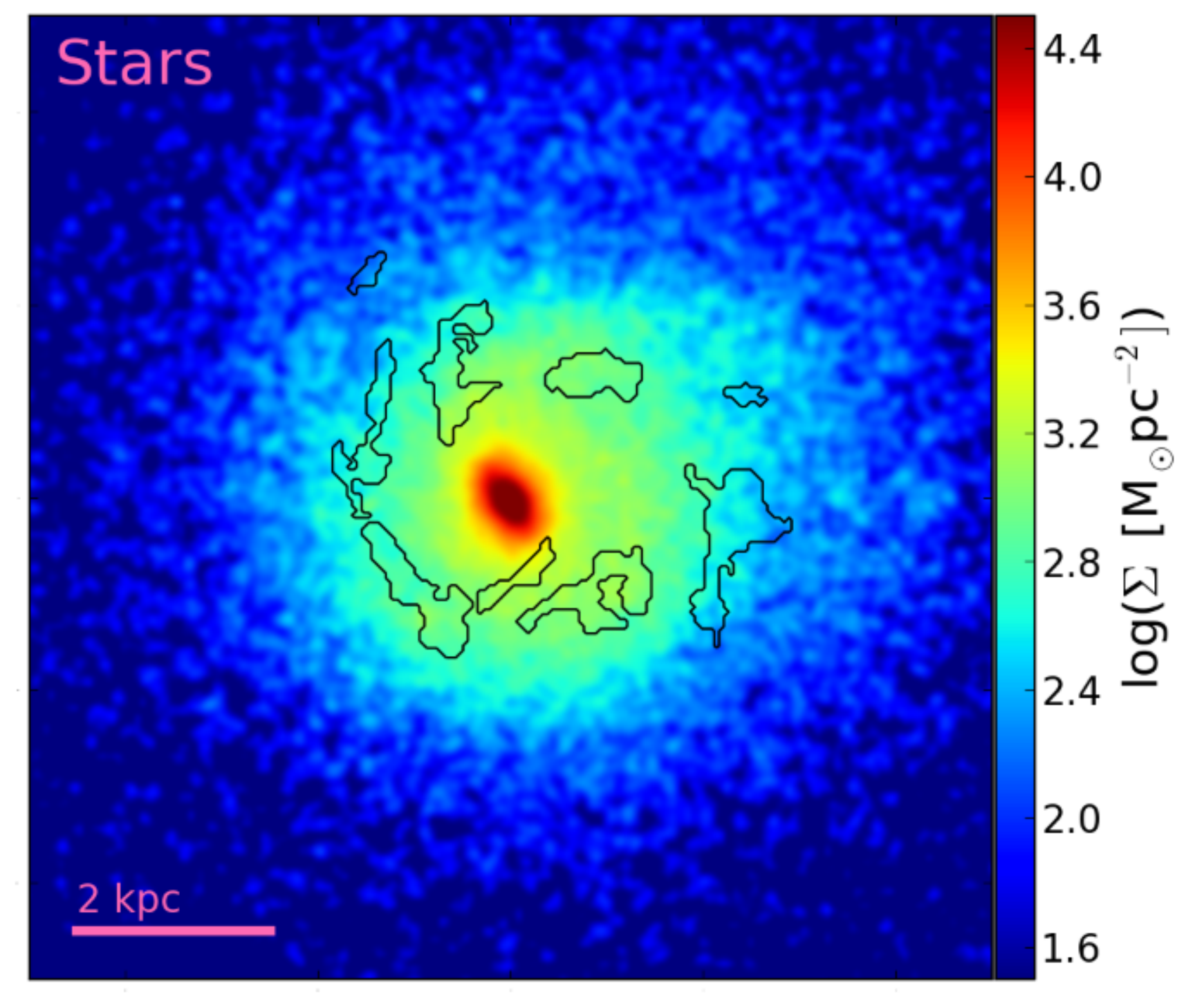}
\caption{Surface density of gas (left and middle) and stars (right) for the \textit{m13} galaxy at $z=2.0$. The size of the region shown is 10 kpc $\times$ 10 kpc (physical). Contours in the face-on projections show gas clumps identified by our clump-finding procedure (ignoring the central clump).}
\label{fig:maps190a}
\end{figure*}

\begin{figure*}
\centering
\includegraphics[width=0.32\textwidth]{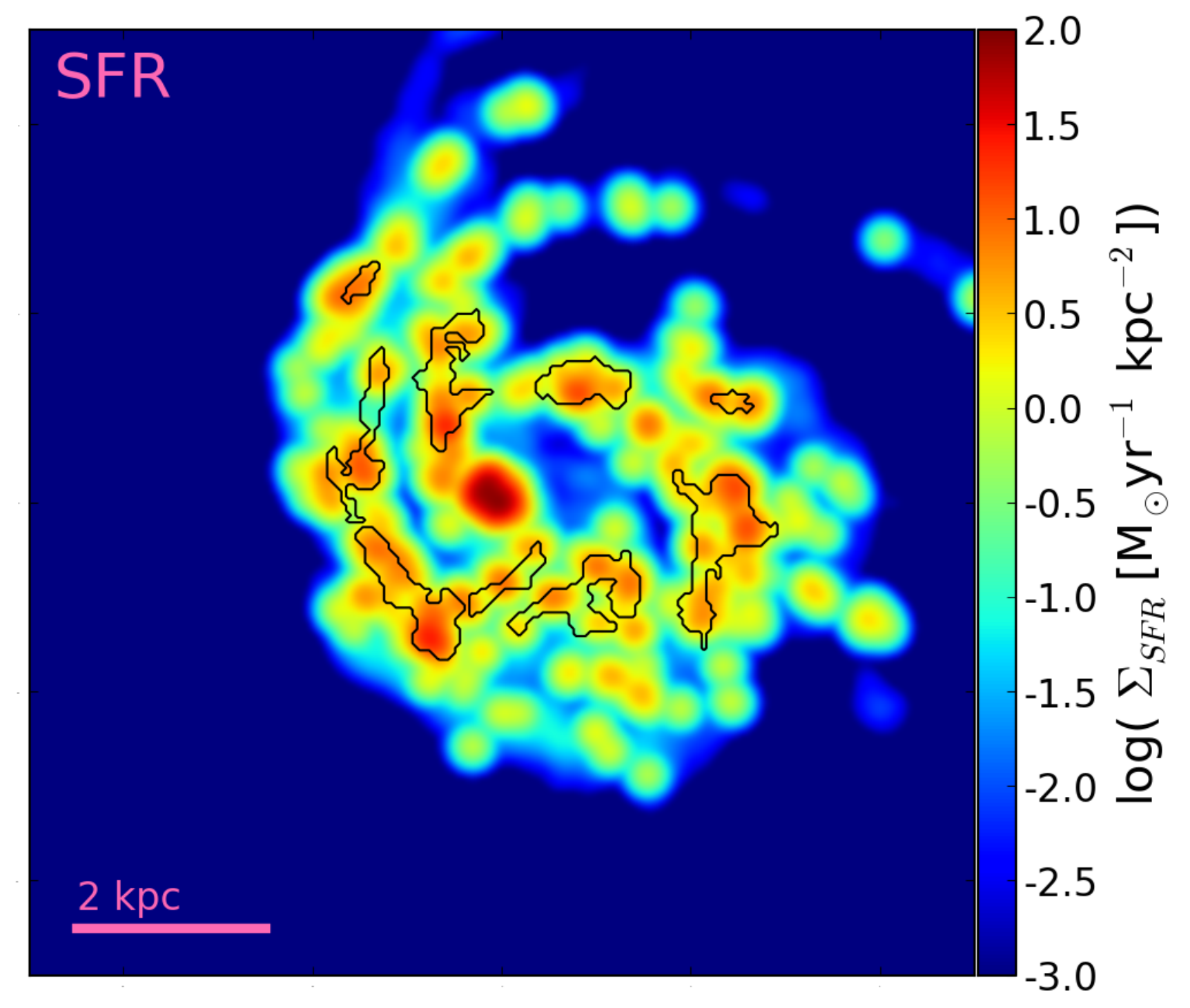}
\includegraphics[width=0.32\textwidth]{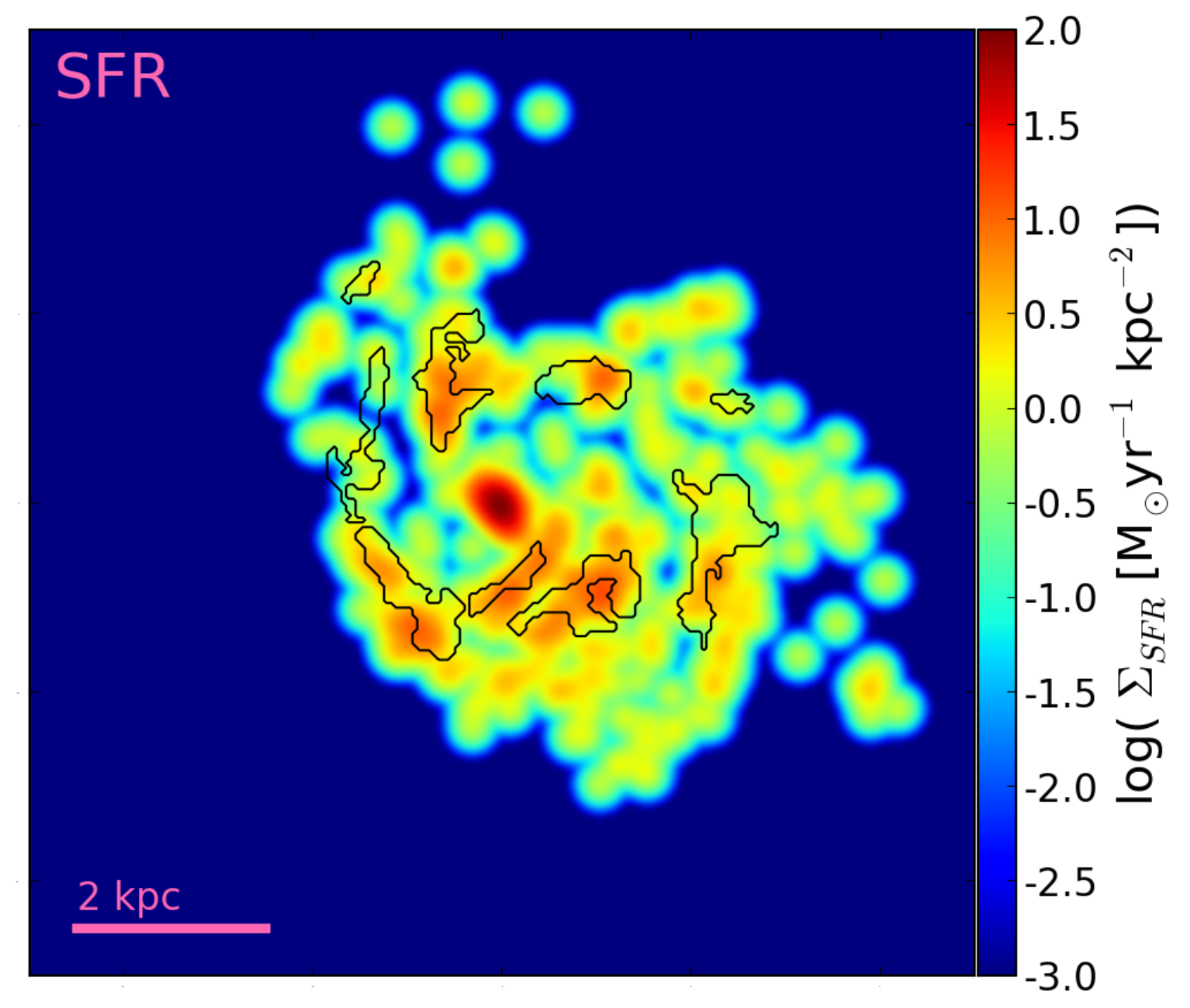}
\includegraphics[width=0.32\textwidth]{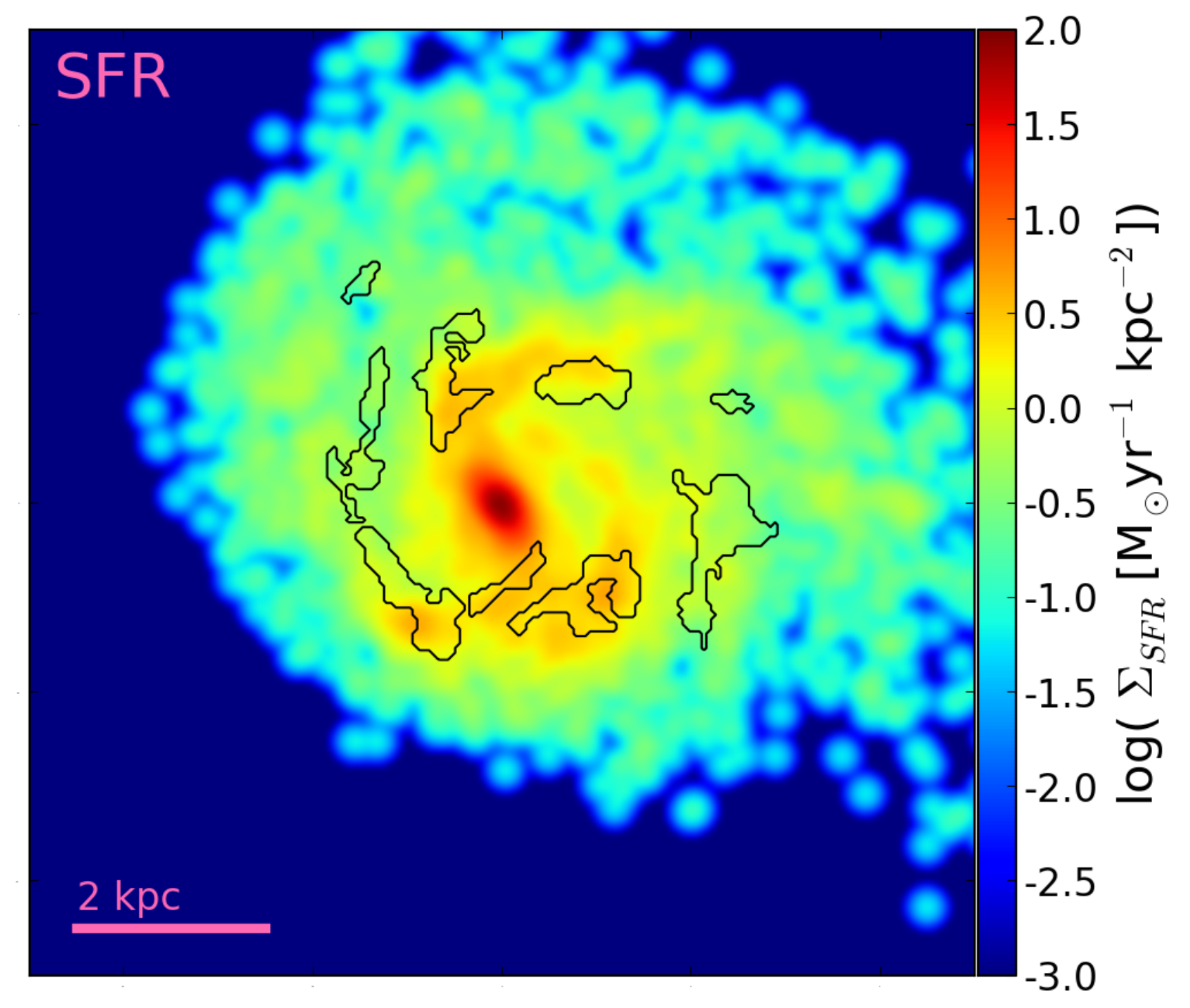}
\caption{Surface density of the star-formation rate calculated over three different time scales (instantaneous on the left, averaged over 10 Myr in the middle, and averaged over 100 Myr in the right) for the \textit{m13} galaxy at $z=2.0$. As in Fig. \ref{fig:maps190a}, the size of the region shown is 10 kpc $\times$ 10 kpc (physical). Contours show gas clumps identified by our clump-finding procedure (ignoring the central clump).}
\label{fig:maps190b}
\end{figure*}

To identify giant star-forming clumps, we assume that they correspond to regions of the disc with high surface density of gas. We choose this two-dimensional criterion for the definition of clumps in order to more easily compare our results with those from observations of real galaxies and mock images created from simulations. We tried using other criteria for the definition of clumps, ones that are more related to the physical (3D) density of gas and we find that they give very similar results when it comes to identifying the largest clumps. In observations of high-redshift galaxies, clumps are usually identified based on UV and optical light, serving as tracers of star formation. Here we use gas surface density to identify clumps, which is a good approximation because maps of the projected star-formation rate (SFR) show that the locations of gas clumps coincide with regions of intense star formation (see Figs. \ref{fig:maps190a} and \ref{fig:maps190b}). In general, we do not observe obvious clumps in the distribution of stars or dark matter coincident with the gas clumps. Discernible star clumps correspond to a few (in total, not per snapshot) smaller galaxies orbiting around \textit{m13}.

As a first step in our clump-finding procedure, we find the total angular momentum of gas particles within 10 (physical) kpc from the centre of mass of the galaxy and determine the Euler angles of the face-on projection in the simulation's coordinate system. Next, we form a two-dimensional 10~kpc$\times$10~kpc grid in the face-on projection, with cell size of 50 pc. We project the location of every particle within 10~kpc radius from the centre of the galaxy onto the grid and for the sake of computational convenience deposit the entire particle into the corresponding cell (i.e. ignoring the particle's SPH kernel). We construct the gas surface density map by dividing the entire gas mass contained in a cell with its surface area. Using all cells with non-zero surface density, we find the mean value of gas surface density and the standard deviation. We use these two values in our definition of clumps -- clumps correspond to regions that are at least one standard deviation above the mean surface density. The threshold values are chosen in this way because they provide the best match to what we visually identify as clumps in the gas surface density maps. Appendix \ref{sec:sensitivity} contains more details on how changing the threshold value may affect our results. Finally, we smooth the gas surface density field using a Gaussian filter with the standard deviation of 50 pc (FWHM of $\sim 120$ pc) and the radius of the Gaussian kernel of 450 pc. The $\sim 100$ pc scale roughly coincides with the lower end of the reported size distribution of observed giant clumps at high redshift. This smoothing scale should be large enough to mitigate the effect of ignoring the SPH kernel in creating gas surface density maps.

To find clumps in the gas surface density map, we make use of a Python package astrodendro\footnote{\url{http://www.dendrograms.org/}} to compute dendrograms -- trees of hierarchical structure in our simulated data. To find clumps, we need to specify three parameters: 1) the minimum value of gas surface density that is used as a threshold to define clumps, which we set to be equal to one standard deviation above the mean surface density of all non-empty (unsmoothed) cells in that snapshot; 2) the minimum difference in the surface density between a large structure and any substructure within it, which we set to $10\%$ of the mean surface density (this parameter is not very important in our particular study, since we primarily care about large clumps, and not so much the substructures within them); 3) the minimum number of pixels within a clump which we set to 20, corresponding to a minimum effective radius of selected clumps of about 125 pc. 

After computing a dendrogram for each snapshot using the described procedure, we obtain a list of clumps that satisfy the criteria we imposed. If a selected region has substructure, we decide what is going to be considered a `clump' (whether the larger structure as a whole or smaller substructures individually) by visual inspection. In most cases we choose the largest structure to represent the clump and we follow the evolution of this object as a whole, without keeping track of individual sub-structures within it. However, in a few instances the largest structure was branching out in such a way that it would be unlikely that it would be recognized as a single clump in real observation. In such cases, we selected the smaller, more compact substructures within that region to be identified as clumps. Many snapshots contain a clump located in the centre of the galaxy, which corresponds to the bulge of the galaxy  -- we disregard those clumps from further analysis.

Using the astrodendro package to select clumps allows us to easily measure the size of each clump in the plane of the disk ($xy$ plane), and express it in terms of its area ($A$), effective major ($a$) and minor ($b$) axes. The axes for each clump are calculated by astrodendro from the contour of the gas clump by fitting an ellipse to it and deriving the parameters of the ellipse. We calculate the effective radius ($R$) of a clump from its area (computed by astrodendro) as $R=\sqrt{A/\pi}$. We assume that the extent of each clump in the $z$-direction (perpendicular to the plane of the disk) equals $2R$ centered on the densest part of the clump. For most clumps, the enclosed mass does not heavily depend on the exact choice of its vertical extent, as long as it is roughly on the order of disc scale height. Even if we extend clumps to 10~kpc in the $z$-direction (while keeping the surface area in the plane of the disc fixed), their mass increases only by a factor of a few (for the most massive clumps that factor is $\lesssim 2$)  

All gas and stellar particles within the volume of space that is identified as a clump, belong to that clump. These selection criteria were devised to match the observational identification of clumps and do not tell us anything about whether clump particles are bound together or not. We do not subtract the underlying disc (stars or gas) from the clumps. Some observers subtract the background before evaluating the clump luminosity and mass \cite[e.g.][]{FSchreiber11, Guo12, Guo15}, while some do not \cite[e.g.][]{Elmegreen09,Wuyts12}. It is important to keep that in mind when comparing results from different groups, as the background subtraction may cause a factor of 2-3 difference in the estimated clump flux \citep{FSchreiber11, Guo15}. 

Fig. \ref{fig:maps190a} shows the gas surface density of \textit{m13} at $z=2$ in the face-on and edge-on  projections (left and middle panels) and the face-on distribution of stars (right panel), which is fairly smooth and does not show prominent overdensities. The contours in the face-on projections mark the gas clumps identified in this snapshot. Fig. \ref{fig:maps190b} shows the surface density of the star-formation rate calculated over three different time scales (instantaneous, averaged over 10 Myr and over 100 Myr). For reference, the contours of gas clumps are also shown in the maps of the SFR. Most identified clumps (and the central clump, which is excluded from our analysis) correspond to regions of intense star formation and would likely be prominent in the rest-frame UV images and other tracers or recent star formation. However, there is no exact one-to-one correspondence between the gas clumps and the regions of elevated SFR, nor are the maps of the SFR calculated over different time scales identical. This is in agreement with the results of \citet{Moody14} who found that clumps identified in the gas maps are not necessarily found in the mock images or stellar maps of the same galaxy. 

\section{Results}
\label{sec:results}

\subsection{Clump properties in the \texttt{m13} simulation}
\label{sec:clump_properties}

\begin{figure*}
\centering
\includegraphics[width=0.9\textwidth]{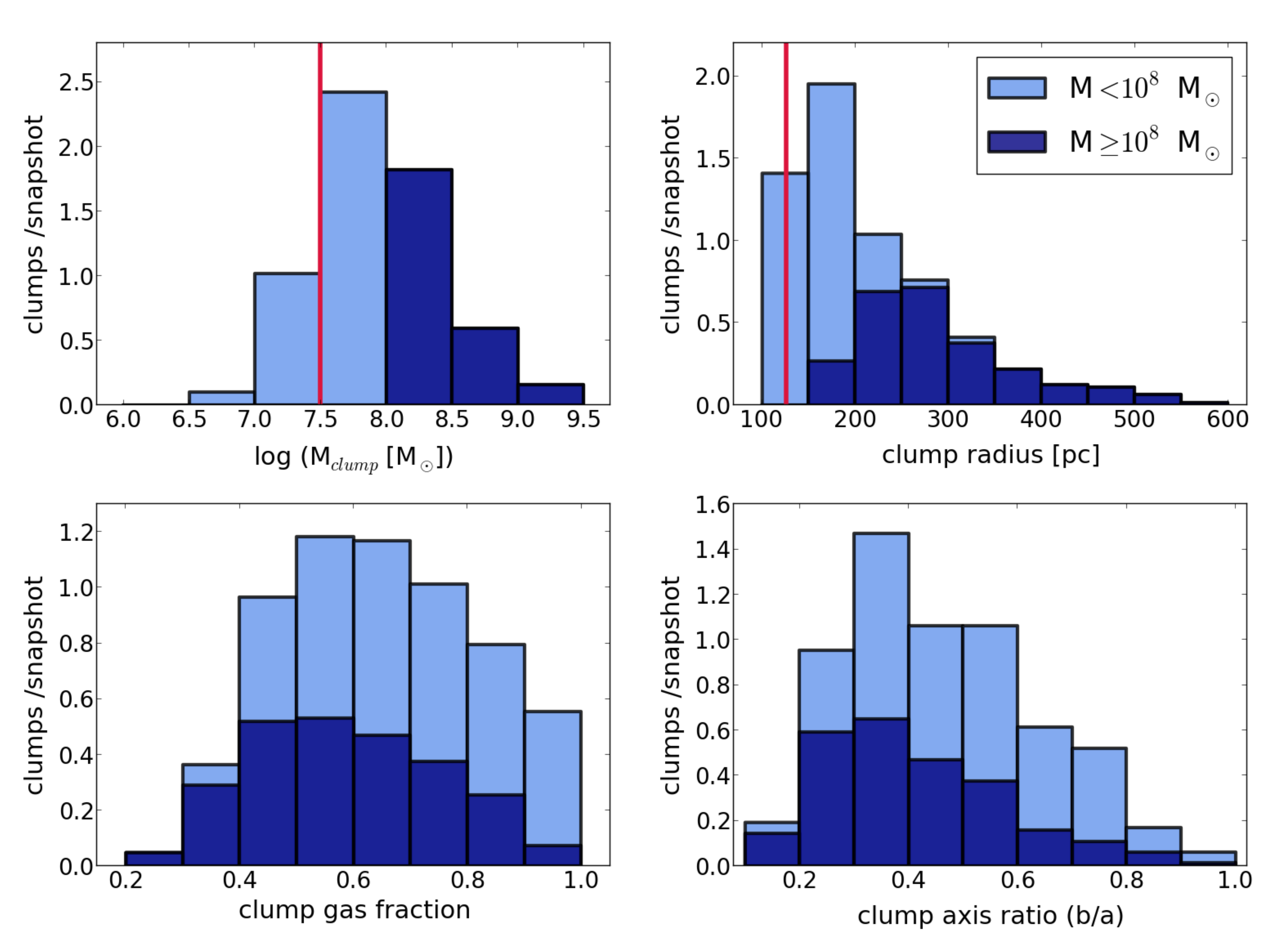}
\caption{ Distribution of clump properties in the \textit{m13} simulation from $z=2.2$ to $z=1.0$. Upper left panel shows the distribution of the total clump baryonic mass. The red line indicates the mass limit above which our clump sample is complete. Clump gas fraction, obtained as the ratio of clump gas mass to the sum of gas and stellar mass is shown in the lower left panel. Upper right panel shows the distribution of clump effective radius, calculated from clump's surface area ($A$) as $R=\sqrt{A/\pi}$. The red line marks $R=125$~pc, the lower limit on the radius, set by our requirement that clumps span the area of at least 20 pixels of size 50~pc $\times$ 50~pc. The distribution of the ratio of minor ($b$) and major ($a$) axes is shown in the lower right panel. Dark blue histograms represent giant clumps with baryonic mass $\geq 10^8$~M$_{\odot}$, and light blue is for less massive clumps (the light blue bars lie on top of the dark ones).}
\label{fig:4figs}
\end{figure*}

The clump-finding procedure identified a total of 506 clumps in the original \textit{m13} series of snapshots at $2.2 \geq z \geq 1.0$ (not counting clumps located in the centre of the galaxy, which we discarded). The average number of identified clumps per snapshot varies with redshift (as discussed in more detail in Section \ref{sec:occurrence}) from about 8 clumps per snapshot at $z\sim 2$, to no clumps at all at $z\sim 1$. Fig. \ref{fig:4figs} shows clump distributions in terms of their baryonic mass, physical size (effective radius), elongation (the ratio of the minor and major axis) and gas fraction (defined as $f_g = M_{gas}/[M_{gas}+M_{stars}$]). Even though the focus of this study is on giant clumps, which typically have masses in the range $10^8-10^9$~M$_{\odot}$, our clump-finding procedure is capable of finding clumps smaller than that, with masses $\sim 10^7$~M$_{\odot}$. The lower limit for clump mass is slightly different for every snapshot because it depends on the gas surface density threshold which varies from snapshot to snapshot. The highest surface density threshold over all snapshots is $\sim 650$\msun pc$^{-2}$, which corresponds to a minimum clump mass of $10^{7.5}$\msun. This is a conservative limit on our mass completeness over the entire analysed redshift range. In many snapshots we are able to identify clumps much smaller than that limit.  

The number of clumps with total baryonic mass greater than $10^8$~M$_{\odot}$ is 155 and their effective radii range from $\sim 150$ to $\sim 600$ pc, which is consistent with the properties of observed giant clumps at high redshift \citep[e.g.][]{FSchreiber11}. Clump gas fractions vary from highly gas-dominated clumps that have over $90\%$ of their mass in gas, to clumps with $f_g \sim 0.3$. Most giant clumps have comparable mass contributions from gas and stars, with the median value for giant clumps of $f_g=0.57$. Morphologies of giant clumps in our sample are very diverse, from nearly spherical clumps with the ratio of minor to major axis close to unity ($b/a\approx 1$), to highly elongated clumps with minor axis almost an order of magnitude smaller than the major ($b/a\sim 0.1$). 

\subsection{Occurrence of clumps over redshifts}
\label{sec:occurrence}

\begin{figure}
\centering
\includegraphics[width=0.45\textwidth]{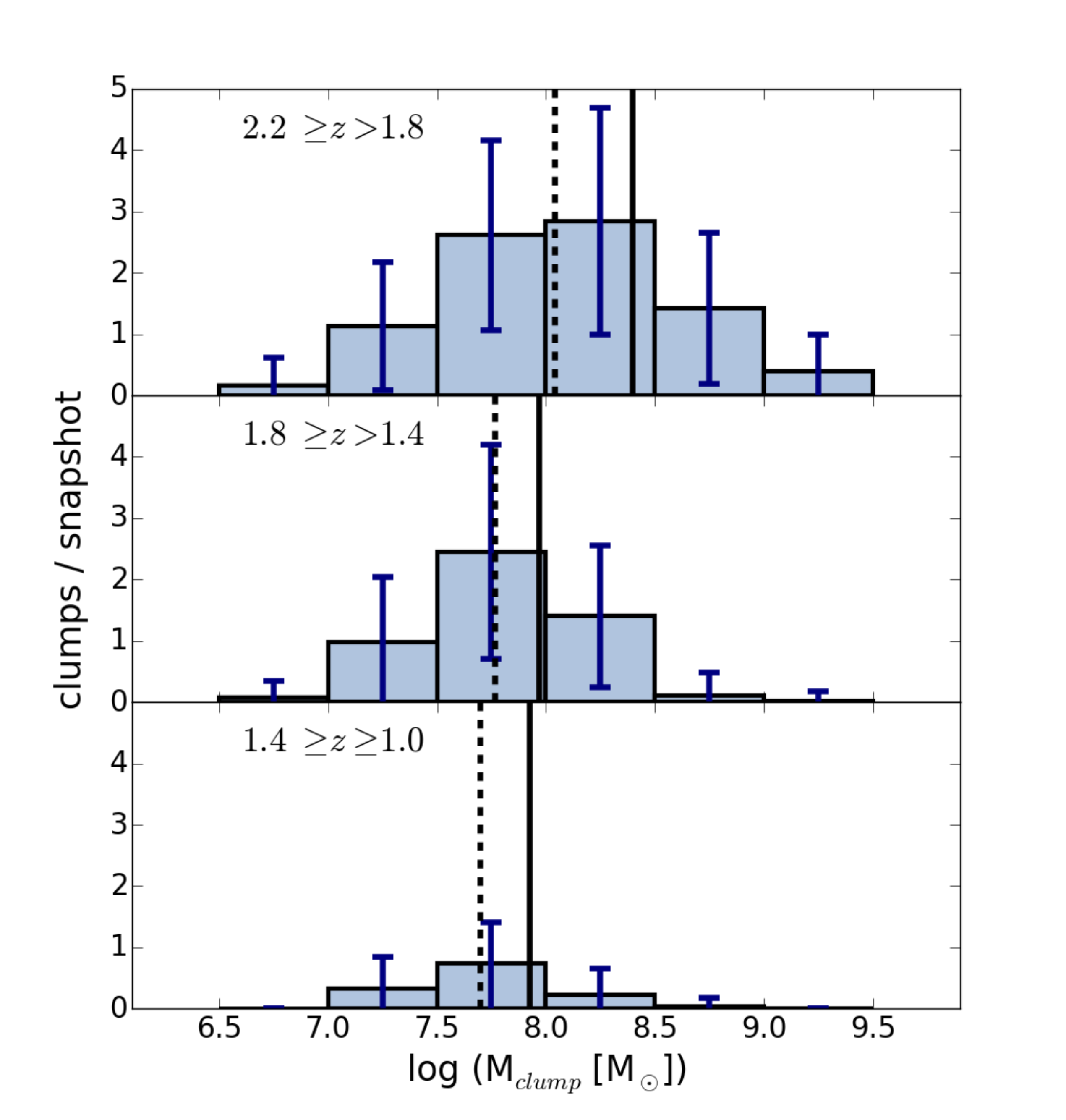}
\caption{Distribution of clump baryonic mass in three redshift bins. Histograms indicate the average number of clumps per snapshot in each bin, whereas the error bars mark the standard deviation. The number of clumps at low redshift is reduced compared to high redshift and the clumps are less massive, as indicated by the mean and median of the distribution (represented by solid and dashed lines, respectively) which move over to lower mass values at lower redshift.}
\label{fig:mass_redshift}
\end{figure}

The number of massive clumps present in the galaxy decreases with time over the redshift range $2.2 \geq z \geq 1.0$ which spans almost 3 Gyr. Fig. \ref{fig:mass_redshift} shows the mass distribution of clumps in three redshift bins. The number of identified clumps clearly decreases with decreasing redshift. The clumps also get smaller with time -- the mean and median of the mass distribution, indicated by solid and dashed lines, respectively, move towards lower masses at lower redshifts.

\begin{figure}
\centering
\includegraphics[width=0.5\textwidth]{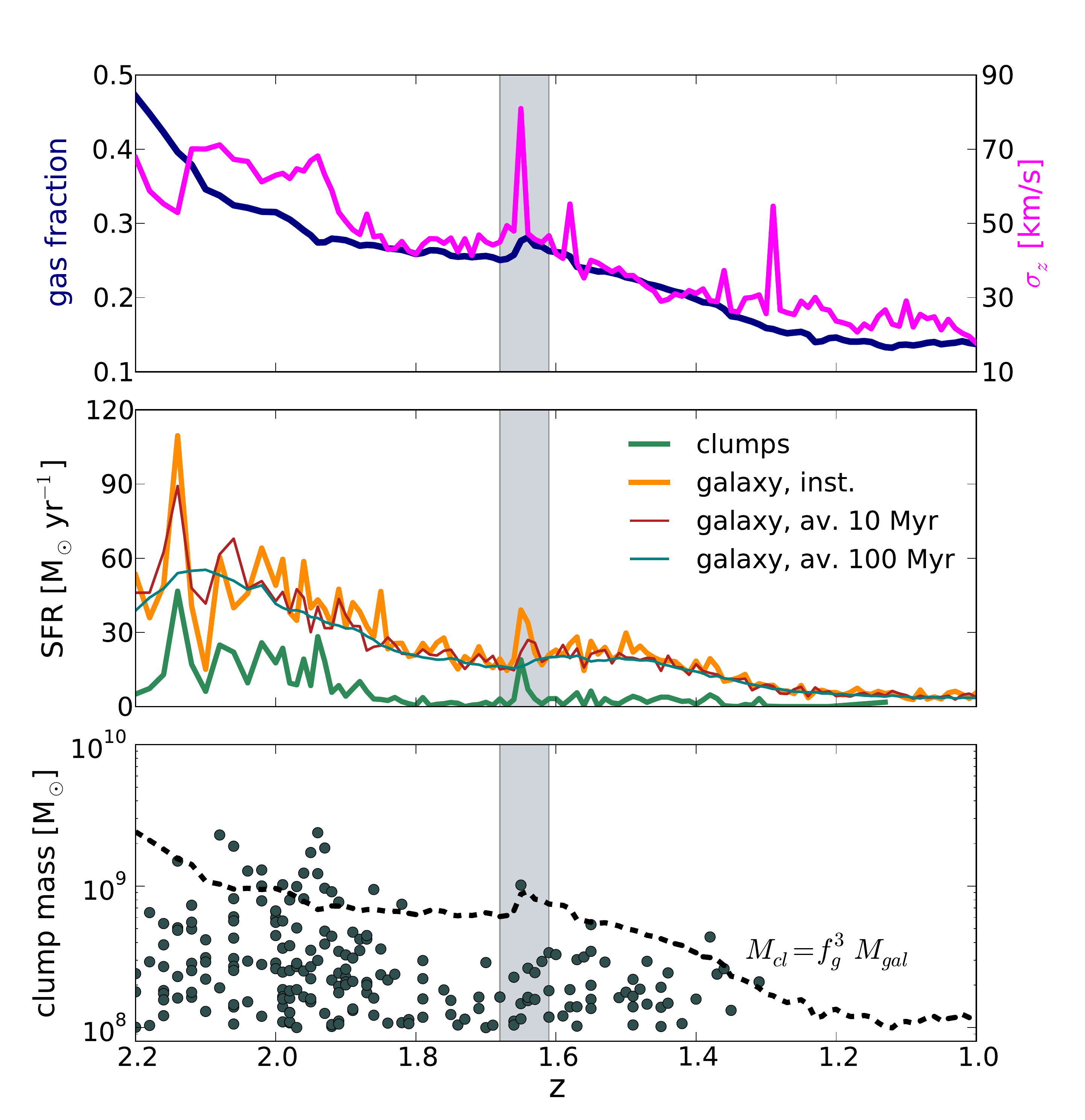}
\caption{\textit{Upper panel:} Blue line represents the gas fraction of the \textit{m13} galaxy as a function of redshift. Gas velocity dispersion perpendicular to the plane of the disc is represented by the violet line. \textit{Middle panel:} Star-formation rates (SFR) in the whole galaxy (orange for instantaneous, red and teal for the SFR calculated by averaging over 10-Myr and 100-Myr time scales, respectively) and in clumps (green) show that $10-60\%$ of the total star formation occurs in clumps at any given time. Averaged over the analysed time period, clumps account for $\sim 20$\% of the total star formation in the galaxy. \textit{Bottom panel:} Mass of individual giant clumps (M$\geqslant 10^8$\msun) at their corresponding redshift. The most massive clumps follow the expected scaling with the gas fraction and mass of the galaxy, $M_{cl}\sim f_g^3M_{gal}$, represented by the gray dashed line. The velocity dispersion, star-formation rate and the occurrence of giant clumps are all high when the gas fraction is high, and decrease with decreasing redshift. The gray band marks the period during which the galaxy experiences a close encounter with a smaller galaxy, resulting in an increase of the gas fraction, star formation rate and number of giant clumps.}
\label{fig:mass_redshift2}
\end{figure}

If large clumps do form due to gravitational instabilities in gas-rich disks, then their occurrence rate might be correlated with the gas fraction of the galaxy. Fig. \ref{fig:mass_redshift2}, top panel, shows the galaxy's gas fraction and the gas velocity dispersion in the $z$ direction, both as functions of redshift. We calculate the gas fraction by taking into account gas in all phases. The line-of-sight velocity dispersion calculation also includes all gas particles -- we do not exclude any inflows/outflows that may be present. The galaxy is very gas-rich at $z>2$, with gas fraction above $30\%$, which is consistent with observations of high-redshift galaxies \citep{Tacconi08, Tacconi10}. With time, the gas fraction decreases, settling at about $10-15\%$ around $z\sim 1$, when the stellar mass of the galaxy is $\sim 6\times 10^{10}$\msun. The gas velocity dispersion follows the same trend, though in a more bursty manner. The middle panel shows the star-formation rate (SFR) in the galaxy as a whole and in clumps. Clumps are sites of intense star formation that account for $10-60\%$ of the total SFR in the galaxy at any given time, even though they make up only a few percent of its total mass. About $10^{10}$\msun\ of stars is formed in clumps during the clumpy phase in the evolution of the galaxy ($2.2 \gtrsim z \gtrsim1.3$), which is $\sim$20\% of the total star formation over that period. The SFR experiences a similar decrease with redshift as the gas fraction. The bottom panel shows the mass of individual giant clumps with mass $\geqslant 10^8$\msun\ as a function of redshift. 
The most massive clumps approximately follow the expected scaling with the gas fraction given in  Eq. \ref{eq:toomre_scale2}$\ (M_{cl}\sim f_g^3M_{gal}$).
There are no clumps with mass above $10^8$\msun\ at $z<1.3$, when the gas fraction falls under $\sim 15\%$. At these redshifts, the clump masses are in the range $10^6-10^8$\msun, i.e. comparable to the mass of `normal' GMCs found in the Milky Way. The properties of clouds in this mass range, found in simulations of isolated galaxies that implement a similar model of stellar feedback as our simulation, are analysed in detail in \cite{Hopkins12}.

There is a relatively short period around $z\sim 1.7$ when the gas fraction slightly increases due to a minor merger with a galaxy of stellar mass $\sim 10^9$~M$_{\odot}$. The gray band in Fig. \ref{fig:mass_redshift2} marks the period during which the small in-falling galaxy is within 30 kpc from the centre of mass of the whole system. 

\subsection{Stellar age in clumps}

\begin{figure}
\centering
\includegraphics[width=0.49\textwidth]{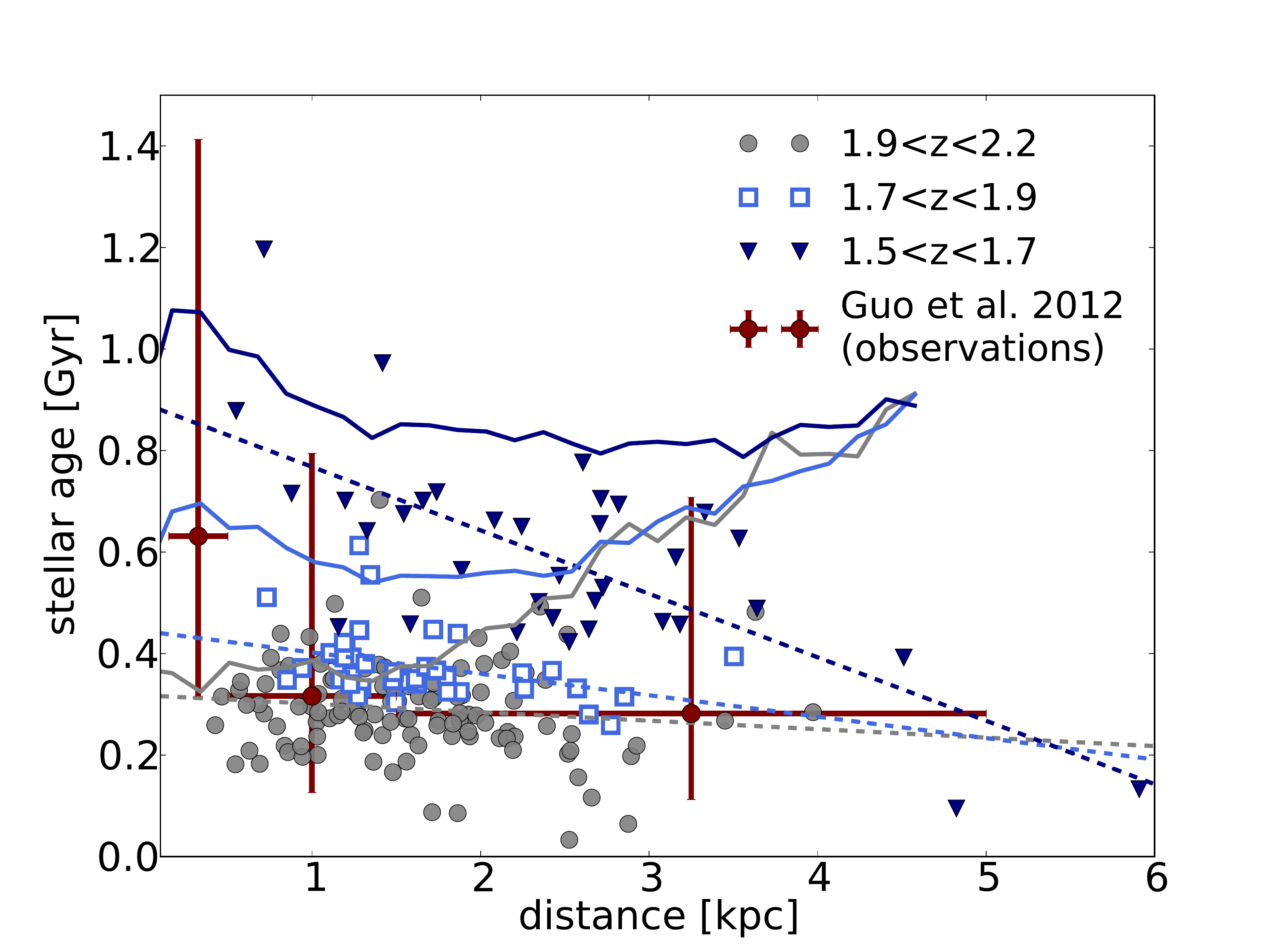}
\caption{Mass-weighted stellar age in individual giant clumps with M$\geqslant 10^8$~M$_{\odot}$ as a function of distance from the centre of the galaxy, color-coded by redshift. For all three redshift bins we show linear fits (dashed) through the data points which indicate that clumps with older stellar populations tend to be located closer to the centre of the galaxy. Average stellar ages of all the stars in the disc (including those in clumps), at the central redshift for each bin, are shown as solid lines. Clumps have slightly younger stellar populations than the disc average because they are regions of intense star formation and hence contain more young stars than the rest of the disc. Observational results from \citet{Guo12} are show for comparison. We multiplied their $x$-axis values by the radius of \textit{our} galaxy ($R= 5$~kpc).}
\label{fig:stellar_age}
\end{figure}

Fig. \ref{fig:stellar_age} shows the mass-weighted age of star particles in giant clumps, as a function of the clump's galactocentric radius, i.e. the distance between the centre of the galaxy and the clump's centre of mass, color-coded by redshift. Mass-weighted stellar ages of clumps are typically a few hundred Myr, however they span a broad range from tens of Myr for a few of the smallest clumps to about a Gyr. These numbers are consistent with the estimated ages of stellar populations in observed high-redshift clumps, which range from $\sim 10$~Myr to several hundred Myr \citep{Elmegreen09,FSchreiber11,Guo12,Wuyts12,Adamo2013}. However, as we show below, stellar ages are not necessarily indicative of clump lifetimes.

\subsection{Radial gradients in clump properties}
\label{sec:gradients}

Radial gradients in various clump properties, most notably in the stellar age of clumps, have been observed and interpreted as evidence of radial migration of clumps and their long survival. In this section, we show that our clumps, although short-lived, can reproduce similar trends with galactocentric radius, thus suggesting that these observational results may not be as informative about clump survival as previously believed. 

Most star-forming clumps in \textit{m13} are centrally located, within $\sim 3$~kpc from the centre, especially at high redshift. For a given redshift bin, there is a tendency for clumps with older stellar populations to be located closer to the centre of the galaxy, as indicated by dashed lines in Fig. \ref{fig:stellar_age} which represent linear fits through data points. This trend is only marginal in the high-redshift bin, but becomes more pronounced at lower redshifts. Anti-correlation of stellar age and distance from the centre (shown in Fig. \ref{fig:stellar_age}) has been observed in high-redshift clumps by \cite{FSchreiber11,Guo12,Adamo2013}. Because clumps are regions of intense star formation, most of them contain slightly younger stellar populations than the disc average at the corresponding redshift (indicated by solid lines in Fig. \ref{fig:stellar_age}), which includes stars in clumps and inter-clump regions. The average stellar age increases with radius in the outskirts of the disc (i.e. beyond $\sim 3$~kpc from the centre). This is similar to the results of \cite{ElBadry15} who analysed stellar migration due to gas outflow/inflow cycles caused by stellar feedback in the FIRE simulations of low-mass galaxies. They found that most stars form in the central parts of the host galaxy and then exhibit systematic outward migration over the course of their lifetime, causing a positive stellar age gradient. 

\begin{figure}
\centering
\includegraphics[width=0.49\textwidth]{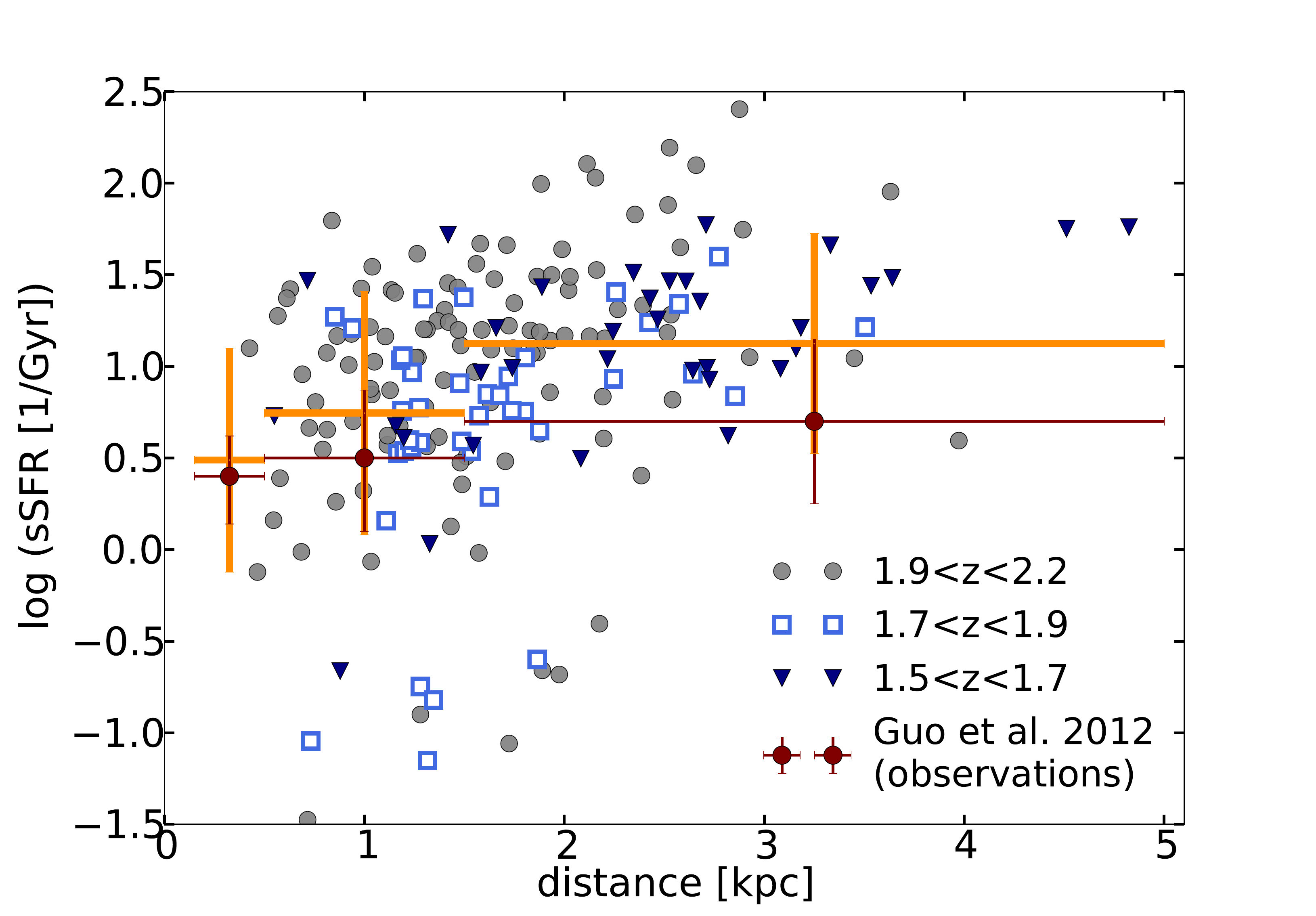}
\includegraphics[width=0.49\textwidth]{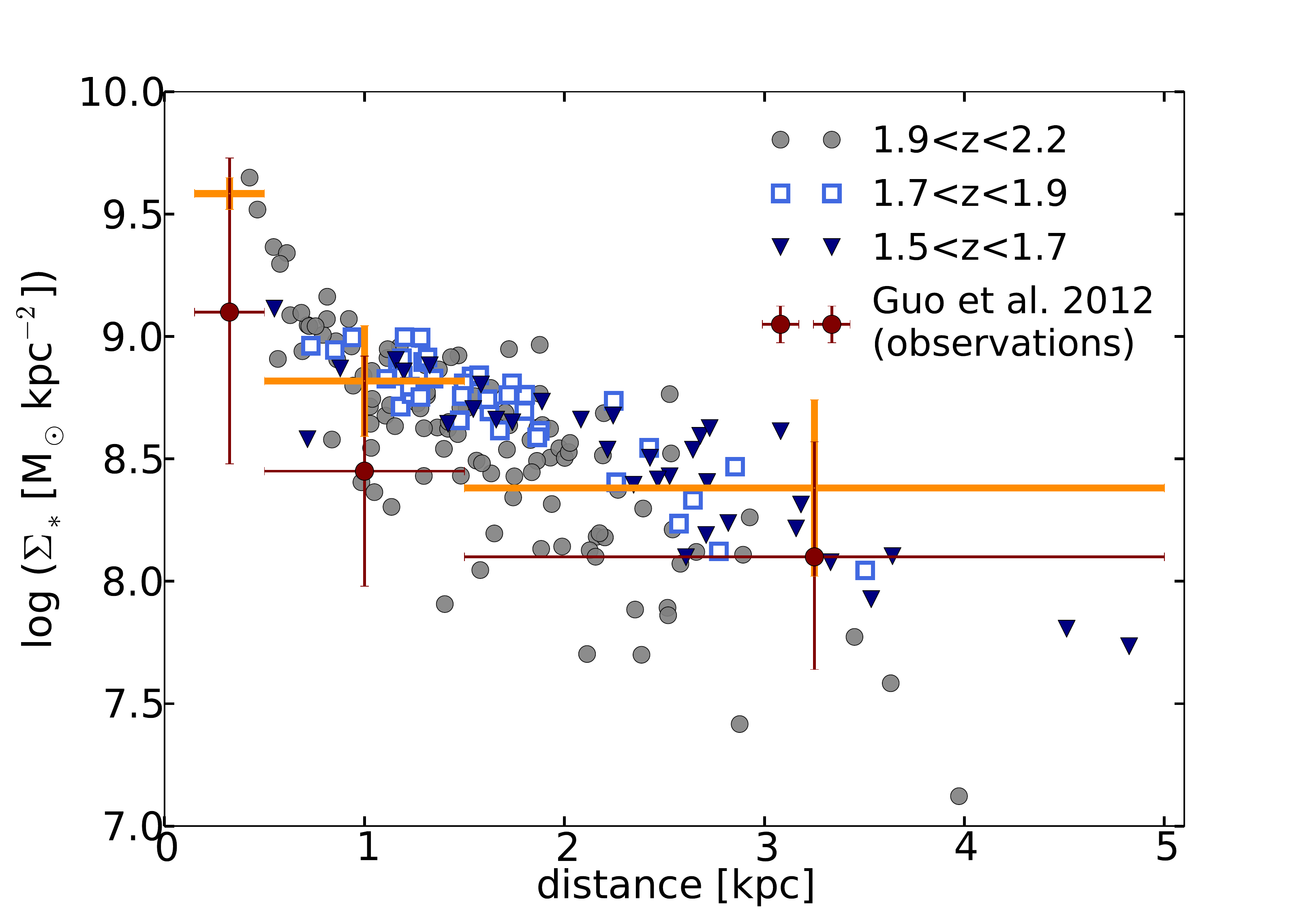}
\includegraphics[width=0.49\textwidth]{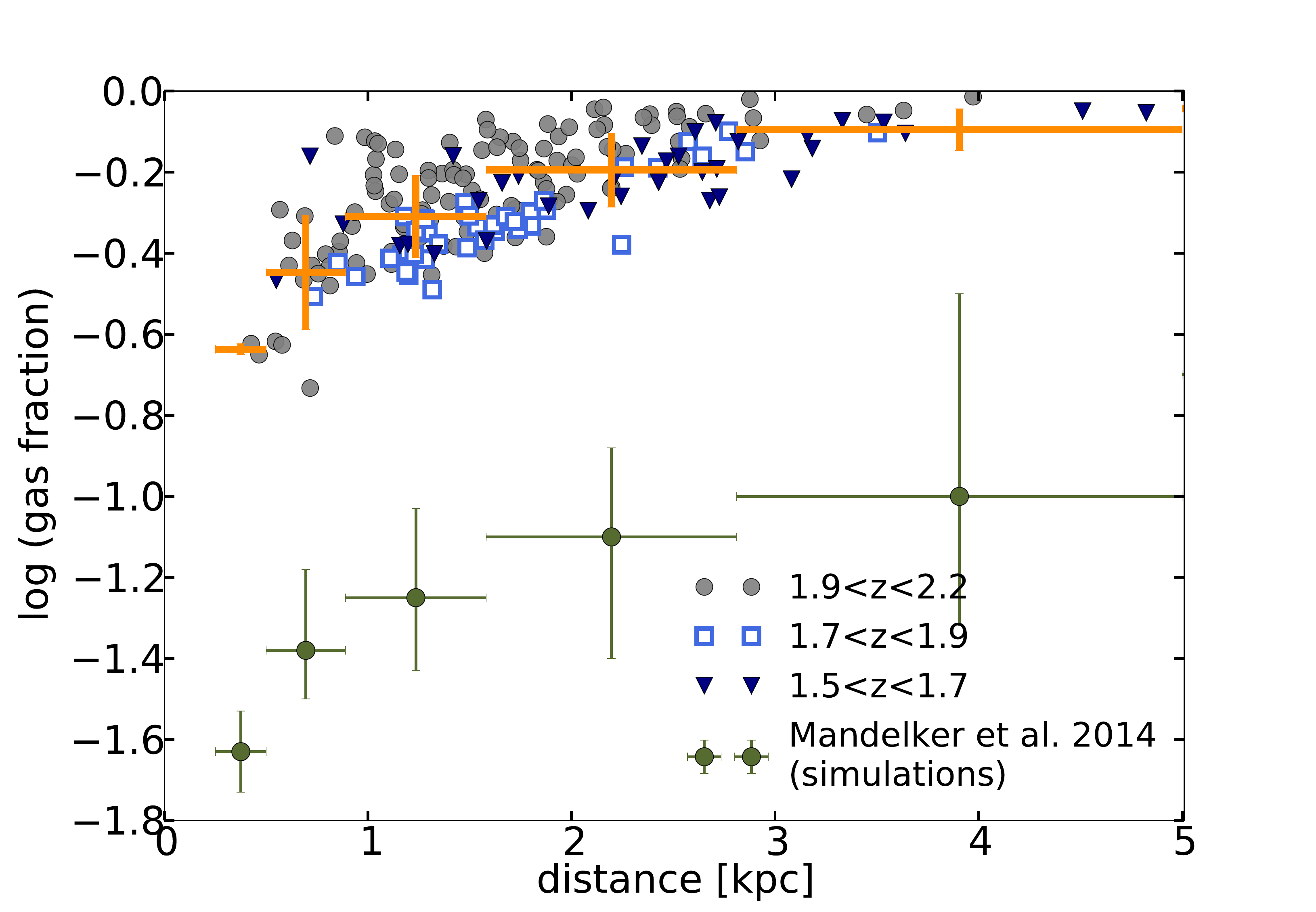}
\caption{Different clump properties shown as functions of the clump's distance from the galactic centre. Top and middle panels show the sSFR and the stellar surface density of our clumps, along with the observational results of \citet{Guo12}. Bottom panel shows the clump gas fractions, compared to simulations of \citet{Mandelker14}. The mean values and standard deviations of our data, binned in the same way as the data we compare them with, are shown in orange.}
\label{fig:gradients}
\end{figure}

In Fig. \ref{fig:gradients} we show the radial dependence of the clump stellar surface density, the specific star-formation rate (sSFR, i.e. the instantaneous SFR of a clump divided by its stellar mass), and gas fraction. We show data for all our clumps, color-coded by redshift as in Fig. \ref{fig:stellar_age}. We compare our results to observations by \cite{Guo12} for the stellar surface density and the sSFR (we use their background-uncorrected results), and to the results from simulations by \cite{Mandelker14} for the clump gas fraction. They report their results in terms of the fractional galactic radius ($d/R$, where $R$ is the radius of the galaxy), hence we multiply their $x$-axis values by $R=5$~kpc, which is the estimated radius of our galaxy. We bin our data in the same way as they did in order to make direct comparison easier. This makes our bins unevenly populated -- while most of them have plenty of clumps, some have only a few members and should be regarded with caution.

Our results for the sSFR of clumps agree very well with the observations by \citet{Guo12} and indicate that the sSFR of clumps slightly increases toward the outer regions of the disc. The scatter is large in both of our data sets, spanning almost the same range (from log(sSFR)$\lesssim$ -1 to log(sSFR)$\gtrsim$ 2). The mean values in each bin are within one standard deviation from each other (caveat: our innermost bin contains only two clumps).

The middle panel of Fig. \ref{fig:gradients} shows the radial dependence of the stellar surface density of clumps. The agreement with the results of \cite{Guo12} is again fairly good.

Finally, we compare our radial dependence of the clump gas fraction with the results from simulations of \citet{Mandelker14} which produced long-lived clumps. Although the radial trends are similar, our clumps have systematically higher gas fractions by $\sim$1~dex. Measuring the absolute gas fractions in high-redshift clumps may be more informative than measuring the radial gradients for the purpose of distinguishing between different (i.e. short- vs. long-lived) clump models. This is consistent with the results of \citet{Mandelker15} who found significantly higher gas fractions in the population of their short-lived clumps, compared to the gas fractions of their long-lived clumps (although the radial dependence of the gas fraction they obtained for the shot-lived clumps is flatter than ours).

\subsection{Clump lifetime}

Snapshots used in the previous section to describe physical properties of clumps in \textit{m13} were not useful for estimating their lifetimes. Typical time differences between two consecutive snapshots in the series are $15-40$~Myr (it varies with redshift), which is too long to follow individual clumps from snapshot to snapshot as the galaxy's morphology can change considerably over that time scale. In order to track individual clumps we re-run our simulation from six starting points and evolve it for 50 Myr in each case, taking snapshots every few Myr\footnote{For re-runs starting at $z=2.0$ and $z=1.5$ we tried out several different time steps ($\Delta t$) ranging from 0.5-5 Myr to find the most appropriate one -- short enough to allow us to track individual clumps, but long enough to reduce the number of needed snapshots to a manageable size. We find that $\Delta t\sim 2$~Myr is a good compromise, and we use that time step for the remaining four re-runs.}.

After identifying gas clumps in all additional snapshots using the same procedure as described in Sec \ref{sec:clump_identification}, we determine the history of each clump from snapshot to snapshot by visual inspection. We do not trace the evolution of clumps by following their member-particles (by their IDs) because we want to avoid making any assumptions regarding how the content of individual clumps may change in time.

Using this method we are able to constrain the lifetime of each clump -- the time difference between the first and last snapshot in which that region of the disc is identified as a clump. For clumps that are already present in the first or still present in the last snapshot in a 50-Myr series, we are only able to set a lower limit on the lifetime using this method. For that reason, we select a sample of clumps -- those that are \textit{formed within the first 10 Myr} of a re-run -- and use only those clumps for the lifetime analysis. In other words, we exclude all clumps that are already present in the first snapshot in a 50 Myr series (because we do not know when these clumps were formed) and all clumps that form later than 10 Myr into a re-run (because they might not have enough time in the remaining period to fully evolve).

\begin{figure}
\centering
\includegraphics[width=0.5\textwidth]{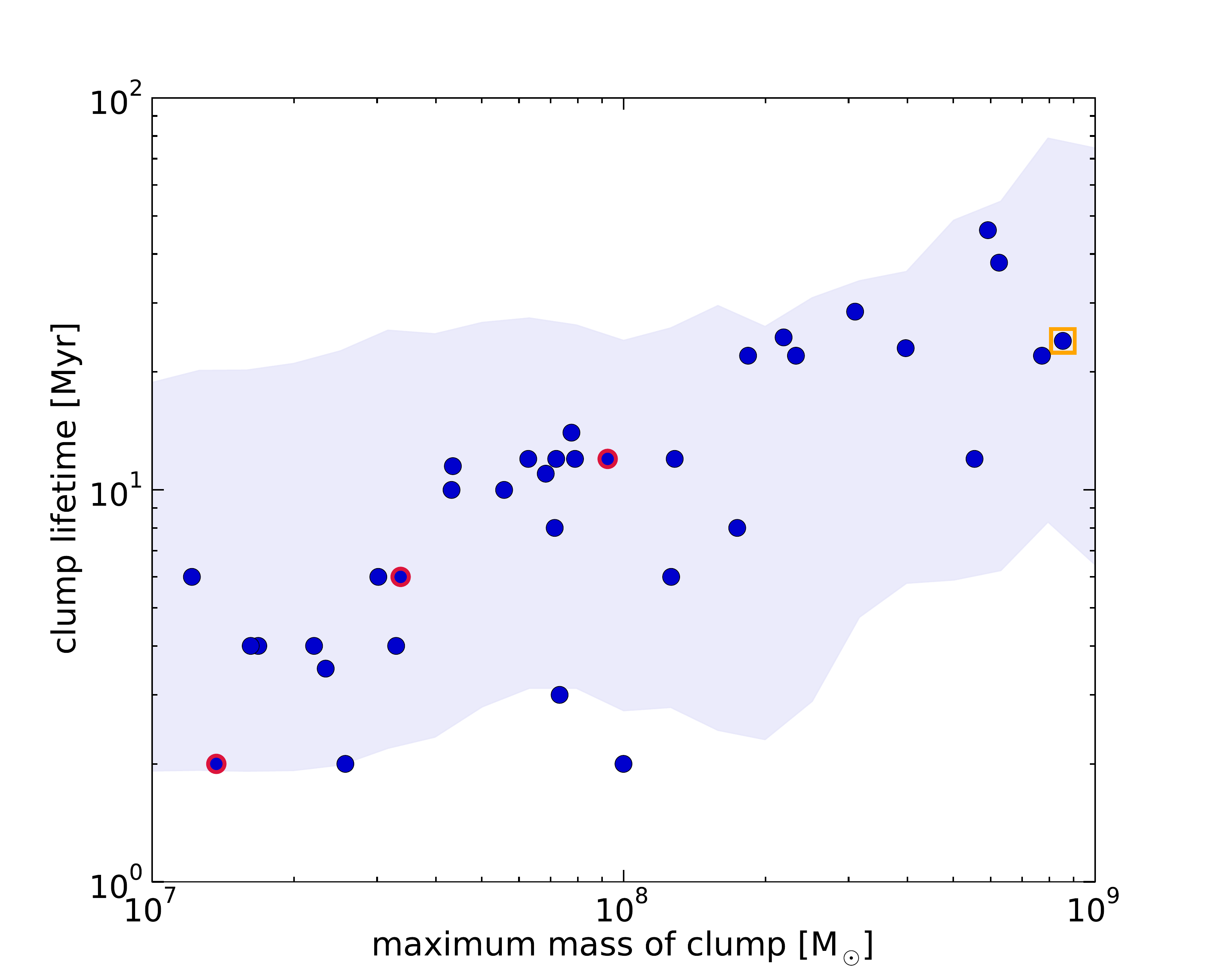}
\caption{Clump lifetime as a function of clump's maximum mass. We show only clumps that are formed within the first 10 Myr of a re-simulated period. Clumps that end their lives by merging with other clumps are indicated in red. The clump marked by a yellow square ended its life by falling to the centre of the galaxy. The shaded region shows the $10-90\%$ range of clump lifetimes as a function of their mass, for clumps found in high-resolution simulations of isolated galaxies by \citet{Hopkins12}.}
\label{fig:lifetime}
\end{figure}

Fig. \ref{fig:lifetime} shows the lifetime of each clump in this sample as a function of its maximum mass. Lifetimes range between 2 and 48 Myr, and the mean lifetime of clumps with mass above $10^8$~M$_{\odot}$ is$\sim 22$ Myr. There were no clumps that formed within the first 10 Myr of a 50 Myr re-simulated period that managed to survive until the very end -- they all got disrupted before the last snapshot in that series (except for one clump, marked by a yellow square in Fig. \ref{fig:lifetime}, that sank to the centre of the galaxy and merged with the central overdensity, which is excluded from our clump sample by definition). Smaller clumps sometimes end their lives by merging with more massive ones. If two or more clumps merge, the newly-formed clump carries the designation of the most massive progenitor, while the smaller ones cease to exist. More massive clumps typically live longer than less massive ones, which is consistent with the results from isolated disc simulations in \cite{Hopkins12}. In Section \ref{sec:clumpstars} we discuss the fate of stars formed inside of a clump after the gas gets disrupted.

\section{Discussion}
\label{sec:discussion}

\subsection{Clump formation and destruction}

\subsubsection{Clump formation in gas-rich systems}

We follow the evolution of a massive, gas-rich, star-forming galaxy from $z=2.2$ to $z=1.0$ and investigate the nature of massive gas clumps formed within it. During the first $\sim 1$~Gyr ($2.2 \gtrsim z \gtrsim 1.5$), when the gas fraction and velocity dispersion are both high (see Fig. \ref{fig:mass_redshift2}), the morphology of the gaseous disc is highly irregular and asymmetric, dominated by a few clumps of increased gas surface density. Some of the clumps look compact and nearly spherical, whereas some are elongated filaments that often break up into smaller sub-clumps. At later times ($z\lesssim1.5$), as the gas fraction and velocity dispersion drop to roughly half of their initial value and the cold gas component stabilizes into a more regular disc of lower surface density, the gas overdensities (i.e. clumps) become smaller and less frequent. This change with redshift is related to the lack of gas-rich mergers and decreased gas accretion. At later times there are no obvious filaments of in-flowing gas and the overall gas content of the galaxy goes down.

\subsubsection{Stellar clumps}

The analysed clumps are identified in the gas surface density maps of the \textit{m13} galaxy. We do not see obvious clumps in stellar maps that include all stars (young and old). If we look at the young stellar population only, we see a certain level of clumpiness, as show in the maps of the SFR (Fig. \ref{fig:maps190b}). Prominent substructures in the total stellar component are associated with smaller galaxies falling into the potential well of \textit{m13}, orbiting around and merging with it. One merger event, with a galaxy of initial stellar mass of $1.6\times 10^9$\msun, starts a few tens of Myr before our first analysed snapshot, and lasts until $z\sim1$. Another galaxy with stellar mass $\sim 10^9$\msun\ passes through the \textit{m13}'s potential well around $z\sim 1.6$. Both of these smaller galaxies lose or exhaust their gas during the first passage through the densest regions of \textit{m13}. After that, they appear only as stellar overdensities and we do not identify them as (gas) clumps. However, their presence might contribute to the disturbed state of the gaseous disc of \textit{m13}.

\subsubsection{Clump destruction and lifetimes}

Giant gas clumps in \textit{m13} have relatively short lives, of order $10^7$~yr. There is a correlation of clump lifetime and mass -- more massive clumps tend to live longer. Smaller clumps can end their lives by merging with more massive ones or by dissolving as their surface density drops below the threshold that is used to define clumps. Massive clumps often end their lives by breaking up into smaller pieces, which can form a ring-like structure, indicating that the clump was disrupted from within by its own stellar feedback. These fragments often merge with other gas clouds to form new clumps. 

The lifetimes of clumps that we measure in \textit{m13} are consistent with the result of \cite{Hopkins12}, as shown in Fig. \ref{fig:lifetime}. They studied clumps in isolated galaxies with similar implementation of stellar feedback as is used in this work but at much higher resolution. Even though their methods for finding clumps and measuring clump lifetime are different from those that we employ, the conclusions derived from their work and ours are very similar. On the other hand, cosmological simulations that include only thermal feedback from supernovae and stellar winds, \citep{Ceverino10, Ceverino2012, Mandelker14} often produce giant clumps that live long lives ($\gtrsim 10^8$ yrs) and migrate to the centre of their discs. Unlike the FIRE simulations, many previous studies do not resolve the evolution of supernovae and hence do not account for the momentum accumulated in the process. Direct comparison between our results and the results presented by these authors is complicated due to many differences in the way we define clumps and measure clump properties, however it seems that including radiation pressure feedback in their simulations (although their treatment of radiation pressure is different from ours) decreases the lifetime of their giant clumps and makes our results qualitatively more similar \citep{Moody14, Mandelker15}.

Another difference between our simulation and some of the previous analyses of clumpy galaxies, especially those using non-cosmological simulations, is our moderate value of the galaxy gas fraction ($f_g\sim 0.3$ at $z=2$, roughly 70\% of which is in the cold phase characterized by temperature $\lesssim 10^4$~K) compared to values of the molecular gas fractions observed in high-redshift galaxies which reach as high as $\sim$0.4 - 0.6 \citep{Tacconi08,Tacconi10}. In their studies, \citet{DekelKrumholz2013, Mandelker14,Bournaud14} argue that clumps can accrete gas from the surrounding gas-rich regions, thus prolonging their lives by replenishing the gas they lost due to stellar feedback and star formation. Since we do see a correlation between high gas fraction and the occurrence of massive clumps, and also between the mass of clumps and their lifetime, it is possible that having higher gas fractions in our simulation might produce some more massive clumps with longer lifetimes. However, it is unlikely that this effect alone can explain the discrepancy in estimates of clump lifetimes, because the results of \citet{Hopkins12}, who studied isolated discs with high gas fractions (initial $f_g \sim 0.6$) found clump lifetimes not too different from ours using a similar implementation of stellar feedback.

Different definitions of clumps may also lead to differences in the estimates of clump lifetimes. Increasing the density threshold (or the minimum clump surface area) would lead to shorter clump lifetimes because any region of the disc would be less likely to meet these more restrictive criteria in order to be identified as a clump. Hence, if we used a more restrictive definition of clumps that identified only very dense structures (more similar to what we call sub-clumps) as clumps, it is likely that this would result in even shorter clump lifetimes than those shown in Fig. \ref{fig:lifetime}.

Comparing the clump lifetime (Fig. \ref{fig:lifetime}) with the age of stellar population in clumps (Fig. \ref{fig:stellar_age}) we find that the latter is about an order of magnitude greater than the former. This indicates that a significant fraction of stars that reside in clumps were not actually formed there, but belong to an older stellar population whose location happens to coincide with the clump or which have been accreted by the clump. This `contamination' by older stars may be more significant in the inner regions of the galaxy, which are more densely populated than the outskirts. This may explain the observed anti-correlation between galactocentric radius and stellar age of clumps. A similar anti-correlation exists in the overall stellar population in the inner parts of \textit{m13} and could simply be translated to clumps. Therefore, measuring stellar ages of clumps might not be a reliable way to establish the absolute age of clumps themselves or the relative age of one clump to another.

\subsubsection{Clump orbits and (lack of) inspiral}

\begin{figure}
\centering
\includegraphics[width=0.48\textwidth]{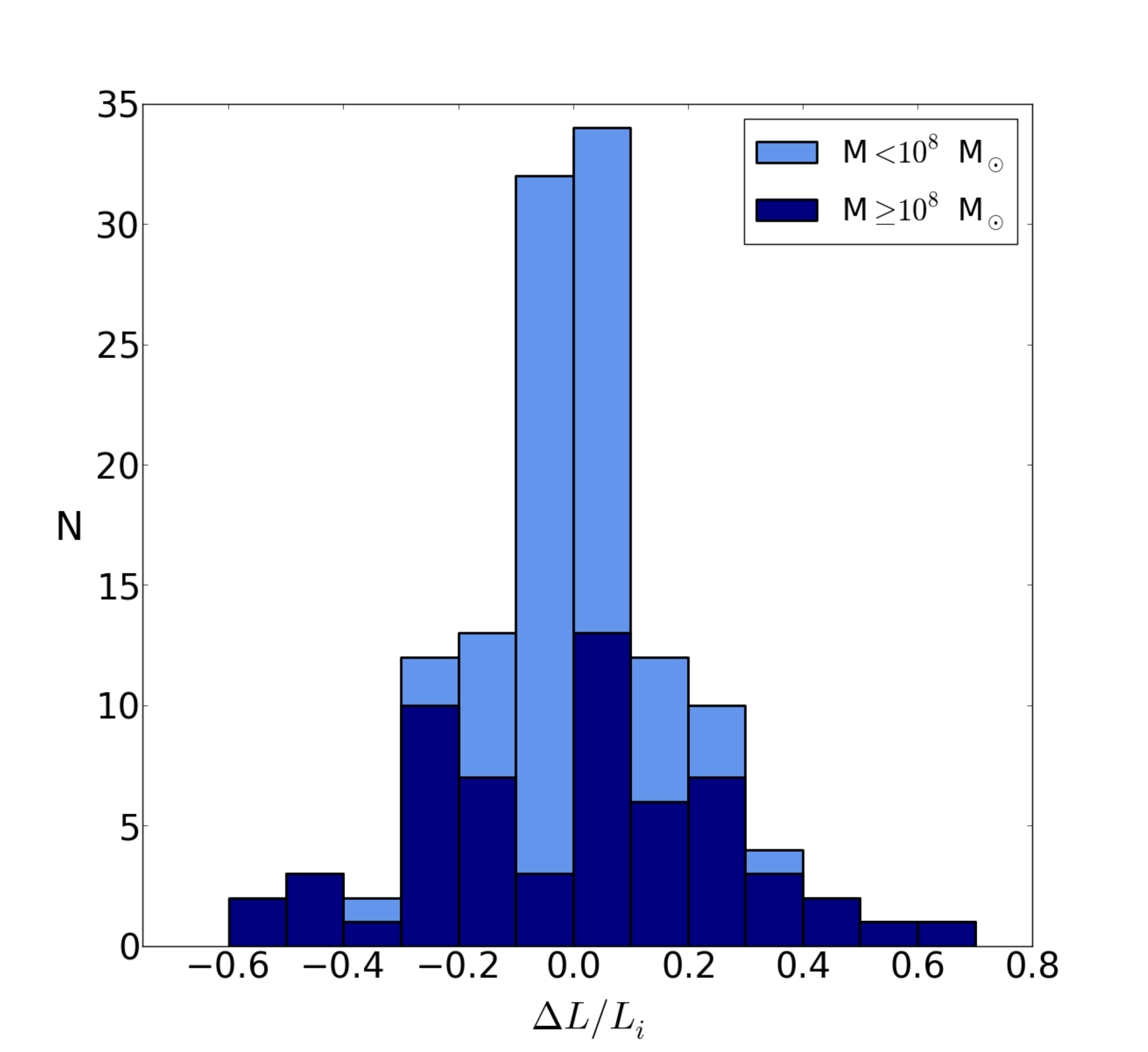}
\caption{Fractional change of clump's orbital angular momentum per unit mass over the course of clumps lifetime. Clumps change their specific orbital angular momentum by up to $\sim 50\%$, and they are equally likely to gain angular momentum as they are to lose it. This result does not support the scenario in which clumps systematically migrate toward the centre of the galaxy.}
\label{fig:deltaL}
\end{figure}

On average, clumps follow the overall rotation of the disc, with some deviations, presumably induced by gravitational torques. Several clumps end up reaching the centre of the galaxy, but interestingly most of those that do start their descent from nearly the same location in the disc. Some other clumps at similar distances from the centre, but at different locations, have very different trajectories and do not end up in the centre. 

The change in the clump's galactocentric radius is related to its loss of orbital angular momentum. To get a sense of how this quantity changes over the lifetime of a clump, in Fig. \ref{fig:deltaL} we show the difference in the specific orbital angular momentum of clumps, i.e. the sum of angular momenta (with respect to the centre of the galaxy) of all clump particles (gas and stars) divided by the mass of the clump
\begin{equation}
L = \frac{\sum_j m_j \boldsymbol{v}_j \times \boldsymbol{r}_j}{\sum_j m_j} \ \mbox{,}
\end{equation} 
between the final and initial snapshot of that clump, expressed as a percentage of its initial angular momentum ($\Delta L /L_i = (L_{f}-L_{i})/L_{i}$). We find that clumps experience a change in their orbital angular momentum by a few tens of percent (up to $\sim 50\%$) and that roughly equal numbers of clumps experience inward migration as outward. Small clumps on average experience smaller changes in their angular momentum compared to more massive ones, due to their shorter lifetimes. 

Our finding that giant clumps are equally likely to gain angular momentum over the course of their lifetime as they are to lose it indicates that clump migration on such short time-scales might be dominated by gravitational torquing and tidal forces due to the asymmetric state of the disc, inflowing gas streams and orbiting companion galaxies, rather than due to dynamical friction which acts on longer times scales and would produce a `one-sided' systematic trend of angular momentum loss, as seen in simulations that produce long-lived clumps \citep[e.g.][]{Ceverino10, Ceverino2012, Bournaud14,Mandelker14}.

\subsection{Are clumps gravitationally bound?}

\begin{figure}
\centering
\includegraphics[width=0.44\textwidth]{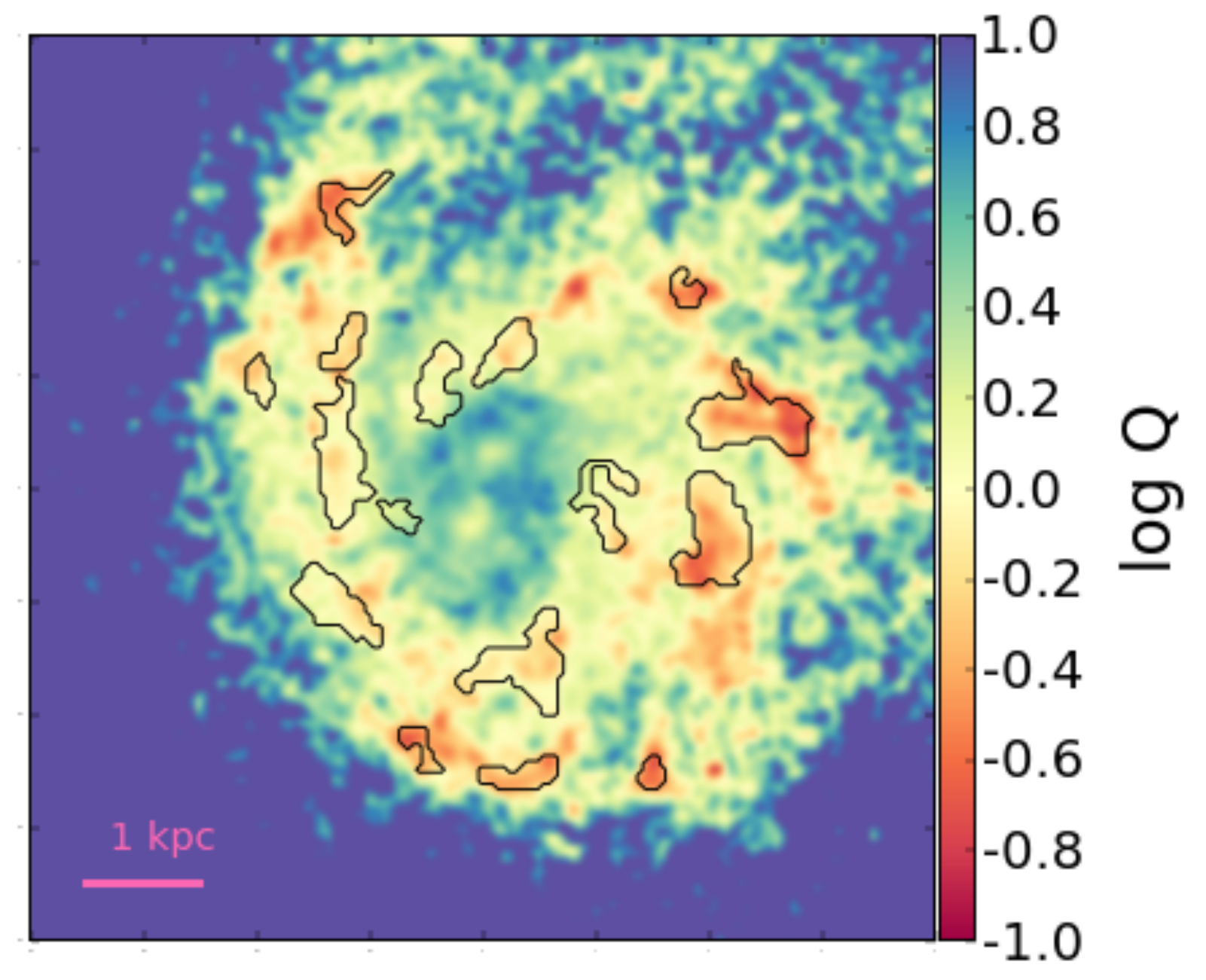}
\caption{Map of Toomre $Q$ parameter of the gas disc at $z\approx 1.9$. Clumps identified in this snapshot are overlaid as black contours. Most clumps, especially those in the early phase of their life, coincide with regions of low $Q$.}
\label{fig:toomre}
\end{figure}

\begin{figure}
\includegraphics[width=0.49\textwidth]{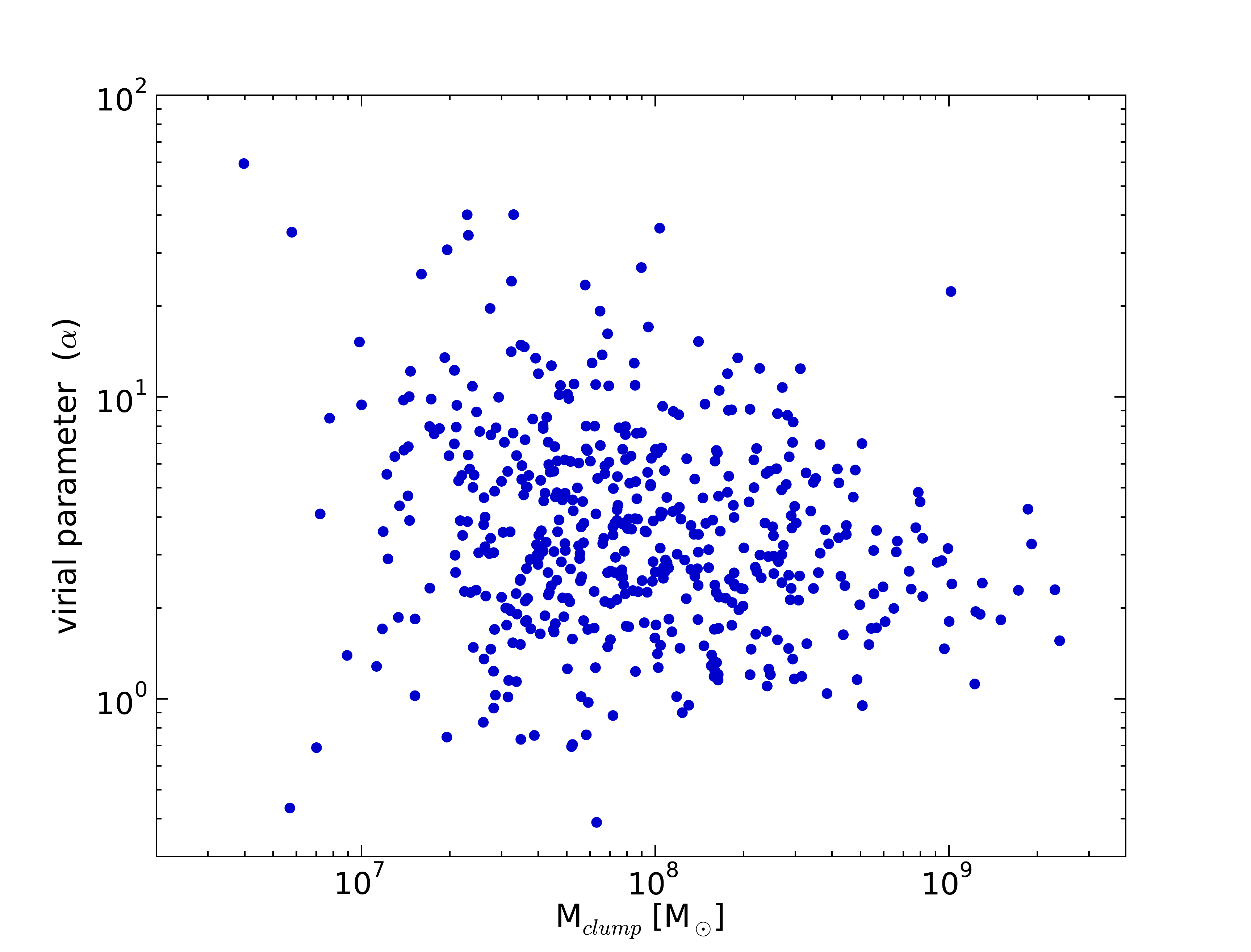}
\caption{Virial parameter $\alpha$ of each clump in the original series of \textit{m13} clumps as a function of their mass. The virial parameter is a ratio of clump's virial mass to its actual mass and can be used to evaluate whether a clump is gravitationally bound or not. Values of $\alpha \gtrsim 1$ indicate that clumps are only marginally bound or unbound.}
\label{fig:virial}
\end{figure}

Empirically determined values of Toomre $Q$ parameter in clumpy galaxies at high redshift \citep{Genzel11} suggest that giant clumps reside in unstable regions of the disc, characterized by $Q$-values below unity. In order to compare our simulation with these observations, we create maps of the `observed' $Q$-parameter (for gas) by locally evaluating Eq. (\ref{eq:toomre}) using the line-of-sight velocity dispersion ($\sigma_z$) and the orbital frequency $\Omega$ (calculated from the local azimuthal velocity) instead of the epicyclic frequency $\kappa$.

Maps of Toomre $Q$ values throughout the disc (Fig. \ref{fig:toomre}) indicate that the locations of many clumps coincide with regions of the disc that are gravitationally unstable ($Q<1$). However, there is large scatter. Often, regions of low $Q$ correspond to areas that are still in the process of turning into clumps. Similarly, `mature' clumps that have already sustained substantial star formation can have values of $Q>1$. This is not surprising because the $Q$ parameter was developed in the context of linear perturbation theory and is only useful as a diagnostic of the onset of gravitational instability and the early phases of fragmentation. When the fragmentation is already under way, we move into non-linear regime and Toomre $Q$ is no longer a valid diagnostic.  

Our results are in agreement with those of \citet{Inoue2016} who performed a detailed study of Toomre Q in simulated high-redshift clumpy galaxies. They found that regions of the disc where clumps form are typically characterized by lower values of Q compared to the interclump regions. However, their clumps often form in regions of $Q\geq 2-3$, indicating that the critical value of $Q=1$ may not necessarily be a good criterion for fragmentation in realistic discs in which many assumptions of the conventional Toomre analysis break down. Their results may also be interpreted as an indication that some mechanism other than the Toomre instability may play a role in the formation of clumps in high-redshift discs.

Some previous studies of simulated clumpy discs \citep[e.g.][]{Ceverino10, Bournaud14,Mandelker14} found clumps to be roughly spherical, gravitationally bound structures able to maintain their compact morphology for hundreds of millions of years. Clumps that we find in \textit{m13} are qualitatively very different. Their morphology is more complex: some clumps seem compact, while others are very elongated, with prominent substructures along the filaments. The shape of clumps can change over the course of their (short) lifetime as clumps stretch, break up, and merge. This behavior does not support the picture of clumps as long-lived, virialized (gravitationally bound and relaxed) structures. The most probable reason for this qualitative difference between our clumps and clumps found in some of the earlier studies is the treatment of stellar feedback. Many simulations so far have used relatively simple models of feedback that mostly include thermal energy from supernovae, but ignore some other types of feedback present in the early phase of stellar evolution. The additional feedback modes that are included in the FIRE simulations, especially radiation pressure, may play an important role in determining clump properties, as discussed in \cite{Hopkins12}.

Fig. \ref{fig:virial} shows the virial parameter ($\alpha$) of each clump in the original series of \textit{m13} snapshots. It is defined as the ratio of the virial mass of each clump and its actual mass \citep{Bertoldi1992}:
\begin{equation}
\alpha = \frac{M_{vir}}{M}= \frac{5\sigma^2R}{GM} \ \mbox{,}
\label{eq:virial}
\end{equation}
where $\sigma$ is the 1D velocity dispersion (in our case $\sigma_z$) of gas particles in the clump and $R$ is the clump radius. The virial parameter of a clump is related to the ratio of its kinetic and potential energy and can be used to gauge whether a clump will collapse under its own gravity. Structures with $\alpha \lesssim 1$ are considered gravitationally bound\footnote{The expression for the virial parameter (Eq. \ref{eq:virial}) can include a constant factor of order unity that depends on the clump density profile. This may change the criterion for virial equilibrium ($\alpha\simeq 1$) by a factor of a few, as discussed by \citet{Mandelker15}.}, whereas those with $\alpha > 1$ are unbound -- they have enough kinetic energy to overcome its self-gravity and expand into the surrounding medium. The results in Fig. \ref{fig:virial} tell us that some clumps might be gravitationally bound. However, a large fraction of them seem not to be, or at least not entirely. Smaller substructures within clumps must be self-gravitating, otherwise they would not be forming stars (that is one of our imposed criteria for star formation; see Section \ref{sec:simulations} for more details). Giant clumps may hence be regarded as massive, fragmented clump complexes, which could explain why they are easily sheared apart and destroyed by feedback. 

Giant clumps, especially the most massive ones at $z\sim 2$, almost always show signs of substructure -- from when they were formed until they get destroyed and broken up into smaller fragments. The remaining fragments often get `recycled' into clumps by coming together or by being accreted onto a larger clump (clumps that end their lives by merging with larger clumps are highlighted in Fig. \ref{fig:lifetime}). Similarly, \cite{Behrendt2015} find evidence of substructure within giant clumps in their high resolution simulation of an isolated galaxy. They suggest that giant clumps seen at high redshifts might just be groups or clusters of smaller ($\sim$ 100 pc) clumps observed with coarse angular resolution. This is in agreement with the results of \citet{Tamburello15} who also studied the properties of clumps in high resolution simulations of isolated discs. They defined clumps as gravitationally bound objects and found that most clumps in their sample have mass of $\sim 10^7$\msun, and that clumps with mass $\geq 10^8$\msun (the mass typically associated with giant clumps) are rare. Because \citet{Tamburello15} use a different definition of clumps compared to our treatment (i.e. we do not assume the gravitationally bound state of clumps a priori), their clump sample might better correspond to what we call sub-clumps in this study.

\subsubsection{Where do the clump stars go?}
\label{sec:clumpstars}

\begin{figure*}
\centering
\includegraphics[height=0.22\textheight]{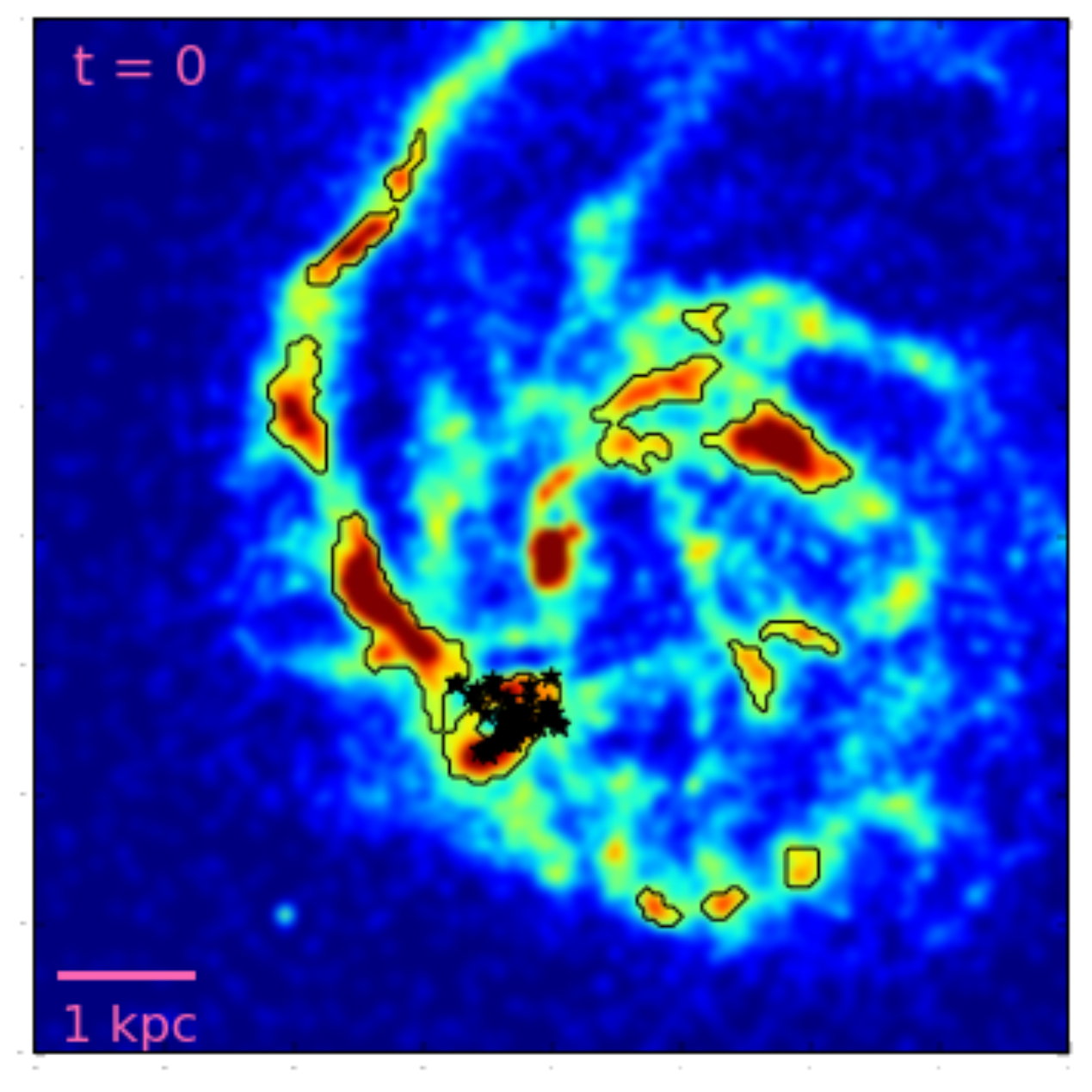}
\includegraphics[height=0.22\textheight]{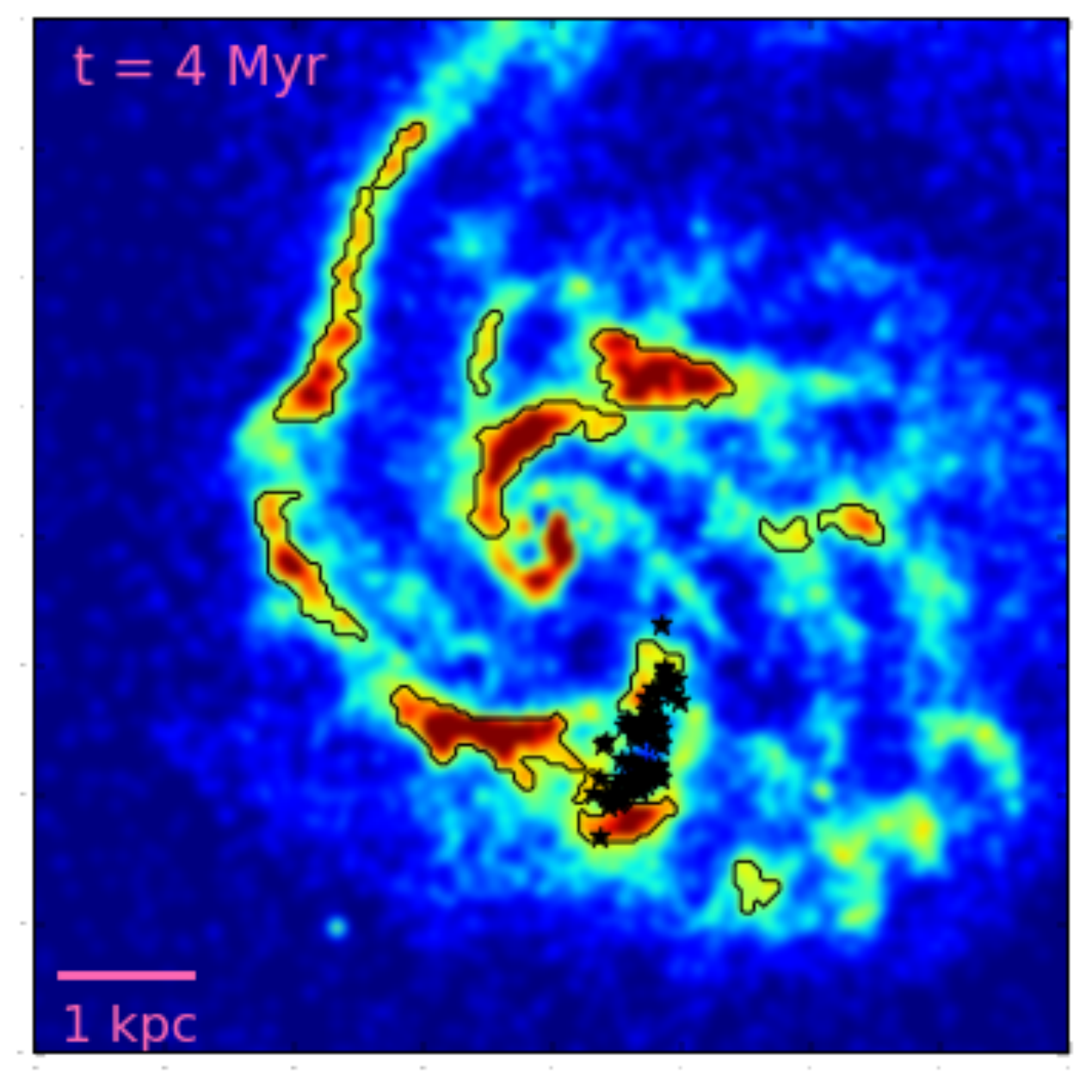}
\includegraphics[height=0.22\textheight]{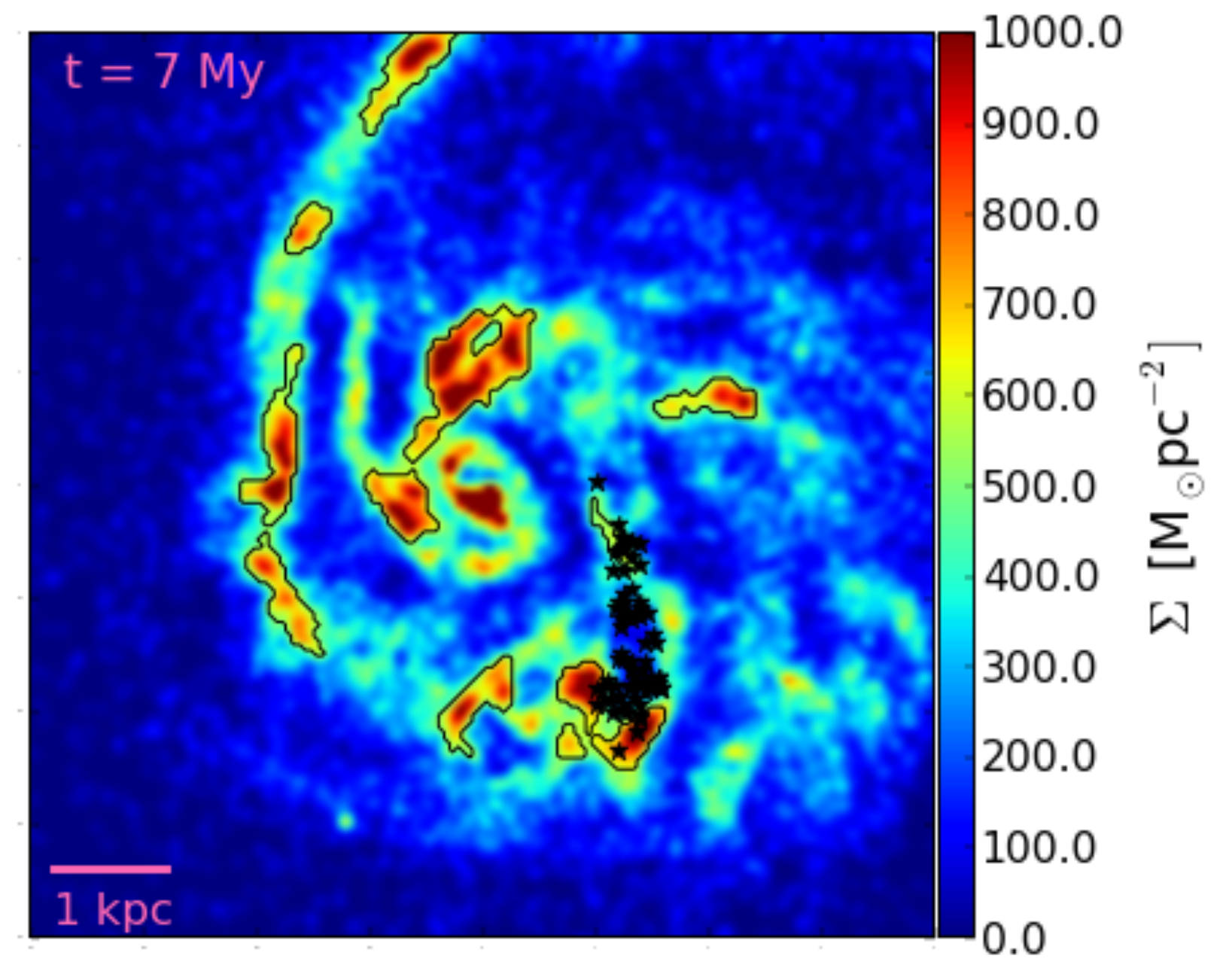}
\includegraphics[height=0.22\textheight]{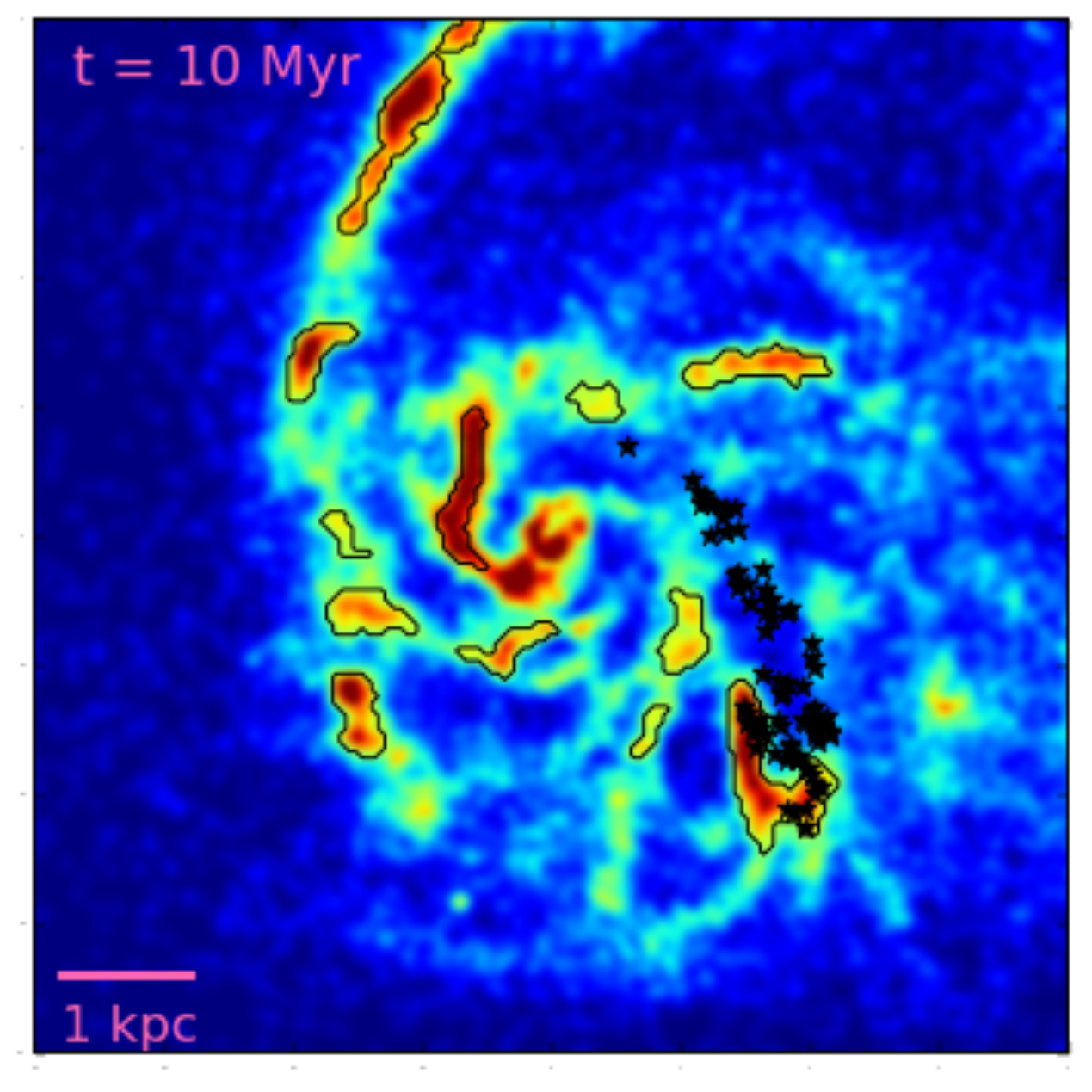}
\includegraphics[height=0.22\textheight]{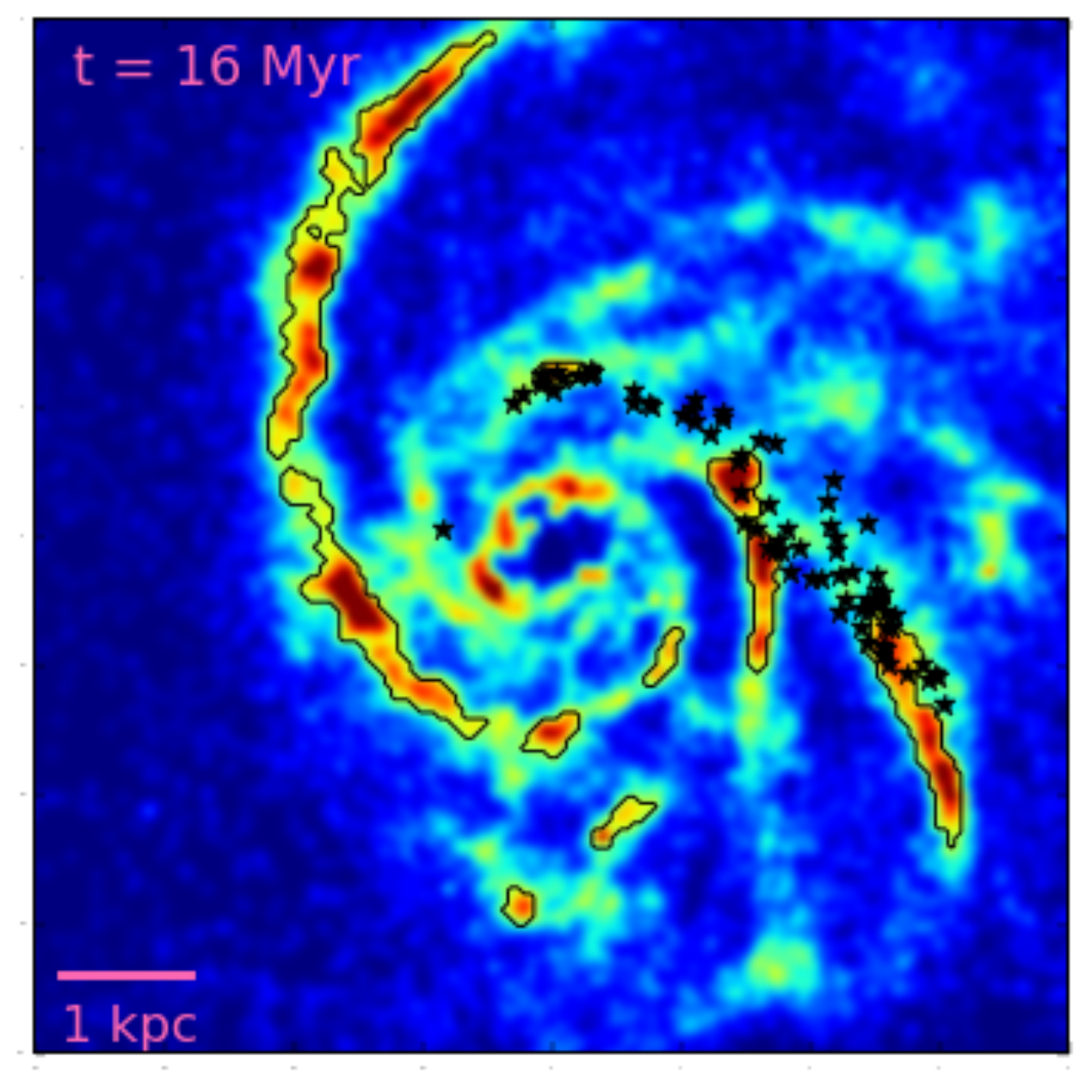}
\includegraphics[height=0.22\textheight]{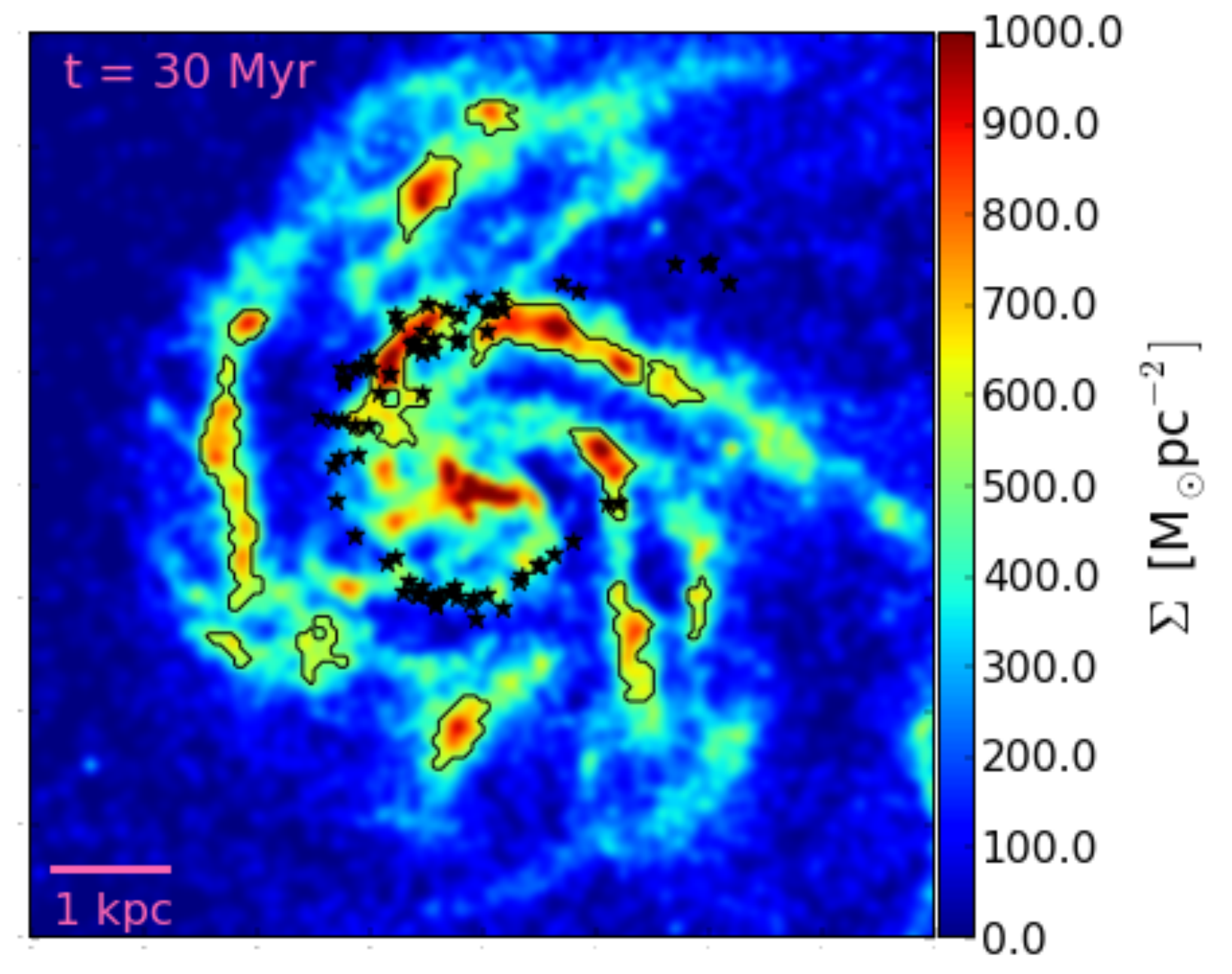}
\caption{A series of snapshots spanning 30 Myr around $z\sim 1.8$. The colormap shows the gas surface density and the black contours represent gas clumps. Black asterisks mark stars that were formed in one of the clumps. The initial snapshot (upper left panel) is taken at the time when the clump starts to break apart. At that point, all the stars are still located in the clump that formed them. Consecutive snapshots were taken at times indicated in the upper left corner with respect to the initial snapshot. Stars that were formed in the same clump are not bound, they gradually drift apart and after 30 Myr form an elongated structure almost 7 kpc long. }
\label{fig:snapseries}
\end{figure*}

Clump lifetime tells us how long the clump exists as an overdensity of gas, but it does not give us insight into what happens to the stars that were formed in the clump after the gas has dispersed. To investigate that, we identify particles in our simulation that were part of a clump as gas particles and then turned into star particles during the lifetime of that clump (particle IDs are preserved as particles turn from gas to stars). Note that each star particle represents a star cluster of mass $\sim 10^5$\msun. We follow the positions of these particles during and after clump's lifetime. We find that stars formed in the same clump do not remain close to each other for a long time -- they get dispersed on time-scales of order few times $10^7$~yr. We demonstrate this in Fig. \ref{fig:snapseries}, which shows a series of snapshots around $z\approx 1.8$. In the first snapshot, marked with `$t=0$', one clump starts to disintegrate. We indicate the positions of star particles that were formed in that clump by black asterisks. In the consecutive images we show the positions of those same particles at later times. After $\sim 30$~Myr, stars that were formed in the same clump are distributed along an arc (or a spiral arm-like feature) $\sim 7$ kpc long.

\subsubsection{Contribution to bulge growth}

One of open questions related to giant clumps in high-redshift galaxies is what is their role in the build-up of the galactic bulge. The \textit{m13} galaxy has a substantial spheroidal component at low redshifts, and its morphology at $z=0$ is that of an elliptical galaxy, with stellar mass of $\sim 10^{11}$\msun. As indicated before, the total mass of stars formed in clumps between $z\approx 2$ and $z\approx 1$ is $\sim 10^{10}$\msun. Even if all that mass ended up in the centre of the galaxy, it would not be enough to account for the entire present-day bulge mass. However, even that is unlikely, as our results do not support the picture in which all clumps migrate to the centre of the galaxy. We find that clumps do not experience very dramatic changes in their orbital angular momentum (up to $\sim 50$\%), and when they do change their radial position, they are equally likely to move inward as they are to move outward.

The results of our simulation therefore do not support the scenario in which giant star-forming clumps form bound stellar clusters in the outer regions of the disc that uniformly sink to the centre of the galaxy, thus building up the bulge. Clumpy morphology of high-redshift galaxies may still play a role in bulge growth by facilitating transport of material across the disc via tidal forces and gravitational torquing. However, that analysis is beyond the scope of this paper.

\subsection{Clumps in comparison with GMCs}

\begin{figure}
\centering
\includegraphics[width=0.48\textwidth]{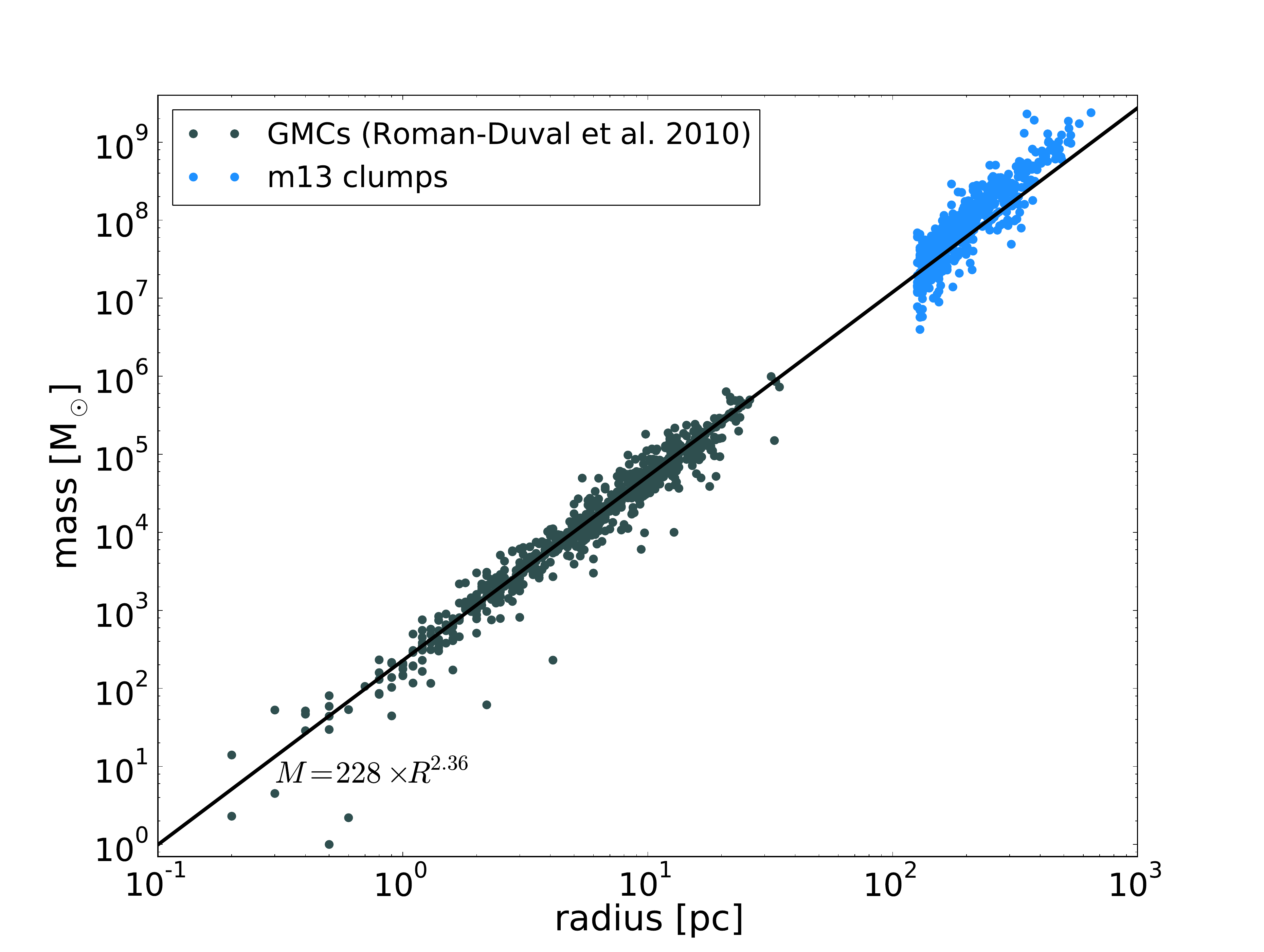}
\caption{Comparison of giant clumps in our simulated high-redshift galaxy and GMCs in the Milky Way. The black line represents the mass-radius relation found for local GMCs by Roman-Duval et al. (2010). Giant clumps identified in the original series of snapshots of \textit{m13} are shown in blue. The simulated giant clumps lie on an extension of the Milky Way GMC relation. The sharp cut-off in radius is a result of the minimum area threshold imposed by our clump-finding procedure. }
\label{fig:compareGMC}
\end{figure}

\begin{figure*}
\centering
\includegraphics[height=0.25\textheight]{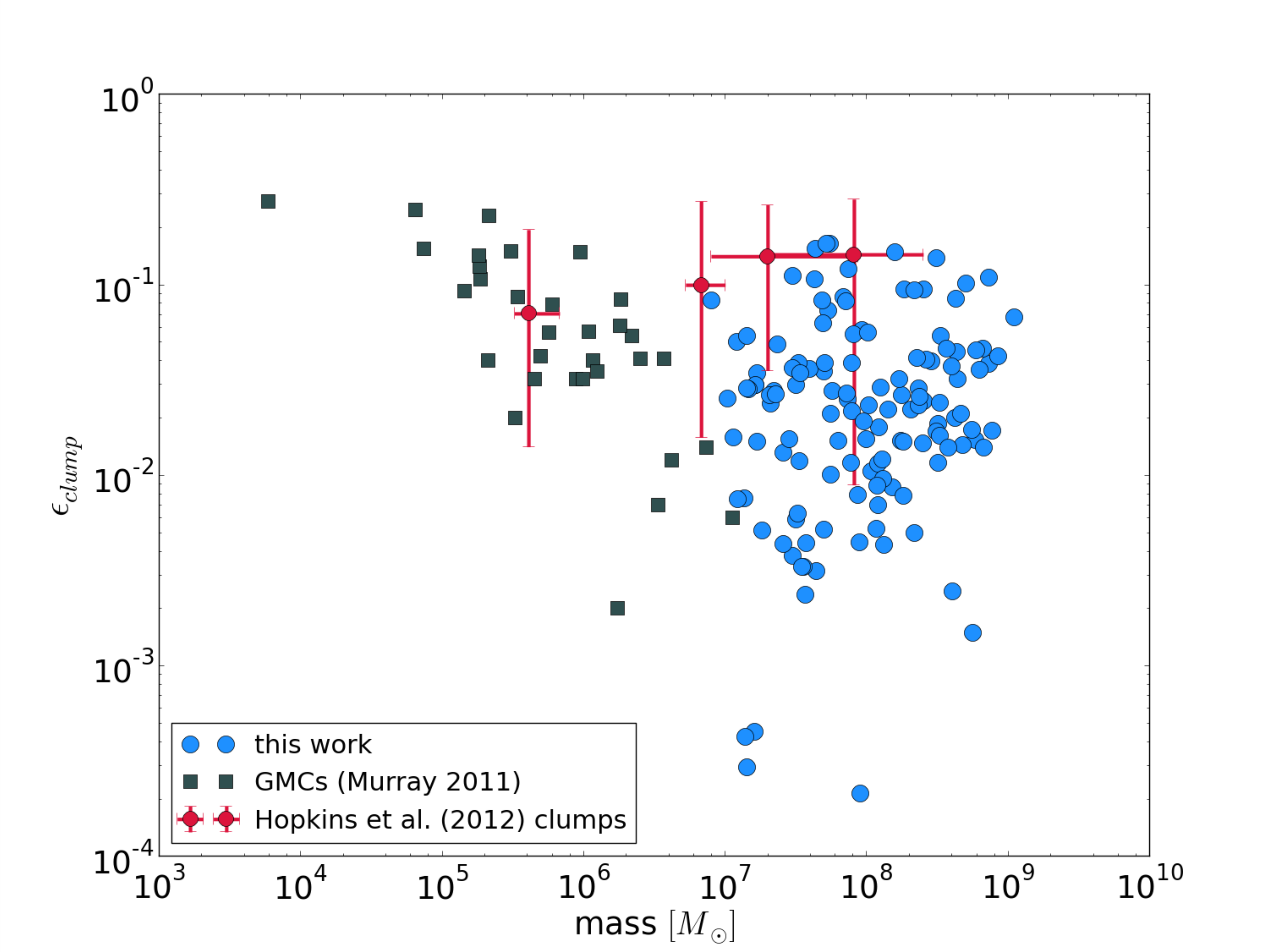}
\includegraphics[height=0.25\textheight]{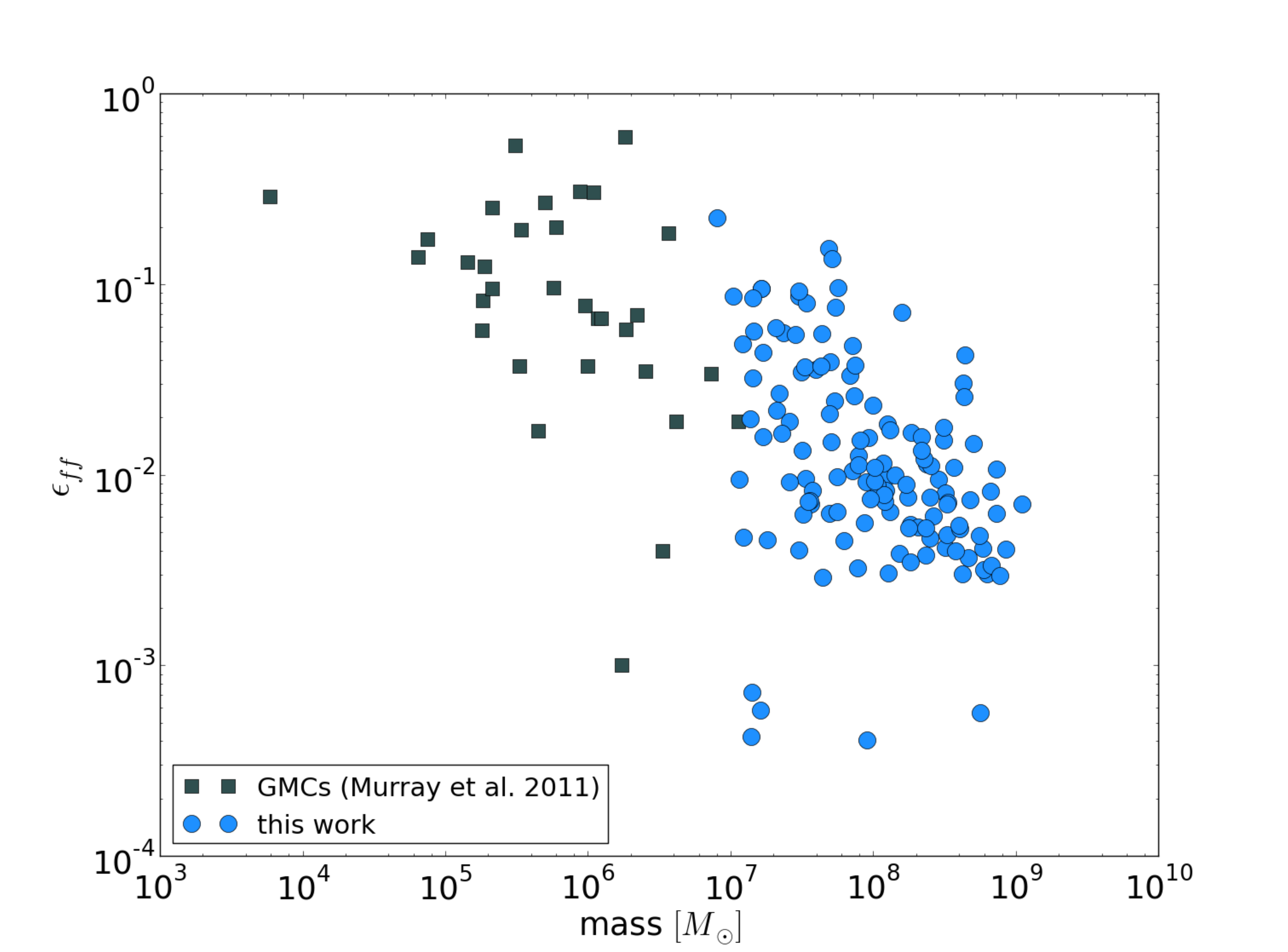}
\caption{Clump star-formation efficiency ($\epsilon_{clump}$, left panel) and star-formation efficiency per free-fall time ($\epsilon_{ff}$, right panel) as functions of mass, compared to the local GMCs (Murray 2011). On the left, we also show the median value and 1$\sigma$ range for clumps found in four high-resolution simulations of isolated galaxies by \citet{Hopkins12}.}
\label{fig:compareGMC2}
\end{figure*}

Finally, we compare giant clumps found in our simulated galaxy with the GMCs observed in the Milky Way. Local GMCs follow a tight relation between their mass and radius \citep{Larson81}, empirically derived to be M[\msun] = 228 R$^{2.36}$[pc] \citep{RomanDuval10}. Fig. \ref{fig:compareGMC} shows that our simulated clumps follow that relation reasonably well, even though their masses and radii are orders of magnitude above those of the GMCs that were used to derive the relation. The sharp cut-off in radius of our clumps is a consequence of our clump identification criteria -- we impose a minimum area of clumps of $5\times 10^4$~pc$^2$, which corresponds to a minimum effective radius of $\sim 125$~pc. 

Next, we compare the star-formation efficiency ($\epsilon_{clump}$) and star-formation efficiency per free-fall time ($\epsilon_{ff}$) of giant clumps to those of local GMCs \citep{Murray11}. To calculate the star-formation efficiency of giant clumps, we use clumps found in six re-runs of our simulation in which we can follow the evolution of individual clumps from snapshot to snapshot. We compute $\epsilon_{clump}$ by calculating the total mass of stars formed in that clump over the course of its lifetime and dividing it by its maximum mass (i.e. total mass of gas and stars in the clump at the time when the clump is at its maximum mass):
\begin{equation}
\epsilon_{clump} = \frac{M_{*, formed}}{M_{clump}} \ \mbox{.}
\end{equation}
To calculate the star formation efficiency per free-fall time $\epsilon_{ff}$, which is more easily obtained in observations than $\epsilon_{clump}$, we multiply $\epsilon_{clump}$ by the ratio of clump's free-fall time ($\tau_{ff}$) and its lifetime. The free-fall time of a clump is given by
\begin{equation}
\tau_{ff}=\sqrt{\frac{3\pi}{32 G \overline{\rho}}} \ \mbox{.}
\end{equation} 
where $\overline{\rho}$ is the mean gas density of the clump calculated at the time when the clump is at its maximum mass. 

The results are shown in Fig. \ref{fig:compareGMC2}. Clumps span a broad range in star-formation efficiency, from less than $10^{-3}$, to a few tens of percent, indicating that giant clumps get destroyed well before they turn all their gas into stars. These values are very similar to star formation efficiency observed in the local GMCs \citep{Murray11}. They are also roughly consistent with the results from four simulations of isolated disc galaxies \citep{Hopkins12} that implement similar feedback processes as this work, but at higher resolution. Note that we use a different clump-finding algorithm and a different definition of clump lifetime than \cite{Hopkins12} did, which may cause a variation, or perhaps a systematic offset, between our results.

Theoretical models by \citet{KrumholzDekel10, DekelKrumholz2013} predict that giant clumps with $M\sim 10^9$\msun\ and characterized by $\epsilon_{ff} \sim 1$\% should be stable and long-lived, which disagrees with our results. Their analytic models differ from our simulations not only in terms of complexity and the ability to treat non-linear effects, but also in assumptions regarding the nature of clumps -- they assume clumps are gravitationally bound objects, whereas we find that clumps are better described as loose complexes of bound sub-clumps (which are easier to disrupt) -- and in the way stellar feedback is implemented (our feedback model gives a higher net momentum flux from stars).

\section{Conclusions}
\label{sec:conclusions}

We use high-resolution cosmological hydrodynamical simulations from the FIRE project that implement explicit models of stellar feedback and ISM physics to investigate giant star-forming clumps in high-redshift galaxies. We present a detailed analysis of a massive ($M_*\sim 10^{10.8}$\msun\ at $z=1$), discy, star-forming galaxy over the redshift range $2.2 \geqslant z \geqslant 1.0$.  Our main findings can be summarized as follows:
\begin{itemize}
\item At any given time around $z\sim 2$ the galaxy hosts a few giant clumps with baryonic masses in the range $\sim 10^8-10^9$\msun\ and radii $\sim 200 -600$~pc. The clumps are identified as overdensities in the gas component of the disc, however most of them also contain significant mass of stars (gas fraction in clumps ranges from 20\% to over 90\%). Clump shape can vary dramatically from clump to clump, from nearly spherical to highly elongated clumps.
\item Maps of stellar surface density in general do not show prominent overdensities at the locations of gas clumps. A few prominent `stellar clumps' found in our simulation belong to smaller galaxies merging with \textit{m13}. 
\item The number of giant clumps decreases with time over the analysed redshift range, and the clumps  become less massive at lower redshift. This follows the Toomre scale ($\sim f_g^3M_{gal}$), given the decreasing galaxy gas fraction over the same time.
\item Individual clump lifetimes are relatively short, typically $\lesssim 20-30$ Myr. More massive clumps on average live longer than less massive ones.  
\item Very massive clumps often end their lives by being disrupted from within and broken apart into smaller pieces by stellar feedback and tidal forces. Less massive clumps gradually fade away below the surface density threshold used to define clumps or merge with larger clumps.
\item  The mass-weighted stellar age of clumps ranges from $\lesssim 100$~Myr to about 1 Gyr. This is much longer than the lifetime of any given gas clump, suggesting that a significant fraction of stars in a clump were formed before the clump itself. They most likely belong to the underlying stellar population of the galaxy that happens to spatially coincide with the gas clump. Therefore, stellar ages of clumps are not necessarily indicative of clumps' own lifetimes. At a given redshift, there is a tendency of clumps with older stellar populations to be located closer to the galactic centre.
\item Most clumps do not seem to be gravitationally bound, based on their often elongated and complex shape and the value of the virial parameter. Smaller regions within clumps are most likely bound and those are the regions were stars are born. The clumps are in fact `complexes' with substantial sub-structure. This facilitates their destruction by stellar feedback. Stars formed in the same clump do not stay together for a long time, they drift apart on a time-scale of $\sim 10^7$ years.
\item During the gas-rich (or `clumpy') phase in the evolution of the galaxy, clumps contain $\sim 10-60$\% of the total star-formation at any given time. They form $\sim 10^{10}$\msun\ of stars over the analysed redshift range, which is $\sim$20\% of the total star formation during that period.
\item Over the course of their short lifetime, clumps can either gain or lose orbital angular momentum. We find that both outcomes are roughly equally likely. There is no evidence of overall inward migration of clumps. Even the most dramatic angular momentum loss causes change only by a factor of $\sim 2$.
\item The lack of evidence for systematic inward migration of clumps, and the total stellar mass produced in clumps, which is only a modest fraction ($\lesssim 20$\%) of the total $z=0$ bulge mass, do not support the scenario in which giant star-forming clumps form bound stellar clusters that sink to the centre of the galaxy, thus forming the bulge. Whether the mere presence of clumps helps the bulge growth by facilitating transport of material across the disc via tidal forces and gravitational torquing is an open question that is not addressed in this work.
\item Giant clumps at high redshift look like scaled-up versions of GMCs in the local Universe. They  lie on the same mass-radius relation and their star-formation efficiencies are similar to those of GMCs. 
\end{itemize}

\section*{Acknowledgements}
We would like to thank the anonymous referee for their comments which helped to improve the quality of the paper. This research made use of astrodendro, a Python package to compute dendrograms of Astronomical data (\url{http://www.dendrograms.org/}). AO thanks Stella Offner for recommending the use of astrodendro for this work. AO is supported by the U.S. Department of Energy, the David \& Lucile Packard Foundation, and the Simons Foundation. Support for PFH was provided by an Alfred P. Sloan Research Fellowship, NASA ATP Grant NNX14AH35G, and NSF Collaborative Research Grant \#1411920 and CAREER grant \#1455342. Numerical calculations were run on the Caltech compute cluster ``Zwicky'' (NSF MRI award \#PHY-0960291) and allocation TG-AST130039 granted by the Extreme Science and Engineering Discovery Environment (XSEDE) supported by the NSF. DK was supported by NSF grant AST-1412153. CAFG was supported by NSF through grants AST-1412836 and AST-1517491, by NASA through grant NNX15AB22G, and by STScI through grant HST-AR-14293.001-A.




\bibliographystyle{mnras}
\bibliography{references} 

\begin{thebibliography}{}
\makeatletter
\relax
\def\mn@urlcharsother{\let\do\@makeother \do\$\do\&\do\#\do\^\do\_\do\%\do\~}
\def\mn@doi{\begingroup\mn@urlcharsother \@ifnextchar [ {\mn@doi@}
  {\mn@doi@[]}}
\def\mn@doi@[#1]#2{\def\@tempa{#1}\ifx\@tempa\@empty \href
  {http://dx.doi.org/#2} {doi:#2}\else \href {http://dx.doi.org/#2} {#1}\fi
  \endgroup}
\def\mn@eprint#1#2{\mn@eprint@#1:#2::\@nil}
\def\mn@eprint@arXiv#1{\href {http://arxiv.org/abs/#1} {{\tt arXiv:#1}}}
\def\mn@eprint@dblp#1{\href {http://dblp.uni-trier.de/rec/bibtex/#1.xml}
  {dblp:#1}}
\def\mn@eprint@#1:#2:#3:#4\@nil{\def\@tempa {#1}\def\@tempb {#2}\def\@tempc
  {#3}\ifx \@tempc \@empty \let \@tempc \@tempb \let \@tempb \@tempa \fi \ifx
  \@tempb \@empty \def\@tempb {arXiv}\fi \@ifundefined
  {mn@eprint@\@tempb}{\@tempb:\@tempc}{\expandafter \expandafter \csname
  mn@eprint@\@tempb\endcsname \expandafter{\@tempc}}}

\bibitem[\protect\citeauthoryear{{Adamo}, {{\"O}stlin}, {Bastian},
  {Zackrisson}, {Livermore}  \& {Guaita}}{{Adamo} et~al.}{2013}]{Adamo2013}
{Adamo} A.,  {{\"O}stlin} G.,  {Bastian} N.,  {Zackrisson} E.,  {Livermore}
  R.~C.,   {Guaita} L.,  2013, \mn@doi [\apj] {10.1088/0004-637X/766/2/105},
  \href {http://adsabs.harvard.edu/abs/2013ApJ...766..105A} {766, 105}

\bibitem[\protect\citeauthoryear{{Agertz} et~al.,}{{Agertz}
  et~al.}{2007}]{Agertz2007}
{Agertz} O.,  et~al., 2007, \mn@doi [\mnras]
  {10.1111/j.1365-2966.2007.12183.x}, \href
  {http://adsabs.harvard.edu/abs/2007MNRAS.380..963A} {380, 963}

\bibitem[\protect\citeauthoryear{{Behrendt}, {Burkert}  \&
  {Schartmann}}{{Behrendt} et~al.}{2016}]{Behrendt2015}
{Behrendt} M.,  {Burkert} A.,   {Schartmann} M.,  2016, \mn@doi [\apjl]
  {10.3847/2041-8205/819/1/L2}, \href
  {http://adsabs.harvard.edu/abs/2016ApJ...819L...2B} {819, L2}

\bibitem[\protect\citeauthoryear{{Bertoldi} \& {McKee}}{{Bertoldi} \&
  {McKee}}{1992}]{Bertoldi1992}
{Bertoldi} F.,  {McKee} C.~F.,  1992, \mn@doi [\apj] {10.1086/171638}, \href
  {http://adsabs.harvard.edu/abs/1992ApJ...395..140B} {395, 140}

\bibitem[\protect\citeauthoryear{{Binney} \& {Tremaine}}{{Binney} \&
  {Tremaine}}{2008}]{BinneyTremaine08}
{Binney} J.,  {Tremaine} S.,  2008, {Galactic Dynamics: Second Edition}.
Princeton University Press

\bibitem[\protect\citeauthoryear{{Bournaud}, {Dekel}, {Teyssier}, {Cacciato},
  {Daddi}, {Juneau}  \& {Shankar}}{{Bournaud} et~al.}{2011}]{Bournaud11}
{Bournaud} F.,  {Dekel} A.,  {Teyssier} R.,  {Cacciato} M.,  {Daddi} E.,
  {Juneau} S.,   {Shankar} F.,  2011, \mn@doi [\apjl]
  {10.1088/2041-8205/741/2/L33}, \href
  {http://adsabs.harvard.edu/abs/2011ApJ...741L..33B} {741, L33}

\bibitem[\protect\citeauthoryear{{Bournaud} et~al.,}{{Bournaud}
  et~al.}{2014}]{Bournaud14}
{Bournaud} F.,  et~al., 2014, \mn@doi [\apj] {10.1088/0004-637X/780/1/57},
  \href {http://adsabs.harvard.edu/abs/2014ApJ...780...57B} {780, 57}

\bibitem[\protect\citeauthoryear{{Brooks}, {Governato}, {Quinn}, {Brook}  \&
  {Wadsley}}{{Brooks} et~al.}{2009}]{Brooks09}
{Brooks} A.~M.,  {Governato} F.,  {Quinn} T.,  {Brook} C.~B.,   {Wadsley} J.,
  2009, \mn@doi [\apj] {10.1088/0004-637X/694/1/396}, \href
  {http://adsabs.harvard.edu/abs/2009ApJ...694..396B} {694, 396}

\bibitem[\protect\citeauthoryear{{Ceverino}, {Dekel}  \& {Bournaud}}{{Ceverino}
  et~al.}{2010}]{Ceverino10}
{Ceverino} D.,  {Dekel} A.,   {Bournaud} F.,  2010, \mn@doi [\mnras]
  {10.1111/j.1365-2966.2010.16433.x}, \href
  {http://adsabs.harvard.edu/abs/2010MNRAS.404.2151C} {404, 2151}

\bibitem[\protect\citeauthoryear{{Ceverino}, {Dekel}, {Mandelker}, {Bournaud},
  {Burkert}, {Genzel}  \& {Primack}}{{Ceverino} et~al.}{2012}]{Ceverino2012}
{Ceverino} D.,  {Dekel} A.,  {Mandelker} N.,  {Bournaud} F.,  {Burkert} A.,
  {Genzel} R.,   {Primack} J.,  2012, \mn@doi [\mnras]
  {10.1111/j.1365-2966.2011.20296.x}, \href
  {http://adsabs.harvard.edu/abs/2012MNRAS.420.3490C} {420, 3490}

\bibitem[\protect\citeauthoryear{{Cowie}, {Hu}  \& {Songaila}}{{Cowie}
  et~al.}{1995}]{Cowie95}
{Cowie} L.~L.,  {Hu} E.~M.,   {Songaila} A.,  1995, \mn@doi [\aj]
  {10.1086/117631}, \href {http://adsabs.harvard.edu/abs/1995AJ....110.1576C}
  {110, 1576}

\bibitem[\protect\citeauthoryear{{Dekel} \& {Krumholz}}{{Dekel} \&
  {Krumholz}}{2013}]{DekelKrumholz2013}
{Dekel} A.,  {Krumholz} M.~R.,  2013, \mn@doi [\mnras] {10.1093/mnras/stt480},
  \href {http://adsabs.harvard.edu/abs/2013MNRAS.432..455D} {432, 455}

\bibitem[\protect\citeauthoryear{{Dekel} et~al.,}{{Dekel}
  et~al.}{2009a}]{Dekel09b}
{Dekel} A.,  et~al., 2009a, \mn@doi [\nat] {10.1038/nature07648}, \href
  {http://adsabs.harvard.edu/abs/2009Natur.457..451D} {457, 451}

\bibitem[\protect\citeauthoryear{{Dekel}, {Sari}  \& {Ceverino}}{{Dekel}
  et~al.}{2009b}]{Dekel09a}
{Dekel} A.,  {Sari} R.,   {Ceverino} D.,  2009b, \mn@doi [\apj]
  {10.1088/0004-637X/703/1/785}, \href
  {http://adsabs.harvard.edu/abs/2009ApJ...703..785D} {703, 785}

\bibitem[\protect\citeauthoryear{{El-Badry}, {Wetzel}, {Geha}, {Hopkins},
  {Kere{\v s}}, {Chan}  \& {Faucher-Gigu{\`e}re}}{{El-Badry}
  et~al.}{2016}]{ElBadry15}
{El-Badry} K.,  {Wetzel} A.,  {Geha} M.,  {Hopkins} P.~F.,  {Kere{\v s}} D.,
  {Chan} T.~K.,   {Faucher-Gigu{\`e}re} C.-A.,  2016, \mn@doi [\apj]
  {10.3847/0004-637X/820/2/131}, \href
  {http://adsabs.harvard.edu/abs/2016ApJ...820..131E} {820, 131}

\bibitem[\protect\citeauthoryear{{Elmegreen}}{{Elmegreen}}{2011}]{Elmegreen11}
{Elmegreen} B.~G.,  2011, \mn@doi [\apj] {10.1088/0004-637X/737/1/10}, \href
  {http://adsabs.harvard.edu/abs/2011ApJ...737...10E} {737, 10}

\bibitem[\protect\citeauthoryear{{Elmegreen}, {Elmegreen}, {Ravindranath}  \&
  {Coe}}{{Elmegreen} et~al.}{2007}]{Elmegreen07}
{Elmegreen} D.~M.,  {Elmegreen} B.~G.,  {Ravindranath} S.,   {Coe} D.~A.,
  2007, \mn@doi [\apj] {10.1086/511667}, \href
  {http://adsabs.harvard.edu/abs/2007ApJ...658..763E} {658, 763}

\bibitem[\protect\citeauthoryear{{Elmegreen}, {Elmegreen}, {Fernandez}  \&
  {Lemonias}}{{Elmegreen} et~al.}{2009}]{Elmegreen09}
{Elmegreen} B.~G.,  {Elmegreen} D.~M.,  {Fernandez} M.~X.,   {Lemonias} J.~J.,
  2009, \mn@doi [\apj] {10.1088/0004-637X/692/1/12}, \href
  {http://adsabs.harvard.edu/abs/2009ApJ...692...12E} {692, 12}

\bibitem[\protect\citeauthoryear{{Faucher-Gigu{\`e}re}, {Kere{\v s}}  \&
  {Ma}}{{Faucher-Gigu{\`e}re} et~al.}{2011}]{FaucherGKeresMa11}
{Faucher-Gigu{\`e}re} C.-A.,  {Kere{\v s}} D.,   {Ma} C.-P.,  2011, \mn@doi
  [\mnras] {10.1111/j.1365-2966.2011.19457.x}, \href
  {http://adsabs.harvard.edu/abs/2011MNRAS.417.2982F} {417, 2982}

\bibitem[\protect\citeauthoryear{{Faucher-Gigu{\`e}re}, {Quataert}  \&
  {Hopkins}}{{Faucher-Gigu{\`e}re} et~al.}{2013}]{FaucherG13}
{Faucher-Gigu{\`e}re} C.-A.,  {Quataert} E.,   {Hopkins} P.~F.,  2013, \mn@doi
  [\mnras] {10.1093/mnras/stt866}, \href
  {http://cdsads.u-strasbg.fr/abs/2013MNRAS.433.1970F} {433, 1970}

\bibitem[\protect\citeauthoryear{{Faucher-Gigu{\`e}re}, {Hopkins}, {Kere{\v
  s}}, {Muratov}, {Quataert}  \& {Murray}}{{Faucher-Gigu{\`e}re}
  et~al.}{2015}]{FaucherG15}
{Faucher-Gigu{\`e}re} C.-A.,  {Hopkins} P.~F.,  {Kere{\v s}} D.,  {Muratov}
  A.~L.,  {Quataert} E.,   {Murray} N.,  2015, \mn@doi [\mnras]
  {10.1093/mnras/stv336}, \href
  {http://adsabs.harvard.edu/abs/2015MNRAS.449..987F} {449, 987}

\bibitem[\protect\citeauthoryear{{Faucher-Gigu{\`e}re}, {Feldmann}, {Quataert},
  {Kere{\v s}}, {Hopkins}  \& {Murray}}{{Faucher-Gigu{\`e}re}
  et~al.}{2016}]{FaucherG16}
{Faucher-Gigu{\`e}re} C.-A.,  {Feldmann} R.,  {Quataert} E.,  {Kere{\v s}} D.,
  {Hopkins} P.~F.,   {Murray} N.,  2016, \mn@doi [\mnras]
  {10.1093/mnrasl/slw091}, \href
  {http://adsabs.harvard.edu/abs/2016MNRAS.461L..32F} {461, L32}

\bibitem[\protect\citeauthoryear{{Feldmann}, {Hopkins}, {Quataert},
  {Faucher-Gigu{\`e}re}  \& {Kere{\v s}}}{{Feldmann} et~al.}{2016}]{Feldmann16}
{Feldmann} R.,  {Hopkins} P.~F.,  {Quataert} E.,  {Faucher-Gigu{\`e}re} C.-A.,
   {Kere{\v s}} D.,  2016, \mn@doi [\mnras] {10.1093/mnrasl/slw014}, \href
  {http://adsabs.harvard.edu/abs/2016MNRAS.458L..14F} {458, L14}

\bibitem[\protect\citeauthoryear{{F{\"o}rster Schreiber} et~al.,}{{F{\"o}rster
  Schreiber} et~al.}{2011}]{FSchreiber11}
{F{\"o}rster Schreiber} N.~M.,  et~al., 2011, \mn@doi [\apj]
  {10.1088/0004-637X/739/1/45}, \href
  {http://adsabs.harvard.edu/abs/2011ApJ...739...45F} {739, 45}

\bibitem[\protect\citeauthoryear{{Genel} et~al.,}{{Genel}
  et~al.}{2012}]{Genel12}
{Genel} S.,  et~al., 2012, \mn@doi [\apj] {10.1088/0004-637X/745/1/11}, \href
  {http://adsabs.harvard.edu/abs/2012ApJ...745...11G} {745, 11}

\bibitem[\protect\citeauthoryear{{Genzel} et~al.,}{{Genzel}
  et~al.}{2008}]{Genzel08}
{Genzel} R.,  et~al., 2008, \mn@doi [\apj] {10.1086/591840}, \href
  {http://adsabs.harvard.edu/abs/2008ApJ...687...59G} {687, 59}

\bibitem[\protect\citeauthoryear{{Genzel} et~al.,}{{Genzel}
  et~al.}{2011}]{Genzel11}
{Genzel} R.,  et~al., 2011, \mn@doi [\apj] {10.1088/0004-637X/733/2/101}, \href
  {http://adsabs.harvard.edu/abs/2011ApJ...733..101G} {733, 101}

\bibitem[\protect\citeauthoryear{{Guo}, {Giavalisco}, {Ferguson}, {Cassata}  \&
  {Koekemoer}}{{Guo} et~al.}{2012}]{Guo12}
{Guo} Y.,  {Giavalisco} M.,  {Ferguson} H.~C.,  {Cassata} P.,   {Koekemoer}
  A.~M.,  2012, \mn@doi [\apj] {10.1088/0004-637X/757/2/120}, \href
  {http://adsabs.harvard.edu/abs/2012ApJ...757..120G} {757, 120}

\bibitem[\protect\citeauthoryear{{Guo} et~al.,}{{Guo} et~al.}{2015}]{Guo15}
{Guo} Y.,  et~al., 2015, \mn@doi [\apj] {10.1088/0004-637X/800/1/39}, \href
  {http://adsabs.harvard.edu/abs/2015ApJ...800...39G} {800, 39}

\bibitem[\protect\citeauthoryear{{Hahn} \& {Abel}}{{Hahn} \&
  {Abel}}{2011}]{HahnAbel11}
{Hahn} O.,  {Abel} T.,  2011, \mn@doi [\mnras]
  {10.1111/j.1365-2966.2011.18820.x}, \href
  {http://adsabs.harvard.edu/abs/2011MNRAS.415.2101H} {415, 2101}

\bibitem[\protect\citeauthoryear{{Hopkins} \& {Hernquist}}{{Hopkins} \&
  {Hernquist}}{2010}]{HopkinsHernquist10}
{Hopkins} P.~F.,  {Hernquist} L.,  2010, \mn@doi [\mnras]
  {10.1111/j.1365-2966.2009.15933.x}, \href
  {http://adsabs.harvard.edu/abs/2010MNRAS.402..985H} {402, 985}

\bibitem[\protect\citeauthoryear{{Hopkins} et~al.,}{{Hopkins}
  et~al.}{2010}]{Hopkins10}
{Hopkins} P.~F.,  et~al., 2010, \mn@doi [\apj] {10.1088/0004-637X/724/2/915},
  \href {http://adsabs.harvard.edu/abs/2010ApJ...724..915H} {724, 915}

\bibitem[\protect\citeauthoryear{{Hopkins}, {Quataert}  \& {Murray}}{{Hopkins}
  et~al.}{2011}]{Hopkins11}
{Hopkins} P.~F.,  {Quataert} E.,   {Murray} N.,  2011, \mn@doi [\mnras]
  {10.1111/j.1365-2966.2011.19306.x}, \href
  {http://adsabs.harvard.edu/abs/2011MNRAS.417..950H} {417, 950}

\bibitem[\protect\citeauthoryear{{Hopkins}, {Kere{\v s}}, {Murray}, {Quataert}
  \& {Hernquist}}{{Hopkins} et~al.}{2012}]{Hopkins12}
{Hopkins} P.~F.,  {Kere{\v s}} D.,  {Murray} N.,  {Quataert} E.,   {Hernquist}
  L.,  2012, \mn@doi [\mnras] {10.1111/j.1365-2966.2012.21981.x}, \href
  {http://adsabs.harvard.edu/abs/2012MNRAS.427..968H} {427, 968}

\bibitem[\protect\citeauthoryear{{Hopkins}, {Cox}, {Hernquist}, {Narayanan},
  {Hayward}  \& {Murray}}{{Hopkins} et~al.}{2013}]{Hopkins13}
{Hopkins} P.~F.,  {Cox} T.~J.,  {Hernquist} L.,  {Narayanan} D.,  {Hayward}
  C.~C.,   {Murray} N.,  2013, \mn@doi [\mnras] {10.1093/mnras/stt017}, \href
  {http://adsabs.harvard.edu/abs/2013MNRAS.430.1901H} {430, 1901}

\bibitem[\protect\citeauthoryear{{Hopkins}, {Kere{\v s}}, {O{\~n}orbe},
  {Faucher-Gigu{\`e}re}, {Quataert}, {Murray}  \& {Bullock}}{{Hopkins}
  et~al.}{2014}]{Hopkins14}
{Hopkins} P.~F.,  {Kere{\v s}} D.,  {O{\~n}orbe} J.,  {Faucher-Gigu{\`e}re}
  C.-A.,  {Quataert} E.,  {Murray} N.,   {Bullock} J.~S.,  2014, \mn@doi
  [\mnras] {10.1093/mnras/stu1738}, \href
  {http://adsabs.harvard.edu/abs/2014MNRAS.445..581H} {445, 581}

\bibitem[\protect\citeauthoryear{{Immeli}, {Samland}, {Westera}  \&
  {Gerhard}}{{Immeli} et~al.}{2004}]{Immeli04}
{Immeli} A.,  {Samland} M.,  {Westera} P.,   {Gerhard} O.,  2004, \mn@doi
  [\apj] {10.1086/422179}, \href
  {http://adsabs.harvard.edu/abs/2004ApJ...611...20I} {611, 20}

\bibitem[\protect\citeauthoryear{{Inoue}, {Dekel}, {Mandelker}, {Ceverino},
  {Bournaud}  \& {Primack}}{{Inoue} et~al.}{2016}]{Inoue2016}
{Inoue} S.,  {Dekel} A.,  {Mandelker} N.,  {Ceverino} D.,  {Bournaud} F.,
  {Primack} J.,  2016, \mn@doi [\mnras] {10.1093/mnras/stv2793}, \href
  {http://adsabs.harvard.edu/abs/2016MNRAS.456.2052I} {456, 2052}

\bibitem[\protect\citeauthoryear{{Jog}}{{Jog}}{1996}]{Jog1996}
{Jog} C.~J.,  1996, \mn@doi [\mnras] {10.1093/mnras/278.1.209}, \href
  {http://adsabs.harvard.edu/abs/1996MNRAS.278..209J} {278, 209}

\bibitem[\protect\citeauthoryear{{Jones}, {Swinbank}, {Ellis}, {Richard}  \&
  {Stark}}{{Jones} et~al.}{2010}]{Jones10}
{Jones} T.~A.,  {Swinbank} A.~M.,  {Ellis} R.~S.,  {Richard} J.,   {Stark}
  D.~P.,  2010, \mn@doi [\mnras] {10.1111/j.1365-2966.2010.16378.x}, \href
  {http://adsabs.harvard.edu/abs/2010MNRAS.404.1247J} {404, 1247}

\bibitem[\protect\citeauthoryear{{Kere{\v s}} \& {Hernquist}}{{Kere{\v s}} \&
  {Hernquist}}{2009}]{Keres09}
{Kere{\v s}} D.,  {Hernquist} L.,  2009, \mn@doi [\apjl]
  {10.1088/0004-637X/700/1/L1}, \href
  {http://adsabs.harvard.edu/abs/2009ApJ...700L...1K} {700, L1}

\bibitem[\protect\citeauthoryear{{Kere{\v s}}, {Katz}, {Weinberg}  \&
  {Dav{\'e}}}{{Kere{\v s}} et~al.}{2005}]{Keres05}
{Kere{\v s}} D.,  {Katz} N.,  {Weinberg} D.~H.,   {Dav{\'e}} R.,  2005, \mn@doi
  [\mnras] {10.1111/j.1365-2966.2005.09451.x}, \href
  {http://adsabs.harvard.edu/abs/2005MNRAS.363....2K} {363, 2}

\bibitem[\protect\citeauthoryear{{Kim} et~al.,}{{Kim} et~al.}{2014}]{Kim14}
{Kim} J.-h.,  et~al., 2014, \mn@doi [\apjs] {10.1088/0067-0049/210/1/14}, \href
  {http://adsabs.harvard.edu/abs/2014ApJS..210...14K} {210, 14}

\bibitem[\protect\citeauthoryear{{Kroupa}}{{Kroupa}}{2002}]{Kroupa02}
{Kroupa} P.,  2002, \mn@doi [Science] {10.1126/science.1067524}, \href
  {http://adsabs.harvard.edu/abs/2002Sci...295...82K} {295, 82}

\bibitem[\protect\citeauthoryear{{Krumholz} \& {Dekel}}{{Krumholz} \&
  {Dekel}}{2010}]{KrumholzDekel10}
{Krumholz} M.~R.,  {Dekel} A.,  2010, \mn@doi [\mnras]
  {10.1111/j.1365-2966.2010.16675.x}, \href
  {http://adsabs.harvard.edu/abs/2010MNRAS.406..112K} {406, 112}

\bibitem[\protect\citeauthoryear{{Larson}}{{Larson}}{1981}]{Larson81}
{Larson} R.~B.,  1981, \mnras, \href
  {http://adsabs.harvard.edu/abs/1981MNRAS.194..809L} {194, 809}

\bibitem[\protect\citeauthoryear{{Leitherer} et~al.,}{{Leitherer}
  et~al.}{1999}]{Leitherer99}
{Leitherer} C.,  et~al., 1999, \mn@doi [\apjs] {10.1086/313233}, \href
  {http://adsabs.harvard.edu/abs/1999ApJS..123....3L} {123, 3}

\bibitem[\protect\citeauthoryear{{Livermore} et~al.,}{{Livermore}
  et~al.}{2012}]{Livermore2012}
{Livermore} R.~C.,  et~al., 2012, \mn@doi [\mnras]
  {10.1111/j.1365-2966.2012.21900.x}, \href
  {http://adsabs.harvard.edu/abs/2012MNRAS.427..688L} {427, 688}

\bibitem[\protect\citeauthoryear{{Livermore} et~al.,}{{Livermore}
  et~al.}{2015}]{Livermore2015}
{Livermore} R.~C.,  et~al., 2015, \mn@doi [\mnras] {10.1093/mnras/stv686},
  \href {http://adsabs.harvard.edu/abs/2015MNRAS.450.1812L} {450, 1812}

\bibitem[\protect\citeauthoryear{{Ma}, {Hopkins}, {Faucher-Gigu{\`e}re},
  {Zolman}, {Muratov}, {Kere{\v s}}  \& {Quataert}}{{Ma} et~al.}{2016}]{Ma15}
{Ma} X.,  {Hopkins} P.~F.,  {Faucher-Gigu{\`e}re} C.-A.,  {Zolman} N.,
  {Muratov} A.~L.,  {Kere{\v s}} D.,   {Quataert} E.,  2016, \mn@doi [\mnras]
  {10.1093/mnras/stv2659}, \href
  {http://adsabs.harvard.edu/abs/2016MNRAS.456.2140M} {456, 2140}

\bibitem[\protect\citeauthoryear{{Mandelker}, {Dekel}, {Ceverino}, {Tweed},
  {Moody}  \& {Primack}}{{Mandelker} et~al.}{2014}]{Mandelker14}
{Mandelker} N.,  {Dekel} A.,  {Ceverino} D.,  {Tweed} D.,  {Moody} C.~E.,
  {Primack} J.,  2014, \mn@doi [\mnras] {10.1093/mnras/stu1340}, \href
  {http://adsabs.harvard.edu/abs/2014MNRAS.443.3675M} {443, 3675}

\bibitem[\protect\citeauthoryear{{Mandelker}, {Dekel}, {Ceverino}, {DeGraf},
  {Guo}  \& {Primack}}{{Mandelker} et~al.}{2016}]{Mandelker15}
{Mandelker} N.,  {Dekel} A.,  {Ceverino} D.,  {DeGraf} C.,  {Guo} Y.,
  {Primack} J.,  2016, \mn@doi [\mnras] {10.1093/mnras/stw2358}, \href
  {http://adsabs.harvard.edu/abs/2016MNRAS.tmp.1460M} {}

\bibitem[\protect\citeauthoryear{{Moody}, {Guo}, {Mandelker}, {Ceverino},
  {Mozena}, {Koo}, {Dekel}  \& {Primack}}{{Moody} et~al.}{2014}]{Moody14}
{Moody} C.~E.,  {Guo} Y.,  {Mandelker} N.,  {Ceverino} D.,  {Mozena} M.,  {Koo}
  D.~C.,  {Dekel} A.,   {Primack} J.,  2014, \mn@doi [\mnras]
  {10.1093/mnras/stu1534}, \href
  {http://adsabs.harvard.edu/abs/2014MNRAS.444.1389M} {444, 1389}

\bibitem[\protect\citeauthoryear{{Muratov}, {Kere{\v s}},
  {Faucher-Gigu{\`e}re}, {Hopkins}, {Quataert}  \& {Murray}}{{Muratov}
  et~al.}{2015}]{Muratov15}
{Muratov} A.~L.,  {Kere{\v s}} D.,  {Faucher-Gigu{\`e}re} C.-A.,  {Hopkins}
  P.~F.,  {Quataert} E.,   {Murray} N.,  2015, \mn@doi [\mnras]
  {10.1093/mnras/stv2126}, \href
  {http://adsabs.harvard.edu/abs/2015MNRAS.454.2691M} {454, 2691}

\bibitem[\protect\citeauthoryear{{Murray}}{{Murray}}{2011}]{Murray11}
{Murray} N.,  2011, \mn@doi [\apj] {10.1088/0004-637X/729/2/133}, \href
  {http://adsabs.harvard.edu/abs/2011ApJ...729..133M} {729, 133}

\bibitem[\protect\citeauthoryear{{Murray}, {Quataert}  \& {Thompson}}{{Murray}
  et~al.}{2010}]{Murray10}
{Murray} N.,  {Quataert} E.,   {Thompson} T.~A.,  2010, \mn@doi [\apj]
  {10.1088/0004-637X/709/1/191}, \href
  {http://adsabs.harvard.edu/abs/2010ApJ...709..191M} {709, 191}

\bibitem[\protect\citeauthoryear{{Noguchi}}{{Noguchi}}{1999}]{Noguchi99}
{Noguchi} M.,  1999, \mn@doi [\apj] {10.1086/306932}, \href
  {http://adsabs.harvard.edu/abs/1999ApJ...514...77N} {514, 77}

\bibitem[\protect\citeauthoryear{{Ostriker} \& {Shetty}}{{Ostriker} \&
  {Shetty}}{2011}]{Ostriker11}
{Ostriker} E.~C.,  {Shetty} R.,  2011, \mn@doi [\apj]
  {10.1088/0004-637X/731/1/41}, \href
  {http://cdsads.u-strasbg.fr/abs/2011ApJ...731...41O} {731, 41}

\bibitem[\protect\citeauthoryear{{Roman-Duval}, {Jackson}, {Heyer}, {Rathborne}
   \& {Simon}}{{Roman-Duval} et~al.}{2010}]{RomanDuval10}
{Roman-Duval} J.,  {Jackson} J.~M.,  {Heyer} M.,  {Rathborne} J.,   {Simon} R.,
   2010, \mn@doi [\apj] {10.1088/0004-637X/723/1/492}, \href
  {http://adsabs.harvard.edu/abs/2010ApJ...723..492R} {723, 492}

\bibitem[\protect\citeauthoryear{{Somerville}, {Primack}  \&
  {Faber}}{{Somerville} et~al.}{2001}]{Somerville01}
{Somerville} R.~S.,  {Primack} J.~R.,   {Faber} S.~M.,  2001, \mn@doi [\mnras]
  {10.1046/j.1365-8711.2001.03975.x}, \href
  {http://adsabs.harvard.edu/abs/2001MNRAS.320..504S} {320, 504}

\bibitem[\protect\citeauthoryear{{Springel}}{{Springel}}{2005}]{Springel05}
{Springel} V.,  2005, \mn@doi [\mnras] {10.1111/j.1365-2966.2005.09655.x},
  \href {http://adsabs.harvard.edu/abs/2005MNRAS.364.1105S} {364, 1105}

\bibitem[\protect\citeauthoryear{{Stewart}, {Bullock}, {Barton}  \&
  {Wechsler}}{{Stewart} et~al.}{2009}]{Stewart09}
{Stewart} K.~R.,  {Bullock} J.~S.,  {Barton} E.~J.,   {Wechsler} R.~H.,  2009,
  \mn@doi [\apj] {10.1088/0004-637X/702/2/1005}, \href
  {http://adsabs.harvard.edu/abs/2009ApJ...702.1005S} {702, 1005}

\bibitem[\protect\citeauthoryear{{Swinbank} et~al.,}{{Swinbank}
  et~al.}{2010}]{Swinbank10}
{Swinbank} A.~M.,  et~al., 2010, \mn@doi [\nat] {10.1038/nature08880}, \href
  {http://adsabs.harvard.edu/abs/2010Natur.464..733S} {464, 733}

\bibitem[\protect\citeauthoryear{{Tacconi} et~al.,}{{Tacconi}
  et~al.}{2008}]{Tacconi08}
{Tacconi} L.~J.,  et~al., 2008, \mn@doi [\apj] {10.1086/587168}, \href
  {http://adsabs.harvard.edu/abs/2008ApJ...680..246T} {680, 246}

\bibitem[\protect\citeauthoryear{{Tacconi} et~al.,}{{Tacconi}
  et~al.}{2010}]{Tacconi10}
{Tacconi} L.~J.,  et~al., 2010, \mn@doi [\nat] {10.1038/nature08773}, \href
  {http://adsabs.harvard.edu/abs/2010Natur.463..781T} {463, 781}

\bibitem[\protect\citeauthoryear{{Tacconi} et~al.,}{{Tacconi}
  et~al.}{2013}]{Tacconi13}
{Tacconi} L.~J.,  et~al., 2013, \mn@doi [\apj] {10.1088/0004-637X/768/1/74},
  \href {http://adsabs.harvard.edu/abs/2013ApJ...768...74T} {768, 74}

\bibitem[\protect\citeauthoryear{{Tamburello}, {Mayer}, {Shen}  \&
  {Wadsley}}{{Tamburello} et~al.}{2015}]{Tamburello15}
{Tamburello} V.,  {Mayer} L.,  {Shen} S.,   {Wadsley} J.,  2015, \mn@doi
  [\mnras] {10.1093/mnras/stv1695}, \href
  {http://adsabs.harvard.edu/abs/2015MNRAS.453.2490T} {453, 2490}

\bibitem[\protect\citeauthoryear{{Toomre}}{{Toomre}}{1964}]{Toomre64}
{Toomre} A.,  1964, \mn@doi [\apj] {10.1086/147861}, \href
  {http://adsabs.harvard.edu/abs/1964ApJ...139.1217T} {139, 1217}

\bibitem[\protect\citeauthoryear{{Wuyts} et~al.,}{{Wuyts}
  et~al.}{2012}]{Wuyts12}
{Wuyts} S.,  et~al., 2012, \mn@doi [\apj] {10.1088/0004-637X/753/2/114}, \href
  {http://adsabs.harvard.edu/abs/2012ApJ...753..114W} {753, 114}

\bibitem[\protect\citeauthoryear{{van den Bergh}, {Abraham}, {Ellis}, {Tanvir},
  {Santiago}  \& {Glazebrook}}{{van den Bergh} et~al.}{1996}]{vdBergh96}
{van den Bergh} S.,  {Abraham} R.~G.,  {Ellis} R.~S.,  {Tanvir} N.~R.,
  {Santiago} B.~X.,   {Glazebrook} K.~G.,  1996, \mn@doi [\aj]
  {10.1086/118020}, \href {http://adsabs.harvard.edu/abs/1996AJ....112..359V}
  {112, 359}

\makeatother
\end{thebibliography}




\appendix
\section{Clump properties in additional FIRE simulations of massive galaxies at redshift 2}
\label{sec:massivefire}

\begin{table}
	\centering
	\caption{Physical parameters for the inner 10 kpc of \textit{m13} and the additional MassiveFIRE galaxies, averaged over $\sim 150$ Myr at $z\sim 2.0$.}
	\label{tab:galaxy_properties}
	\begin{tabular}{|lcccc|} 
		\hline
		name& M$_*$ [\msun]& M$_{gas}$ [\msun]& $f_g$ &  SFR [\msun yr$^{-1}$]\\
		\hline
		m13&  $3.8\times 10^{10}$ & $1.6\times 10^{10}$ & 0.29 & 44\\
		MFz2\_A1 & $2.3\times 10^{10}$ & $0.4\times 10^{10}$ & 0.13 &  5\\
		MFz2\_A3 & $1.1\times 10^{10}$ & $0.6\times 10^{10}$ & 0.32 & 4\\
		\hline
	\end{tabular}
\end{table}

\begin{figure*}
\centering
\includegraphics[width=0.45\textwidth]{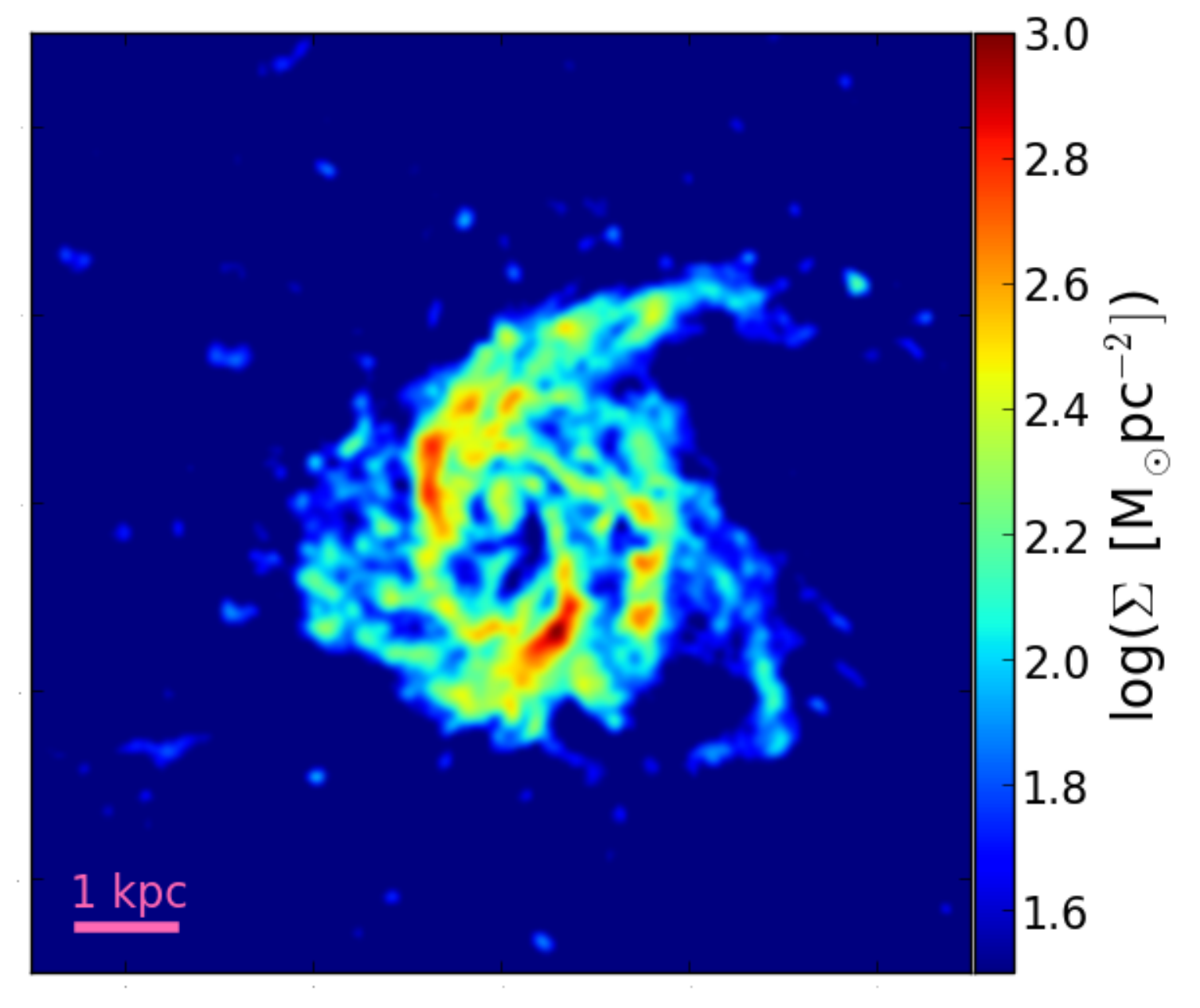}
\includegraphics[width=0.45\textwidth]{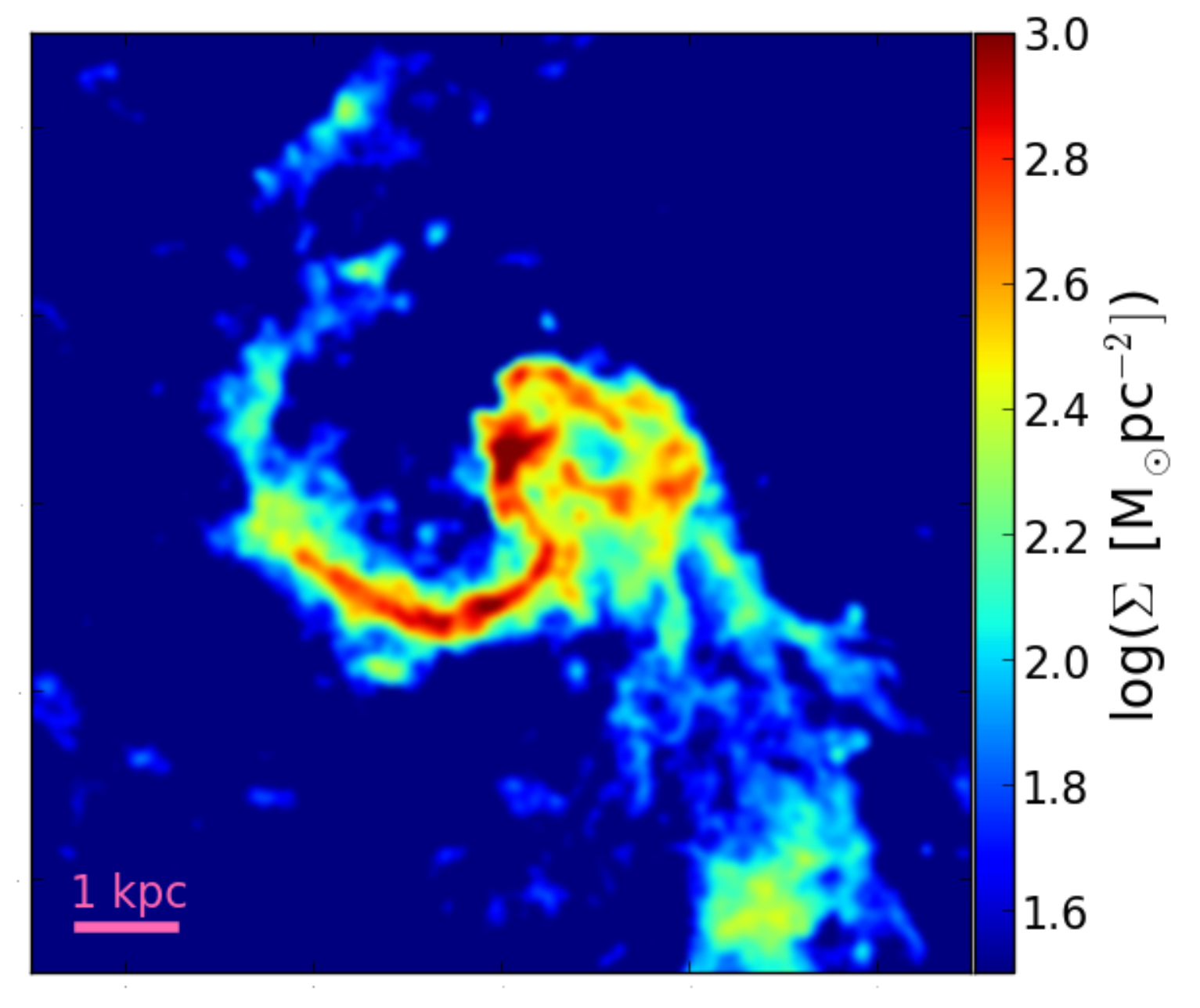}
\caption{Gas surface density in the face-on projection at $z\sim 2$ for the two additional simulations, \textit{MFz2\_A1} on the left and \textit{MFz2\_A3} on the right panel.}
\label{fig:massive_gas_sd}
\end{figure*}

\begin{figure*}
\centering
\includegraphics[width=0.9\textwidth]{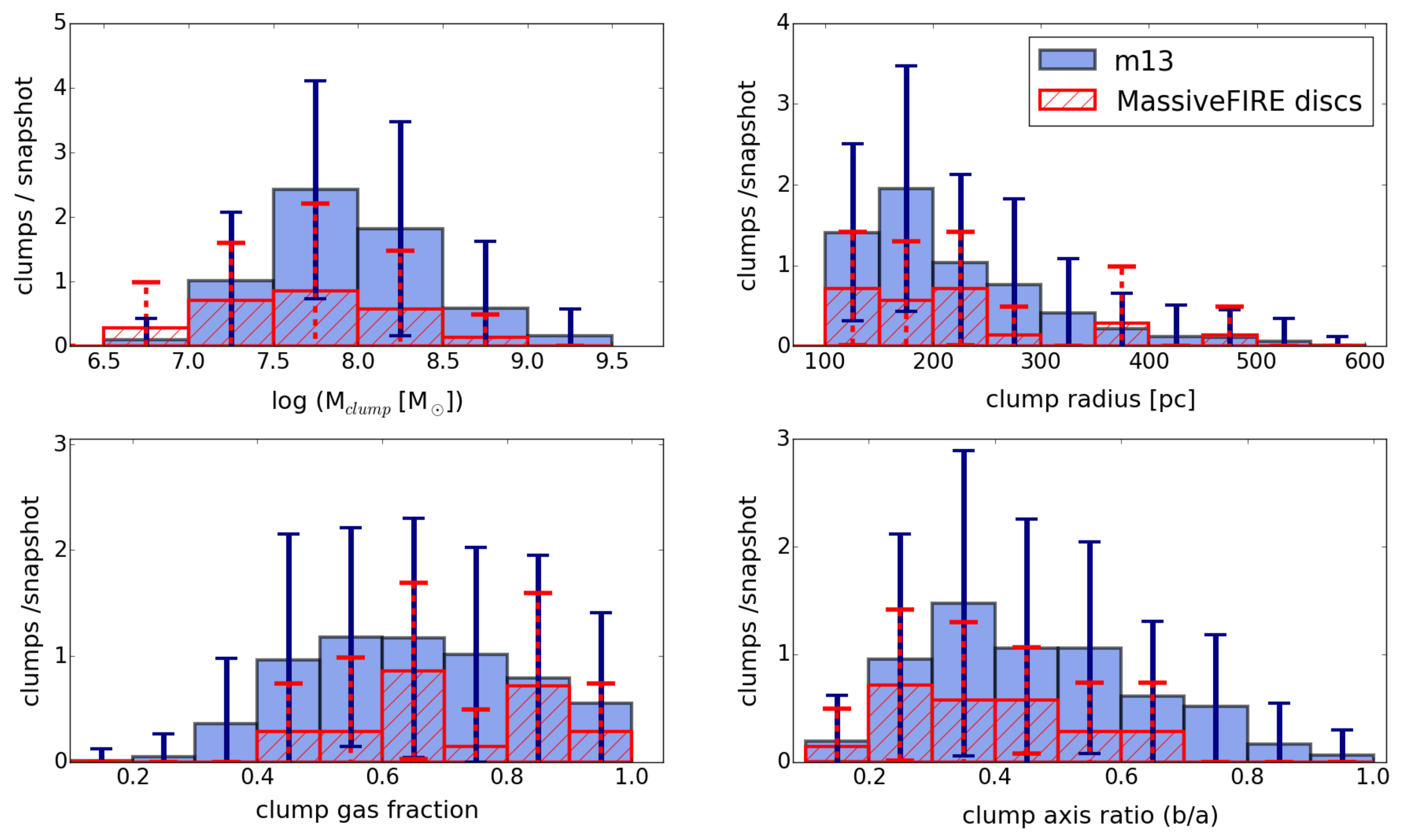}
\caption{Distribution of clump properties, same as in Fig. \ref{fig:4figs}. Blue histograms show the average number of clumps in each bin, calculated using all clumps from the \textit{m13} simulation. Error bars mark the standard deviation for each bin. The same is shown in red for clumps in $z\sim 2$ \textit{MassiveFIRE} snapshots with disc-like gas distributions. The mean number of \textit{MassiveFIRE} clumps in each bin falls within $1\sigma$ of the \textit{m13} result.}
\label{fig:4figs_massive}
\end{figure*}

Simulation \textit{m13} is the most massive galaxy in the original FIRE suite of simulations that were evolved all the way to $z=0$. These simulations produce many galaxy properties that match observations across a broad redshift range \citep{Hopkins14}. Simulation \textit{m13} is the only galaxy in that sample with mass similar to the mass of observed clumpy galaxies at high redshift (M$_* \sim$ $10^{10}-10^{11}$\msun). To evolve such massive systems with high resolution down to $z=0$ is computationally demanding, so \textit{m13} was run at a lower resolution than the other original FIRE zoom-in simulations.

Because our analysis is based on one galaxy, there is a possibility that this system is a peculiar outlier from the rest of the population of similar galaxies. To test this, we compare the properties of \textit{m13} at $z=2.0$ with the properties of a couple of additional massive galaxies evolved down to $z\sim 2$, using the same code and the same implementation of stellar feedback, but at higher resolution. These galaxies (called \textit{MFz2\_A1} and \textit{MFz2\_A1}) are part of the \textit{MassiveFIRE} sample \citep{Feldmann16}, a suite of high-resolution simulations of massive galaxies at high-redshift ($z \gtrsim 1.7$). Their mass resolution is 16 times better (i.e. particle mass is 16 times smaller) than that of \textit{m13}. Table \ref{tab:galaxy_properties} contains some of the basic properties of \textit{m13} and the additional galaxies at $z=2$.

These particular simulations were chosen because at $z\sim 2$ they do not show obvious signs of ongoing major mergers and appear to have discy morphologies. The gas surface densities of these two galaxies at $z=2$ are shown in Fig. \ref{fig:massive_gas_sd}.  Upon further investigation, we noticed that in many $z\sim 2$ snapshots these galaxies do not have well-defined gas discs (unlike \textit{m13} which has a substantial gas disc throughout the redshift range analysed in this work) and their gaseous material seems much more erratic than that of \textit{m13}. Hence, to make the comparison of clump properties more robust, we only use snapshots in which the additional simulated galaxies have disc-like gas distributions. This criterion however reduces our sample of `good' snapshots to the total of seven (four from \textit{MFz2\_A3} and three from \textit{MFz2\_A1}).

We apply the clump-finding procedure described in Section \ref{sec:clump_identification}
\footnote{We had to slightly modify the way we compute the gas surface density threshold used in the clump-finding procedure to account for higher resolution (i.e. smaller particle mass) of \textit{MassiveFIRE} simulations. In order to achieve the surface density threshold values in the \textit{MassiveFIRE} snapshots similar to those we use for \textit{m13}, in the high-resolution simulations we calculate the threshold using all cells with eight or more gas particles, instead of all cells with at least one gas particle. } to the snapshots of additional simulations and compare the clump properties to those of \textit{m13}. The resulting distributions are shown in Fig. \ref{fig:4figs_massive}. Blue histograms show the average number of clumps per bin for all \textit{m13} snapshots that contain gas clumps. The error bars show the standard deviation ($1\sigma$) for each bin. Results from the seven disc-like snapshots of \textit{MassiveFIRE} simulations are shown in red. Even though we have a fairly small sample of suitable \textit{MassiveFIRE} snapshots taken over a small redshift range, the average number of \textit{MassiveFIRE} clumps per snapshot in each bin falls within $1\sigma$ of the \textit{m13} average, suggesting that clumps properties in the \textit{m13} and \textit{MassiveFIRE} simulations are similar.

Although we do not include them in Fig. \ref{fig:4figs_massive}, galaxies with non-discy gas distributions might be an interesting topic for future studies in the context of high-redshift clumpy galaxies. Applying the same analysis as for the snapshots with disc-like gas morphology, we found that some of these highly irregular gas distributions break up into gas clumps. These clumps might be a result of outflows and winds, unlike clumps in discs which most likely form via gravitational instability. Preliminary analysis of clumps in non-discy snapshots indicate that they might be more gas-dominated compared to clumps found in discs. A detailed investigation into the properties of clumps in non-discy galaxies, and how they compare to clumps observed in high-redshift galaxies is left for future work.

\section{Sensitivity to the clump finder parameters}
\label{sec:sensitivity}

In this section we analyze the sensitivity of clump properties on the gas surface density threshold used to define clumps in our clump-finding procedure. There is no single definition of a galaxy clump that is universally adopted, although attempts have been made to come up with one \citep[see e.g.][]{Guo15}. Different groups use different criteria to define clumps and to measure their properties. Here we show that our choice of parameters used in the definition of clumps, although arbitrary to some extent, does not significantly affect the main result of our study.
 
The threshold value of the gas surface density that we use to define clumps is calculated for each snapshot as a 1$\sigma$ above the mean surface density of non-empty grid cells in an \textit{non-smoothed} 10-kpc map centered on \textit{m13}. This definition of the threshold value was chosen after trying out many different options and evaluating each one by visual assessment. In other words, we chose the definition of clumps that seemed to work best at identifying regions of the disc that look like clumps in the maps of the gas surface density. This method is similar to what is done in observations, where clumps are usually identified by visual assessment.

In order to quantify how strongly our clumps are contrasted with respect to the rest of the disc, we compare the used threshold values to surface densities of smoothed maps on different scales. The used surface density threshold values correspond to regions that are overdense by factors of $\sim 10-15$ or $\sim 5 \sigma$ above the mean surface density of pixels in the \textit{smoothed} 10 kpc maps. However, if we consider a smaller region around the galaxy, both the mean and the standard deviation of the gas surface density increase and the nominal overdensity becomes lower. For example, our used threshold values correspond to a factor of $\sim 4-8$ or $\sim 3 \sigma$ overdensities when the average values and sigmas are computed on a smoothed 5 kpc map around the center of the galaxy. This ambiguity in the density contrast of clumps is expected to be present not only in simulations, but also in observations of high-redshift galaxies, in which the definition of a clump may change depending on the sensitivity of observations.

To test whether a small change in the clump definition would have a significant impact on the derived properties of clumps, we re-analyze the original snapshots in the redshift range $2.0 \geq z > 1.8$ (20 snapshots in total) with different clump-finding criteria. We change the parameters of our clump finder in order to identify clumps that are 4, 5 and 6 standard deviations above the mean value, in the smoothed 10 kpc maps of the gas surface density. The average difference between the threshold values used in this comparison is about $18\%$ (i.e. $5\sigma$ threshold values are higher(lower) than the $4\sigma$($6\sigma$) values by $\sim 18\%$).

\begin{figure}
\centering
\includegraphics[width=0.45\textwidth]{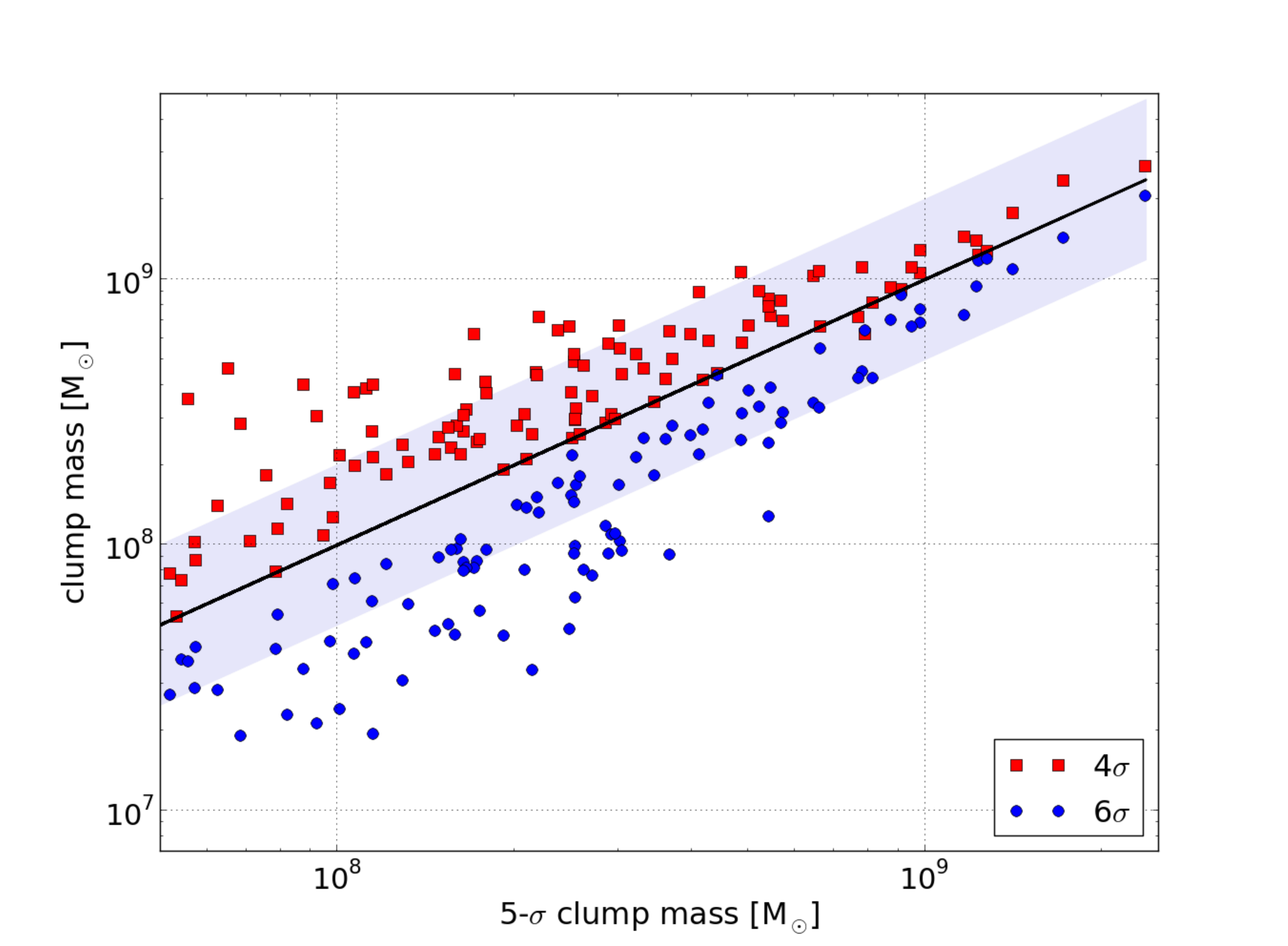}
\caption{Comparison of clump mass found using different parameters in the clump-finding algorithm. Masses of clumps defined as 4$\sigma$ and 6$\sigma$ overdensities in the smoothed gas surface density maps are shown against the mass of the same clump found using the 5$\sigma$ definition. If the masses were equal, they would fall on the black line. The colored band marks the region where the 4$\sigma$ and 6$\sigma$ masses are within a factor of 2 from the 5$\sigma$ mass of the same clump. Most clumps populate this region, indicating that our estimates of clump mass are fairly robust to changing the clump definition.}
\label{fig:mass_sigmas}
\end{figure}

Fig. \ref{fig:mass_sigmas} compares the 5$\sigma$ baryonic mass of each clump to the mass of the corresponding clump identified with the 4$\sigma$ and 6$\sigma$ definitions, shown in red and blue, respectively. For most clumps, especially the most massive ones, the total mass varies by less than a factor of two between the 5$\sigma$ definition and the 4$\sigma$ and 6$\sigma$ definitions.

\begin{figure}
\centering
\includegraphics[width=0.4\textwidth]{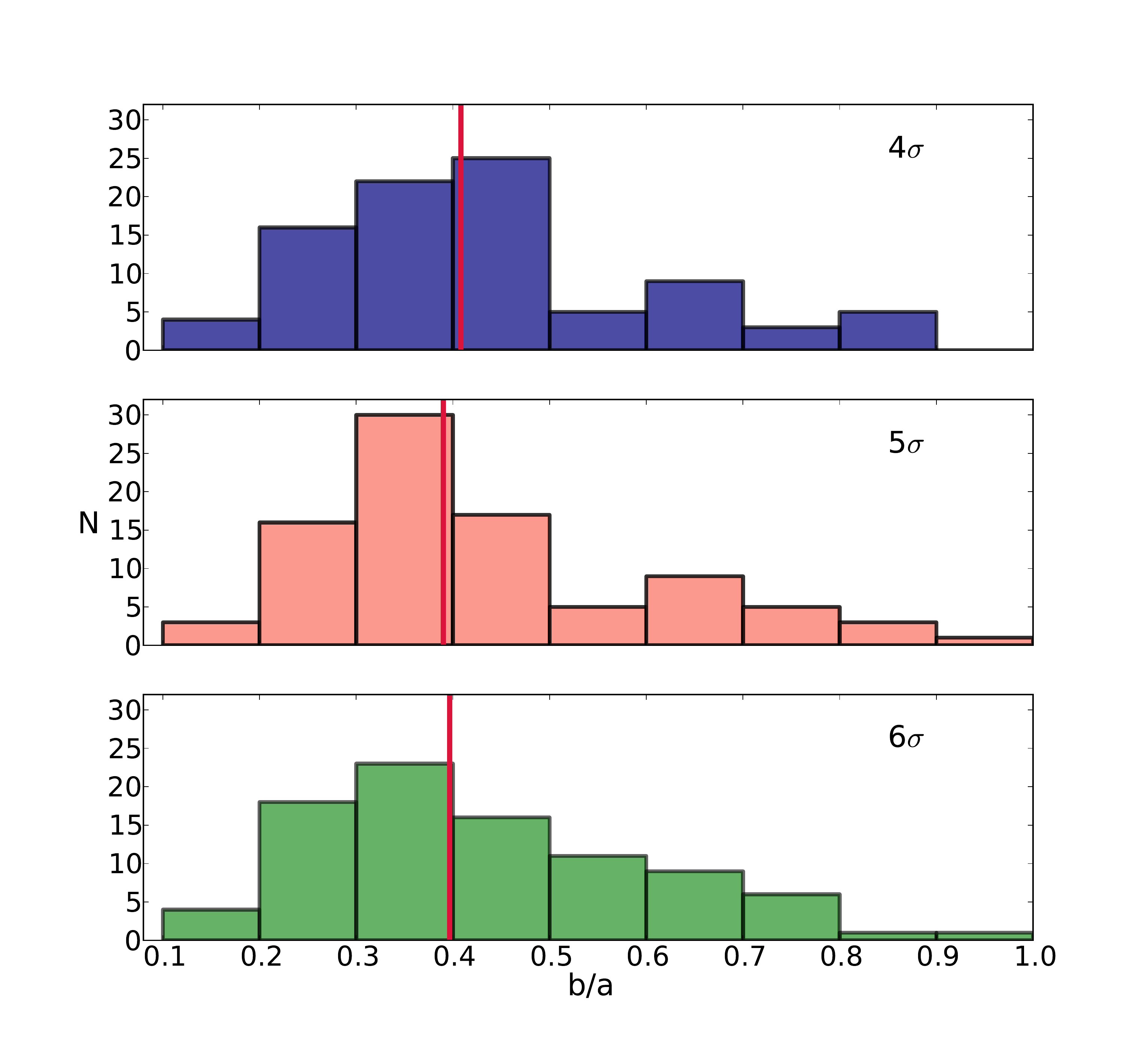}
\includegraphics[width=0.4\textwidth]{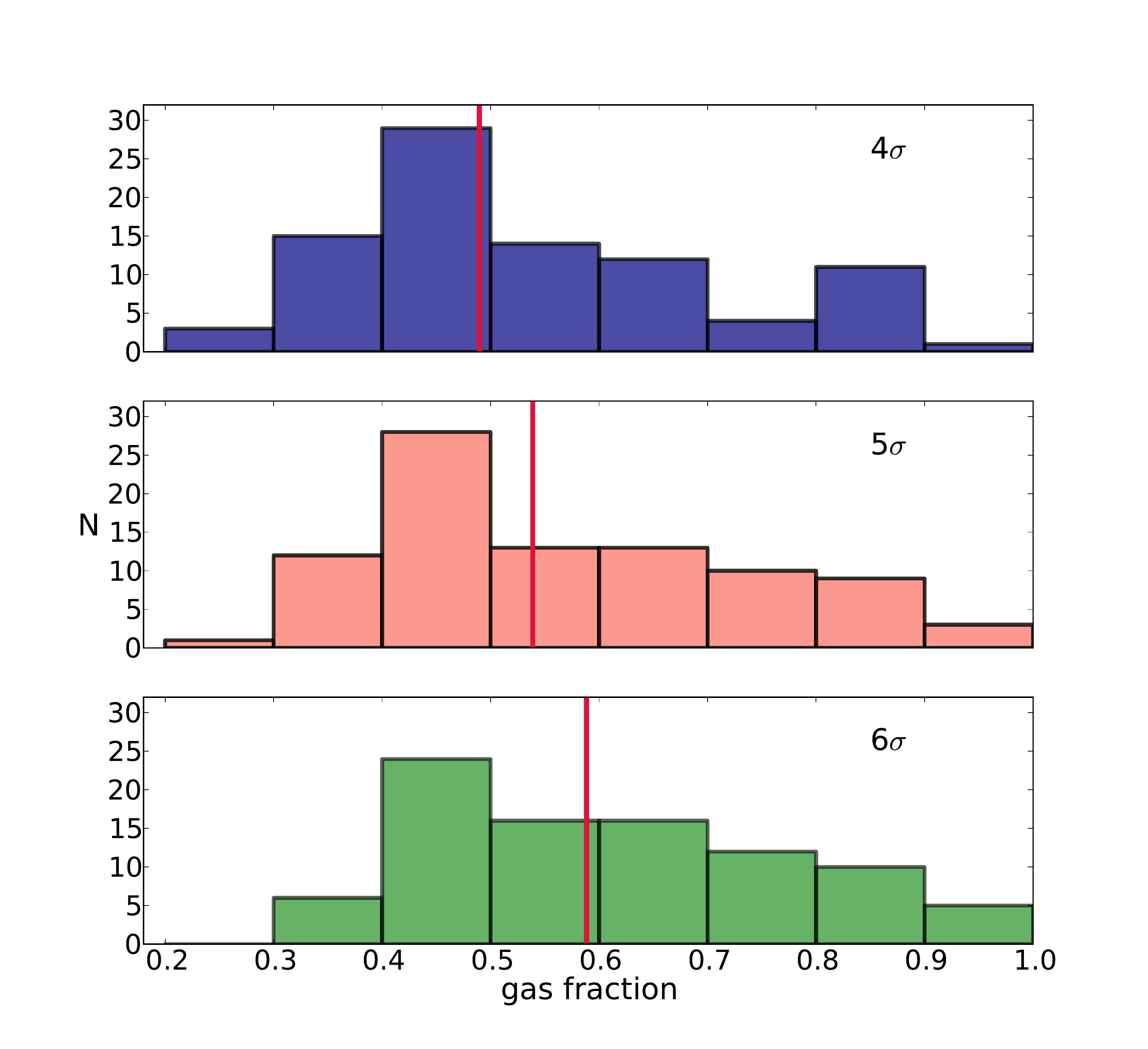}
\includegraphics[width=0.4\textwidth]{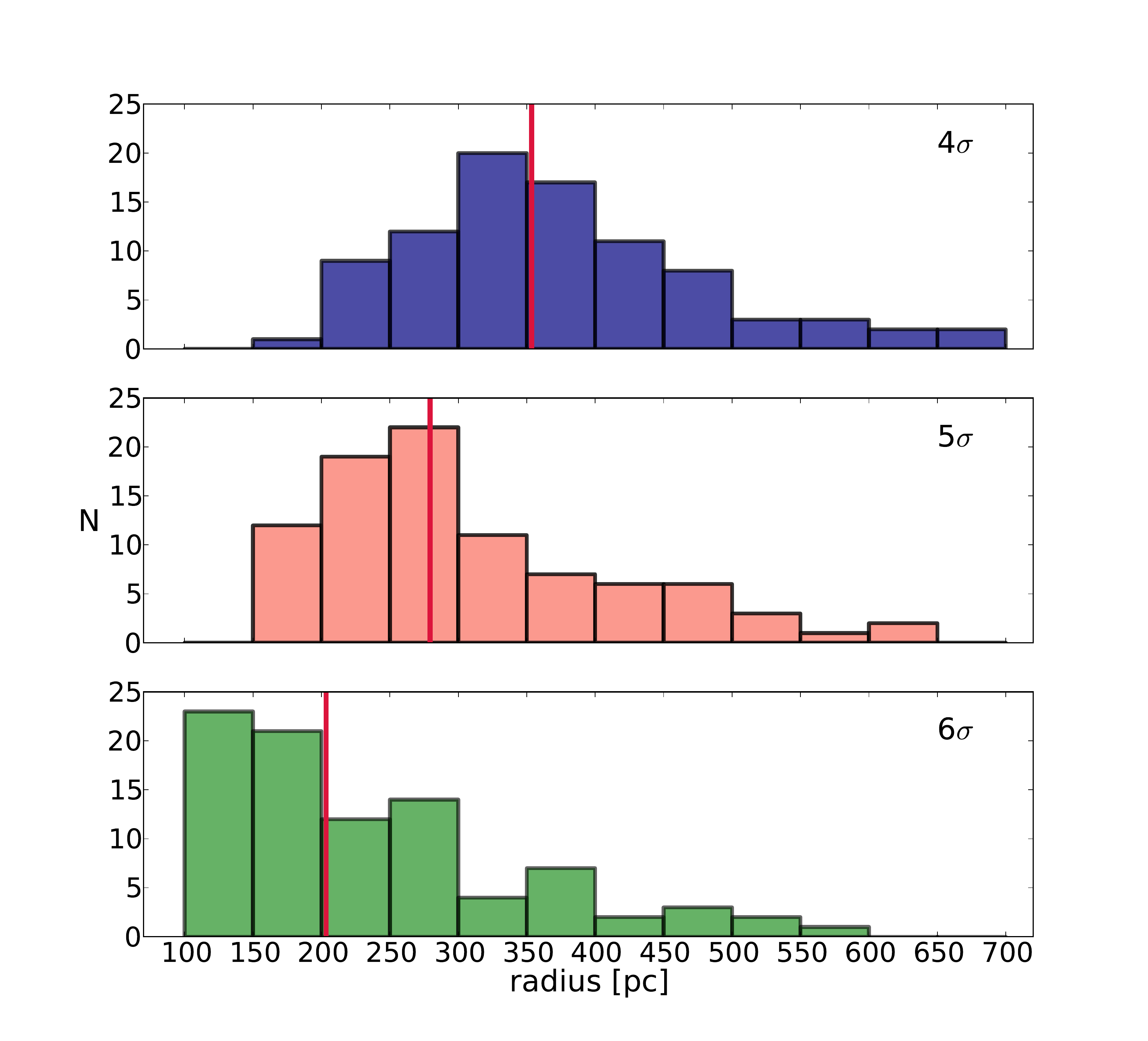}
\caption{Distributions of clump properties (minor-to-major axis ratio, gas fraction, and clump radius) resulting from using different parameters (4$\sigma$, 5$\sigma$, and 6$\sigma$ thresholds in the gas surface density) in the clump definition. Red lines show the median values for each case.}
\label{fig:other_sigmas}
\end{figure}

\begin{figure}
\centering
\includegraphics[width=0.45\textwidth]{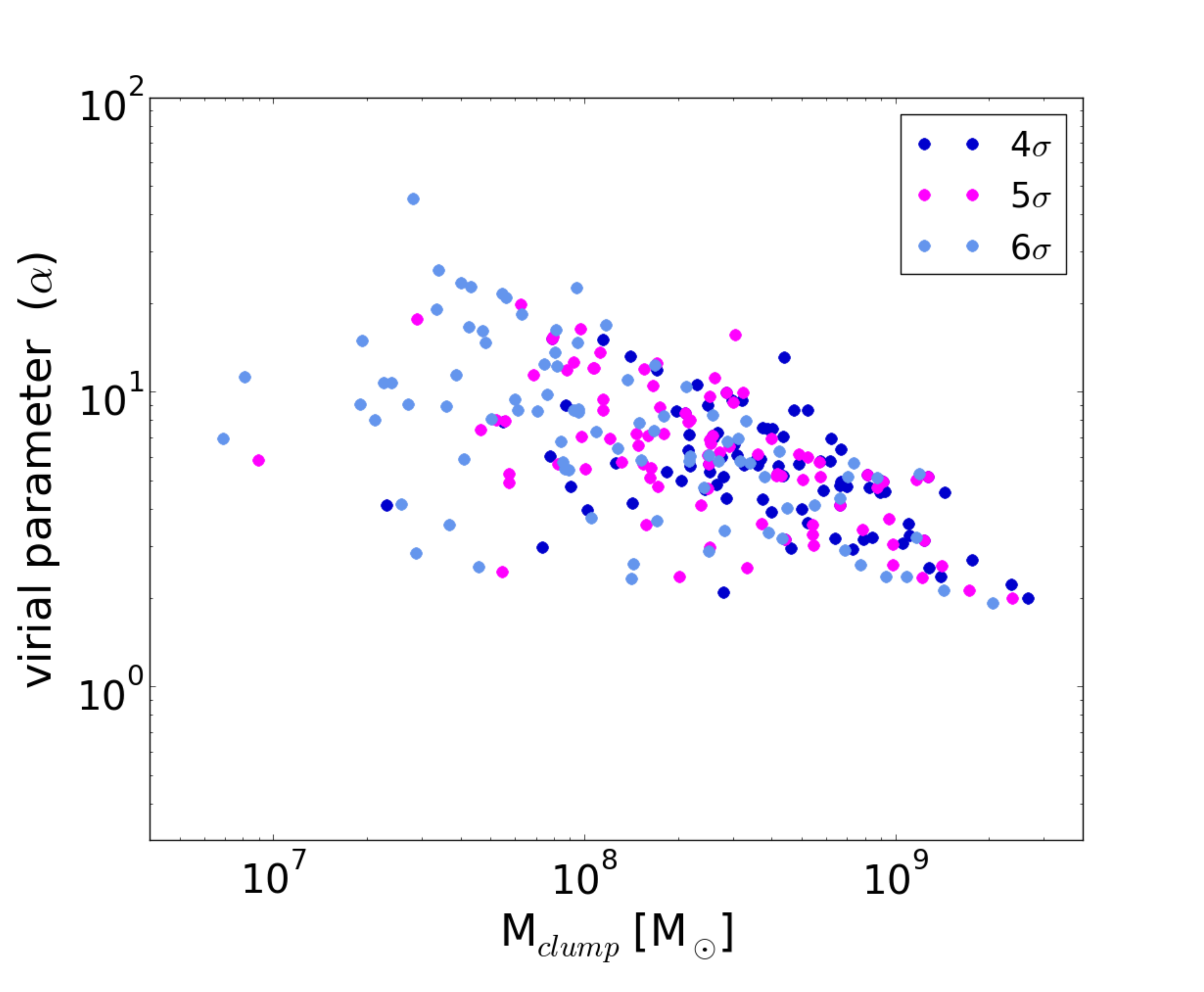}
\caption{Virial parameter of a clump as a function of its mass, found using three different parameters in the clump-finding procedure (gas surface density thresholds set to 4$\sigma$, 5$\sigma$, and 6$\sigma$ above the mean value). There is no systematic difference in the clump virial parameter caused by using different definitions.}
\label{fig:virial_sigmas}
\end{figure}

Fig. \ref{fig:other_sigmas} shows how distributions of different clump properties change with changing the clump definition. Properties like the minor-to-major axis ratio and the gas fraction change very little. The distribution of clump radii seems to be more affected by changing the clump definition, however the median values remain within a few tens of percent. In Fig. \ref{fig:virial_sigmas} we compare the virial parameter of clumps found using these three definitions, and we find that they mostly overlap. Hence, changes in the clump definition do not affect the (un-)bound state of clumps in a significant way.

In conclusion, although specific properties of any individual clump may change with changing the clump definition, these tests suggest that our main findings are not sensitive to small changes in the clump definition.


\bsp	
\label{lastpage}
\end{document}